\newif\ifpdf
\ifx\pdfoutput\undefined
\pdffalse % we are not running PDFLaTeX
\else
\pdfoutput=1 % we are running PDFLaTeX
\pdftrue \fi

\documentclass[envcountsame,envcountsect,11pt]{svmultmodified}

\usepackage{makeidx}         % allows index generation
\usepackage{graphicx}        % standard LaTeX graphics tool
\usepackage{multicol}        % used for the two-column index
\usepackage[bottom]{footmisc}% places footnotes at page bottom
\usepackage{amssymb,amsfonts,amsmath}
\usepackage{epsf}
\usepackage{calc,xcolor,amscd,amsfonts,epsfig,amsmath,amssymb,enumerate}
\usepackage{verbatim}
\usepackage{mathpazo}
\usepackage{mathrsfs}
\usepackage{graphicx}

\DeclareFontFamily{OT1}{cmss}{} \DeclareFontShape{OT1}{cmss}{m}{n}
{<5> <6> <7> <8> <9> <10> <11> <12> <13> <14.4> cmss10}{}
\DeclareMathAlphabet{\cmss}{OT1}{cmss}{m}{n}

\DeclareFontFamily{OT1}{fraktura}{}
\DeclareFontShape{OT1}{fraktura}{m}{n} {<5> <6> <7> <8> <9> <10>
<11> <12> <13> <14.4> [1.1] eufm10}{}
\DeclareMathAlphabet{\fraktura}{OT1}{fraktura}{m}{n}

\newcommand{\textd}{\text{\rm d}\mkern0.5mu}
\newcommand{\texti}{\text{\rm i}\mkern0.7mu}

\newcommand{\Var}{{\text{\rm Var}}}

\renewcommand{\AA}{\mathcal A}
\newcommand{\BB}{\mathcal B}

\newcommand{\DD}{\mathcal D}
\newcommand{\EE}{\mathcal E}

\newcommand{\II}{\mathcal I}

\newcommand{\NN}{\mathcal N}

\newcommand{\CalS}{\mathcal S}

\newcommand{\A}{\mathbb A}

\newcommand{\C}{\mathbb C}
\renewcommand{\E}{\mathbb E}

\renewcommand{\I}{\mathbb I}

\newcommand{\N}{\mathbb N}

\newcommand{\BbbP}{\mathbb P}

\newcommand{\R}{\mathbb R}
\newcommand{\BbbS}{\mathbb S}
\newcommand{\T}{\mathbb T}

\newcommand{\Z}{\mathbb Z}

\newcommand{\scrS}{\mathscr{S}}

\newcommand{\bk}{\boldsymbol k}

\def\myffrac#1#2 in #3{\raise 2.6pt\hbox{$#3 #1$}\mkern-1.5mu\raise 0.8pt\hbox{$#3/$}\mkern-1.1mu\lower 1.5pt\hbox{$#3 #2$}}
\newcommand{\ffracB}[2]{\mathchoice%
	{\myffrac{#1}{#2} in \scriptstyle}
	{\myffrac{#1}{#2} in \scriptstyle}
	{\myffrac{#1}{#2} in \scriptscriptstyle}
	{\myffrac{#1}{#2} in \scriptscriptstyle}
}

\newcommand{\uemph}[1]{\underline{\emph{#1}}}
\newcommand{\betac}{\beta_\cc}
\newcommand{\betat}{\beta_{\text{\rm t}}}
\newcommand{\frakG}{\mathfrak G}

\newcommand{\hate}{{\text{\rm \^e}}}

\newcommand{\hatv}{{\text{\rm \^v}}}
\newcommand{\hatw}{{\text{\rm \^w}}}

\newcommand{\hatb}{{\text{\rm \^b}}}

\newcommand{\twoeqref}[2]{(\ref{#1}--\ref{#2})}

\newcommand{\cc}{{\text{\rm c}}}

\newcommand{\bigtimes}{\operatornamewithlimits{\displaystyle\times}}

\newcommand{\texte}{\text{\rm e}}

\begin{document}

\title*{Reflection Positivity and Phase Transitions\\in Lattice Spin Models}
\titlerunning{Phase transitions in lattice models} 

\author{Marek Biskup}

\institute{Department of Mathematics, UCLA, Los Angeles, CA 90095-1555\\
\texttt{biskup@math.ucla.edu}}

\maketitle

% THIS IS TO OVERRIDE THEIR DEFINITIONS WITH SECRET VARIABLES
\def\theequation{\arabic{section}.\arabic{equation}}
\newenvironment{proofsectB}[1]
 {\vskip0.1cm\noindent{\rmfamily\itshape#1.}}{\qed\vspace{0.15cm}}%{\newline\vspace{0.15cm}}

\numberwithin{equation}{section}
\def\theequation{\arabic{section}.\arabic{equation}}

%\tableofcontents

\subsection*{Contents}
\setcounter{minitocdepth}{2}
\dominitoc

\section{Introduction}
\addcontentsline{toc}{section}{{\numberline {}Introduction}{}}

Phase transitions are one of the most fascinating, and also most perplexing, phenomena in equilibrium statistical mechanics. On the physics side, many approximate methods to explain or otherwise justify phase transitions are known but a complete mathematical understanding is available only in a handful of simplest of all cases. One set of tractable systems consists of the so called \emph{lattice spin models}. Originally, these came to existence as simplified versions of (somewhat more realistic) models of crystalline materials in solid state physics but their versatile nature earned them a life of their own in many other disciplines where complex systems are of interest.

The present set of notes describes one successful \emph{mathematical} approach to phase transitions in lattice spin models which is based on the technique of \emph{reflection positivity}. This technique was developed in the late 1970s in the groundbreaking works of F.~Dyson, J.~Fr\"ohlich, R.~Israel, E.~Lieb, B.~Simon and T.~Spencer who used it to establish phase transitions in a host of physically-interesting classical and quantum lattice spin models; most notably, the classical Heisenberg ferromagnet and the quantum XY model and Heisenberg antiferromagnet. Other powerful techniques --- e.g., Pirogov-Sinai theory, lace expansion or multiscale analysis in field theory --- are available at present that can serve a similar purpose in related contexts, but we will leave their review to experts in those areas.

The most attractive feature of reflection positivity --- especially, compared to the alternative techniques --- is the simplicity of the resulting proofs. There are generally two types of arguments one can use: The first one is to derive the so called \emph{infrared bound}, which states in quantitative terms that the fluctuations of the spin variables are dominated by those of a lattice Gaussian free field. In systems with an internal symmetry, this yields a proof of a symmetry-breaking phase transition by way of a spin-condensation argument. Another route goes via the so called \emph{chessboard estimates}, which allow one to implement a Peierls-type argument regardless of whether the model exhibits an internal symmetry or not. 

Avid users of the alternative techniques are often quick to point out that the simplicity of proofs has its price: As a rather restrictive condition, reflection positivity applies only to a small (in a well defined sense) class of systems. Fortunately for the technique and mathematical physics in general,  the models to which it does apply constitute a large portion of what is interesting for physics, \emph{and} to physicists. Thus, unless one is exclusively after universal statements --- i.e., those robust under rather arbitrary perturbations --- the route via reflection positivity is often fairly satisfactory.

The spectacular success of reflection positivity from the late 1970s was followed by many interesting developments. For instance, in various joint collaborations, R.~Dobrushin, R.~Koteck\'y and S.~Shlosman showed how chessboard estimates can be used to prove a phase transition in a class of systems with naturally-defined ordered and disordered components; most prominently, the $q$-state Potts model for $q\gg1$. Another neat application came in the papers of M.~Aizenman from early 1980s in which he combined the infrared bound with his random-current representation to conclude mean-field critical behavior in the nearest-neighbor Ising ferromagnet above 4 dimensions. Yet another example is the proof, by L.~Chayes, R.~Koteck\'y and S.~Shlosman, that the Fisher-renormalization scheme in annealed-diluted systems may be substituted by the emergence of an intermediate phase.

These notes discuss also more recent results where their author had a chance to contribute to the field. The common ground for some of these is the use of reflection positivity to provide mathematical justification of ``well-known'' conclusions from physics folklore. For instance, in papers by N.~Crawford, L.~Chayes and the present author, the infrared bound was shown to imply that, once a model undergoes a field or energy driven first-order transition in mean-field theory, a similar transition will occur in the lattice model provided the spatial dimension is sufficiently high or the interaction is sufficiently spread-out (but still reflection positive). Another result --- due to L.~Chayes, S.~Starr and the present author --- asserts that if a reflection-positive quantum spin system undergoes a phase transition at intermediate temperatures in its classical limit, a similar transition occurs in the quantum system provided the magnitude of the quantum spin is sufficiently~large. 

There have also been recent cases where reflection positivity brought a definite end to a controversy that physics arguments were not able to resolve. One instance concerned certain non-linear vector and liquid-crystal models; it was debated whether a transition can occur already in 2 dimensions. This was settled in recent work of A.~van Enter and S.~Shlosman. Another instance involved spin systems whose (infinite) set of ground states had a much larger set of symmetries than the Hamiltonian of the model; two competing physics reasonings argued for, and against, the survival of these states at low temperatures. Here, in papers of L.~Chayes, S. Kivelson, Z.~Nussinov and the present author, spin-wave free energy calculations were combined with chessboard estimates to construct a rigorous proof of phase coexistence of only a \emph{finite} number of low-temperature states.

These recent activities show that the full potential of reflection positivity may not yet have been fully exhausted and that the technique will continue to play an important role in mathematical statistical mechanics. It is hoped that the present text will help newcomers to this field learn the essentials of the subject before the need arises to plow through the research papers where the original derivations first appeared.

\subsection*{Organization}
This text began as class notes for nine hours of lectures on reflection positivity at the 2006 Prague School and gradually grew into a survey of (part of) this research area. The presentation opens with a review of basic facts about lattice spin models and then proceeds to study two applications of the infrared bound: a spin-condensation argument and a link to mean-field theory. These are followed by the classical derivation of the infrared bound from reflection positivity. The remainder of the notes is spent on applications of a by-product of this derivation, the chessboard estimate, to proofs of phase coexistence. 

The emphasis of the notes is on a \emph{pedagogical} introduction to reflection positivity; for this reason we often sacrifice on generality and rather demonstrate the main ideas on the simplest case of interest. To compensate for the inevitable loss of generality, each chapter is endowed with a section ``Literature remarks'' where we attempt to list the references deemed most relevant to the topic at hand. The notes are closed with a short section on topics that are not covered as well as some open problems that the author finds worthy of some thought.

\subsection*{Acknowledgments}
\addcontentsline{toc}{section}{{\numberline {}Acknowledgments}{}}
A review naturally draws on the work of many authors. We will do our best to give proper credit to their contribution in the sections ``Literature remarks'' that appear at the end of each chapter and, of course, in the list of references. I have personally learned a great deal of the subject from Lincoln Chayes with whom I have subsequently coauthored more than half-a-dozen papers some of which will be discussed here. During this time I also benefited from collaborations with N.~Crawford, S.~Kivelson, R.~Koteck\'y, Z.~Nussinov and S.~Starr, and from various enlightening discussions involving A.~van Enter, B.~Nachtergaele and S.~Shlosman.

The text would presumably never exist were it not for Roman Koteck\'y's summer school; I wish to thank him for allowing me to speak on this subject. I am indebted to the participants of the school for comments during the lectures and to T. Bodineau, A.~van Enter, E.~Lieb and S.~Shlosman for suggestions on the first draft of the notes. My presence at the school was made possible thanks to the support from the ESF-program ``Phase Transitions and Fluctuation Phenomena for Random Dynamics in Spatially Extended Systems'' and from the National Science Foundation under the grant~DMS-0505356.

%%%%%%%%%%%%%%%%%%%%%%%%%%

\section[Lattice spin models]{Lattice spin models: Crash course}

This section prepares the ground for the rest of the course by introducing the main concepts from the theory of Gibbs measures for lattice spin models\index{spin model}. The results introduced here are selected entirely for the purpose of these notes; readers wishing a more comprehensive --- and in-depth --- treatment should consult classic textbooks on the subject. 

\subsection{Basic setup}
Let us start discussing the setup of the models to which we will direct our attention throughout this course. The basic ingredients are as follows:
\begin{itemize}
\item
\emph{Lattice}: We will take the $d$-dimensional hypercubic lattice $\Z^d$ as our underlying graph. This is the graph with vertices at all points in $\R^d$ with integer coordinates and edges between any \emph{nearest neighbor} pair of vertices; i.e., those at Euclidean distance one. We will use $\langle x,y\rangle$ to denote an (unordered) nearest-neighbor pair.
\item
\emph{Spins}: At each $x\in\Z^d$ we will consider a spin\index{spin} $S_x$, by which we will mean a random variable taking values in a closed subset~$\Omega$ of~$\R^\nu$, for some~$\nu\ge1$. We will use~$S_x\cdot S_y$ to denote a scalar product between~$S_x$ and~$S_y$ (Euclidean or otherwise).
\item
\emph{Spin configurations}: For $\Lambda\subset\Z^d$, we will refer to~$S_\Lambda:=(S_x)_{x\in\Lambda}$ as the spin configuration\index{spin configuration} in~$\Lambda$. We will be generically interested in describing the statistical properties of such spin configurations with respect to certain (canonical) measures.
\item
\emph{Boundary conditions}:
To describe the law of~$S_\Lambda$, we will not be able to ignore that some spins are also outside~$\Lambda$. We will refer to the configuration~$S_{\Lambda^\cc}$ of these spins as the boundary condition\index{boundary condition}. The latter will usually be fixed and may often even be considered a parameter of the game. When both $S_\Lambda$ and~$S_{\Lambda^\cc}$ are known, we will write
\begin{equation}
%\label{}
S:=(S_\Lambda,S_{\Lambda^\cc})
\end{equation}
to denote their concatenation on all of~$\Z^d$.
\end{itemize}
The above setting incorporates rather varied physical contexts. The spins may be thought of as describing magnetic moments of atoms in a crystal, displacement of atoms from their equilibrium positions or even orientation of grains in nearly-crystalline granular materials.

To define the dynamics of spin systems, we will need to specify the energetics. This is conveniently  done by prescribing the \emph{Hamiltonian} \index{Hamiltonian} which is a function on the spin-configuration space $\Omega^{\Z^d}$ that tells us how much energy each spin configuration has. Of course, to have all quantities well defined we need to fix a \emph{finite} volume $\Lambda\subset\Z^d$ and compute only the energy in $\Lambda$. The most general formula we will ever need is
\begin{equation}
%\label{}
H_\Lambda(S):=\sum_{\begin{subarray}{c}
A\subset\Z^d\text{ finite}\\A\cap\Lambda\ne\emptyset
\end{subarray}}
\Phi_A(S)
\end{equation}
where $\Phi_A$ is a function that depends only on~$S_A$. To make everything well defined, we require, e.g., that~$\Phi_A$ is translation invariant and that $\sum_{A\ni0}\Vert\Phi_A\Vert_\infty<\infty$. (The infinity norm may be replaced by some other norm; in particular, should the need arise to talk about unbounded spins.) It is often more convenient to write the above as a formal sum:
\begin{equation}
%\label{}
H(S):=\sum_A\Phi_A(S)
\end{equation}
with the above specific understanding whenever a precise definition is desired.

The energy is not sufficient on its own to define the statistical mechanics of such spin systems; we also need to specify the \emph{a priori measure} on the spins. This will be achieved by prescribing a Borel measure~$\mu_0$ on~$\Omega$ (which may or may not be finite). Before the interaction is ``switched on,'' the spin configurations will be ``distributed'' according to the product measure, i.e., the \emph{a priori} law of~$S_\Lambda$ is~$\bigotimes_{x\in\Lambda}\mu_0(\textd S_x)$. The full statistical-mechanical law is then given by a \emph{Gibbs measure} which (in finite volume) takes the general form $\texte^{-\beta H(S)}\prod_x\mu_0(\textd S_x)$; cf Sect.~\ref{sec1.3} for more details.

\subsection{Examples}
\label{sec1.2}
Here are a few examples of spin systems:

\bigskip\noindent
(1) \uemph{$O(n)$-model}: \index{$O(n)$ model} Here $\Omega:=\BbbS^{n-1}=\{z\in\R^n\colon|z|_2=1\}$ with $\mu_0$ := surface measure on $\BbbS^{n-1}$. The Hamiltonian is
\begin{equation}
\label{H-1.4}
H(S):=-J\sum_{\langle x,y\rangle}S_x\cdot S_y
\end{equation}
where the dot denotes the usual (Euclidean) dot-product in~$\R^n$ and~$J\ge0$. (Note that this comes at no loss as the sign of~$J$ can be changed by reversing the spins on the odd sublattice of~$\Z^d$.) 

Note that if~$A\in O(n)$ --- i.e., $A$ is an $n$-dimensional orthogonal matrix --- then
\begin{equation}
%\label{}
AS_x\cdot AS_y=S_x\cdot S_y
\end{equation}
and so $H(AS)=H(S)$. Since also~$\mu_0\circ A^{-1}=\mu_0$, the model possesses a \emph{global rotation invariance} --- with respect to a simultaneous rotation of all spins. (For $n=1$ this reduces to the invariance under the flip $+1\leftrightarrow-1$.)

Two instances of this model are known by other names: $n=2$ is the \emph{rotor model} while $n=3$ is the (classical) \emph{Heisenberg ferromagnet}.\index{Heisenberg model}

\bigskip\noindent
(2) \uemph{Ising model}: \index{Ising model} Formally, this is the $O(1)$-model. Explicitly, the spin variables~$\sigma_x$ take values in~$\Omega:=\{-1,+1\}$ with uniform \emph{a priori} measure; the Hamiltonian is
\begin{equation}
%\label{}
H(\sigma):=-J\sum_{\langle x,y\rangle}\sigma_x\sigma_y
\end{equation}
Note that the energy is smaller when the spins at nearest neighbors align and higher when they antialign. (A similar statement holds, of course, for all~$O(n)$ models.) This is due to the choice of the sign~$J\ge0$ which makes these models \emph{ferromagnets}.

\bigskip\noindent
(3) \uemph{Potts model}:\index{Potts model}
This is a generalization of the Ising model beyond two spin states. Explicitly, we fix~$q\in\N$ and let $\sigma_x$ take values in $\{1,\dots,q\}$ (with a uniform \emph{a priori} measure). The Hamiltonian is
\begin{equation}
\label{H-1.7}
H(\sigma):=-J\sum_{\langle x,y\rangle}\delta_{\sigma_x,\sigma_y}
\end{equation}
so the energy is $-J$ when $\sigma_x$ and $\sigma_y$ ``align'' and zero otherwise. The $q=2$ case is the Ising model and $q=1$ may be related to bond percolation on~$\Z^d$ (via the so called \emph{Fortuin-Kasteleyn representation} leading to the so called \emph{random-cluster model}\index{random-cluster model}).

It turns out that the Hamiltonian \eqref{H-1.7} can be brought to the form \eqref{H-1.4}. Indeed, let $\Omega$ denote the set of $q$ points uniformly spread on the unit sphere in~$\R^{q-1}$; we may think of these as the vertices of a $q$-simplex (or a regular $q$-hedron). The cases $q=2,3,4$ are depicted in this figure:

\vglue0.3cm
\centerline{\,\,\includegraphics[width=4.1in]{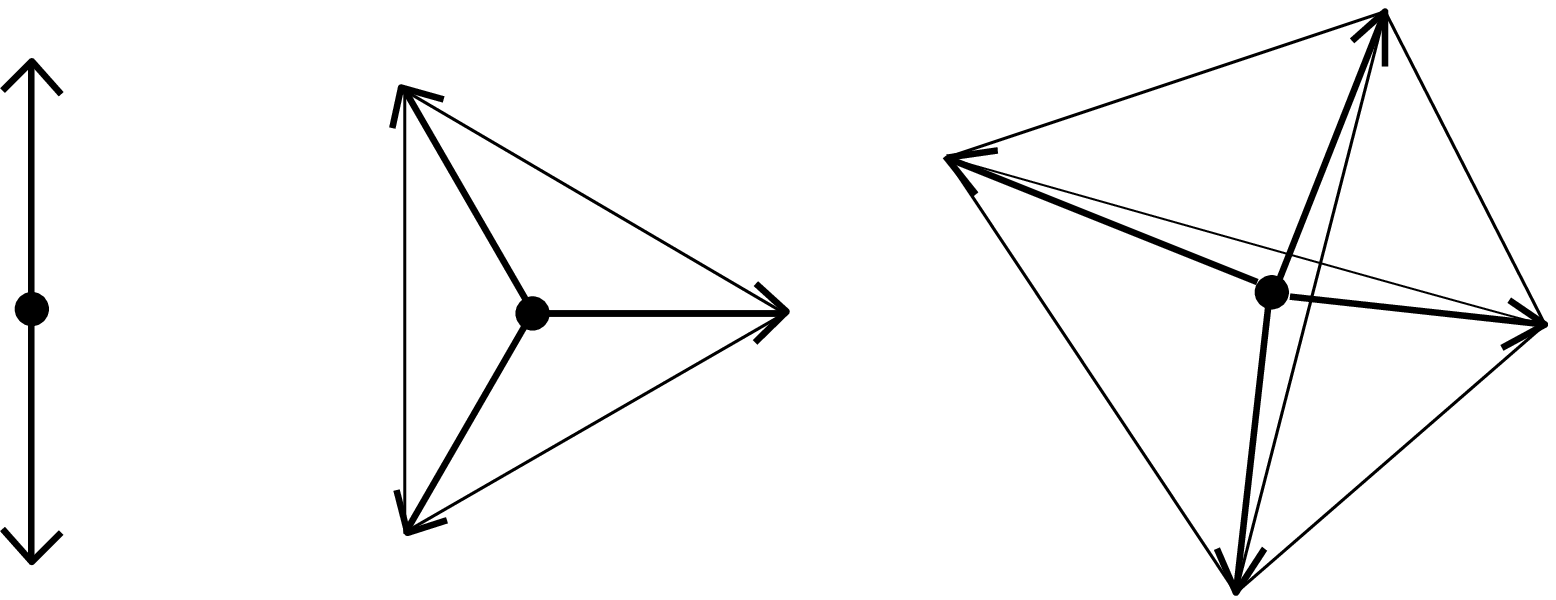}}

\noindent
More explicitly, the elements of~$\Omega$ are vectors~$\hatv_\alpha$, $\alpha=1,\dots,q$, such that
\begin{equation}
\hatv_\alpha\cdot\hatv_\beta=\begin{cases}1,\qquad&\text{if
}\alpha=\beta,\\
-\frac1{q-1},\qquad&\text{otherwise.}
\end{cases}
\end{equation}
The existence of such vectors can be proved by induction on~$q$. Clearly, if~$S_x$ corresponds to~$\sigma_x$ and~$S_y$ to~$\sigma_y$, then
\begin{equation}
%\label{}
S_x\cdot S_y=\frac q{q-1}\delta_{\sigma_x,\sigma_y}-\frac1{q-1}
\end{equation}
and so the Potts Hamiltonian is to within an additive constant of
\begin{equation}
%\label{}
H(S):=-\tilde J\sum_{\langle x,y\rangle}S_x\cdot S_y
\end{equation}
with~$\tilde J:=J\frac{q-1}q$. This form, sometimes referred to as \emph{tetrahedral representation}, will be far more useful for our purposes than \eqref{H-1.7}.

\bigskip\noindent
(4) \uemph{Liquid-crystal model}: \index{liquid-crystal model} There are many models that describe certain granular materials known to many of us from digital displays: liquid crystals. A distinguished feature of such materials is the presence of \emph{orientational long-range order} where a majority of the grains align with one another despite the fact that the system as a whole is rotationally invariant. One of the simplest models capturing this phenomenon is as follows: Consider spins $S_x\in\BbbS^{n-1}$ with a uniform \emph{a priori} measure. The Hamiltonian is
\begin{equation}
%\label{}
H(S):=-J\sum_{\langle x,y\rangle}(S_x\cdot S_y)^2
\end{equation}
The interaction features global rotation invariance and the square takes care of the fact that reflection of any spin does not change the energy (i.e., only the \emph{orientation} rather than the \emph{direction} of the spin matters).

As for the Potts model, the Hamiltonian can again be brought to the form reminiscent of the $O(n)$-model. Indeed, given a spin~$S\in\BbbS^{n-1}$ with Cartesian components~$S^{(\alpha)}$, $\alpha=1,\dots,n$, define an $n\times n$ matrix $Q$ by
\begin{equation}
\label{MB-Qmatrix}
Q_{\alpha\beta}:=S^{(\alpha)}S^{(\beta)}-\frac1n\delta_{\alpha\beta}
\end{equation}
(The subtraction of the identity is rather arbitrary and more or less unnecessary; its goal is to achieve zero trace and thus reduce the number of independent variables characterizing~$Q$ to $n-1$; i.e., as many degrees of freedom as~$S$ has.)
As is easy to check, if $Q\leftrightarrow S$ and $\tilde Q\leftrightarrow\tilde S$ are related via the above formula, then
\begin{equation}
%\label{}
\text{Tr}(Q\tilde Q)=(S\cdot\tilde S)^2-\frac1n
\end{equation}
Since $Q$ is symmetric, the trace evaluates to
\begin{equation}
%\label{}
\text{Tr}(Q\tilde Q)=\sum_{\alpha,\beta}Q_{\alpha\beta}\tilde Q_{\alpha\beta}
\end{equation}
which is the canonical scalar product on $n\times n$ matrices. In this language the Hamiltonian takes again the form we saw in the $O(n)$ model.

\bigskip
At this point we pause to remark that all of the above Hamiltonians are of the following rather general form:
\begin{equation}
\label{H-spins}
H(S)=+\frac12\sum_{x,y}J_{x,y}|S_x-S_y|^2
\end{equation}
where $(J_{xy})$ is a collection of suitable \emph{coupling constants} and $|\cdot|$ denotes the Euclidean norm in~$\R^n$. This is possible because, in all cases, the (corresponding) norm of~$S_x$ is constant and so adding it to the Hamiltonian has no effect on the probability measure. The model thus obtained bears striking similarity to our last example:

\bigskip\noindent
(5) \uemph{Lattice Gaussian free field}:\index{Gaussian!free field}
Let $\Omega:=\R$, $\mu_0$ := Lebesgue measure and let $\cmss P(x,y)$ be the transition kernel of a symmetric random walk on~$\Z^d$; i.e.,~$\cmss P(x,y)=\cmss P(0,y-x)=\cmss P(0,x-y)$. \index{transition kernel} In this case we will denote the variables by~$\phi_x$; the Hamiltonian is
\begin{equation}
\label{H-GFF}
H(\phi):=\frac12\sum_{x,y}\cmss P(x,y)(\phi_y-\phi_x)^2
\end{equation}
This can be rewritten as
\begin{equation}
%\label{}
H(\phi)=\bigl(\phi,(1-\cmss P)\phi\bigr)_{L^2(\Z^d)} =\colon\,\EE_{1-\cmss P}(\phi,\phi)
\end{equation}
where experts on harmonic analysis of Markov chains will recognize $\EE_{1-\cmss P}(\phi,\phi)$ to be the \emph{Dirichlet form} \index{Dirichlet form} associated with the generator $1-\cmss P$ of the above random walk. In the Gibbs measure, the law of the~$\phi_x$'s will be Gaussian with grad-squared interactions; hence the name of the model.

\bigskip
The sole difference between \eqref{H-spins} and \eqref{H-GFF} is that, unlike the~$\phi_x$'s, the spins~$S_x$ are generally confined to a subset of a Euclidean space and/or their \emph{a priori} measure is not Lebesgue --- which will ultimately mean their law is \emph{not} Gaussian. One purpose of this course is to show how this formal similarity can nevertheless be exploited to provide information on the models \eqref{H-spins}.

\subsection{Gibbs formalism}
\label{sec1.3}\noindent
Now we are ready to describe the statistical-mechanical properties of the above models for which we resort to the formalism of Gibbs-Boltzmann distributions. First we define these in finite volume: Given a finite set~$\Lambda\subset\Z^d$ and a boundary condition~$S_{\Lambda^\cc}$ we define the \emph{Gibbs measure} \index{Gibbs measure} in $\Lambda$ to be the probability measure on~$\Omega^\Lambda$~given~by
\begin{equation}
%\label{}
\mu_{\Lambda,\beta}^{(S_{\Lambda^\cc})}(\textd S_\Lambda)
:=\frac{\texte^{-\beta H_\Lambda(S)}}{Z_{\Lambda,\beta}(S_{\Lambda^\cc})}\prod_{x\in\Lambda}\mu_0(\textd S_x)
\end{equation}
Here~$\beta\ge0$ is the \emph{inverse temperature} \index{inverse temperature} --- in physics terms, $\beta:=\frac1{\text{k}_{\text{B}}T}$ where~$\text{k}_{\text{B}}$ is the Boltzmann constant and~$T$ is the temperature measured in Kelvins --- and~$Z_{\Lambda,\beta}(S_{\Lambda^\cc})$ is the normalization constant called the \emph{partition function}. \index{partition function}
 
To extend this concept to infinite volume we have two options:
%\settowidth{\leftmargini}{(11)}
\begin{enumerate}
\item[(1)]
Consider all possible weak cluster points of the family $\{\mu_{\Lambda,\beta}^{(S_{\Lambda^\cc})}\}$ as~$\Lambda\uparrow\Z^d$ (with the boundary condition possibly varying with~$\Lambda$) and all convex combinations thereof.
\item[(2)]
Identify a distinguishing property of Gibbs measures and use it to define infinite volume objects directly.
\end{enumerate}
While approach (1) is ultimately very useful in practical problems, option (2) is more elegant at this level of generality. The requisite ``distinguishing property'' is as follows:

\begin{lemma}[DLR condition]
\label{lemma1.1}\index{DLR condition}
Let~$\Lambda\subset\Delta\subset\Z^d$ be finite sets and let~$S_{\Delta^\cc}\in\Omega^{\Delta^\cc}$. Then (for $\mu_{\Delta,\beta}^{(S_{\Delta^\cc})}$-a.e.\ $S_{\Lambda^\cc}$),
\begin{equation}
%\label{}
\mu_{\Delta,\beta}^{(S_{\Delta^\cc})}\bigl(\,\cdot\,\big|S_{\Lambda^\cc}\bigr)=
\mu_{\Lambda,\beta}^{(S_{\Lambda^\cc})}(\cdot)
\end{equation}In simple terms, conditioning the Gibbs measure in~$\Delta$ on the configuration in~$\Delta\setminus\Lambda$, we get the Gibbs measure in~$\Lambda$ with the corresponding boundary condition.
\end{lemma}

This leads to:

\begin{definition}[DLR Gibbs measures]
\index{Gibbs measure!in infinite volume}
A probability measure on $\Omega^{\Z^d}$ is called an \emph{infinite volume Gibbs measure} for interaction~$H$ and inverse temperature~$\beta$ if for all finite~$\Lambda\subset\Z^d$ and $\mu$-a.e.\ $S_{\Lambda^\cc}$,
\begin{equation}
%\label{}
\mu\bigl(\,\cdot\,\big|S_{\Lambda^\cc}\bigr)=
\mu_{\Lambda,\beta}^{(S_{\Lambda^\cc})}(\cdot)
\end{equation}
where~$\mu_{\Lambda,\beta}^{(S_{\Lambda^\cc})}$ is defined using the Hamiltonian~$H_\Lambda$. 
\end{definition}

We will use~$\frakG_\beta$ to denote the set of all infinite volume Gibbs measures at inverse temperature~$\beta$ (assuming the model is clear from the context).

Here are some straightforward, nonetheless important consequences of these definitions:
%\settowidth{\leftmargini}{(11)}
\begin{enumerate}
\item[(1)]
By Lemma~\ref{lemma1.1}, any weak cluster point of $(\mu_{\Lambda,\beta}^{(S_{\Lambda^\cc})})$ belongs to~$\frakG_\beta$.
\item[(2)]
By the Backward Martingale Convergence Theorem, if $\Lambda_n\uparrow\Z^d$ and $\mu\in\frakG_\beta$, then for $\mu$-a.e.\ spin configuration~$S$ the sequence $\mu_{\Lambda_n,\beta}^{(S_{\Lambda_n^\cc})}$ has a weak limit, which then belongs to $\frakG_\beta$.
\item[(3)]
$\mathfrak G_\beta$ is a convex set (and is closed in the topology of weak convergence). Moreover, $\mu\in\frakG_\beta$ is extremal in $\frakG_\beta$ iff $\mu_{\Lambda_n,\beta}^{(S_{\Lambda_n^\cc})}\overset{\text{w}}\longrightarrow\mu$ for $\mu$-almost every spin configuration~$S$.
\end{enumerate}
Similarly direct is the proof of the following ``continuity'' property:
%\settowidth{\leftmargini}{(11)}
\begin{enumerate}
\item[(4)]
Let~$H_n$ be a sequence of Hamiltonians converging --- in the sup-norm on the potentials $\Phi_A$ --- to Hamiltonian~$H$, and let~$\beta_n$ be a sequence with~$\beta_n\to\beta<\infty$. Let~$\mu_n$ be the sequence of the corresponding Gibbs measures. Then every (weak) cluster point of~$(\mu_n)$ is an infinite-volume Gibbs measure for the Hamiltonian~$H$ and inverse temperature~$\beta$. 
\end{enumerate}
Note that the fact that $\mathfrak G_\beta$ is closed and convex ensures that each element can be written as a unique convex combination of extreme points (by the Krein-Millman theorem). The DLR condition permits to extract the corresponding decomposition probabilistically by conditioning on the $\sigma$-algebra of tail events.

\smallskip
Now we give a meaning to the terms that are frequently (though sometimes vaguely) employed by physicists:

\begin{definition}[Phase coexistence]\index{phase coexistence}\index{phase transition}
We say that the model is at \emph{phase coexistence} (or undergoes a \emph{first-order phase transition}) whenever the parameters are such that $|\frakG_\beta|>1$.
\end{definition}

The simplest example where this happens is the Ising model. Let 
\begin{equation}
%\label{}
\Lambda_L:=\{1,\dots,L\}^d
\end{equation}
and consider the Ising model in~$\Lambda_L$ with all boundary spins set to~$+1$. (This is the so called \emph{plus boundary condition}\index{boundary condition!plus}.) As a consequence of stochastic domination --- which we will not discuss here --- $\mu_{\Lambda_L,\beta}^+$ tends weakly to a measure $\mu^+$ as~$L\to\infty$. Similarly, for the \emph{minus} boundary condition\index{boundary condition!minus}, $\mu_{\Lambda_L,\beta}^-\overset{\text{w}}\longrightarrow\mu^-$. It turns out that, in dimensions $d\ge2$ there exists~$\betac(d)\in(0,\infty)$ such that
\begin{equation}
\label{MB2.22}
\beta>\betac(d)\quad\Rightarrow\quad \mu^+\ne\mu^-
\end{equation}
i.e, the model is at phase coexistence, \index{Ising model!phase transition in}
while for~$\beta<\betac(d)$, the set of all infinite volume Gibbs measures is a singleton --- which means that the model is in the uniqueness regime. One of our goals is to prove similar statements in all of the models introduced above.

\subsection{Torus measures}
In the above, we always put a boundary condition in the complement of the finite set~$\Lambda$. However, it is sometimes convenient to consider other boundary conditions. One possibility is to ignore the existence of $\Lambda^\cc$ altogether --- this leads to the so called \emph{free boundary condition}. \index{boundary condition!free} Another possibility is to wrap~$\Lambda$ into a graph without a boundary --- typically a torus. This is the case of \emph{periodic} or \emph{torus boundary conditions}. \index{boundary condition!periodic}

Consider the torus $\T_L$, which we define as a graph with vertices $(\Z/L\Z)^d$, endowed with the corresponding (periodic) nearest-neighbor relation. For nearest-neighbor interactions, the corresponding Hamiltonian is defined easily, but some care is needed for interactions that can be of arbitrary range. If~$S\in\Omega^{\T_L}$ we define the \emph{torus Hamiltonian} $H_L(S)$ by
\begin{equation}
%\label{}
H_L(S):=H_{\Lambda_L}(\text{periodic extension of~$S$ to~$\Z^d$})
\end{equation}
where we recall~$\Lambda_L:=\{1,\dots,L\}^d$.
For $H(S):=-\frac12\sum_{x,y}J_{x,y}S_x\cdot S_y$ we get
\begin{equation}
%\label{}
H_L(S)=-\frac12\sum_{x,y\in\T_L}J_{x,y}^{(L)}\,S_x\cdot S_y
\end{equation}
where $J_{x,y}^{(L)}$ are the periodized coupling constants
\begin{equation}
\label{JL}
J_{x,y}^{(L)}:=\sum_{z\in\Z^d}J_{x,y+Lz}
\end{equation}The Gibbs measure on~$\Omega^{\T_L}$ is then defined accordingly:
\begin{equation}
%\label{}
\mu_{L,\beta}(\textd S):=\frac{\texte^{-\beta H_L(S)}}{Z_{L,\beta}}\prod_{x\in\T_L}\mu_0(\textd S_x)
\end{equation}
where~$Z_{L,\beta}$ is the torus partition function. The following holds:

\begin{lemma}
%\label{lemma}
Every (weak) cluster point of $(\mu_{L,\beta})_{L\ge1}$ lies in~$\frakG_\beta$.
\end{lemma}

Note that there \emph{is} something to prove here because, due to \eqref{JL}, the interaction depends on~$L$.

\subsection{Some thermodynamics}
Statistical mechanics combines, in its historical development, molecular theory with empirical thermodynamics. Many mathematically rigorous accounts of statistical mechanics thus naturally start the exposition with the notion of the \emph{free energy}. \index{free energy} We will need this notion only tangentially --- it suffices to think of the free energy as a cumulant generating function --- in the proofs of phase coexistence. The relevant statement is as follows:

\begin{theorem}
\label{thm-exist}
For~$x\in\Z^d$ let~$\tau_x$ be the shift-by-$x$ defined by $(\tau_xS)_y:=S_{y-x}$. Let~$g\colon\Omega^{\Z^d}\to\R$ be a bounded, local function --- i.e., one that depends only on a finite number of spins --- and recall that~$\mu_{L,\beta}$ denote the torus Gibbs measures. Then:
%\settowidth{\leftmargini}{(11)}
\begin{enumerate}
\item[(1)]
The limit
\begin{equation}
\label{MB1.26}
f(h):=\lim_{L\to\infty}\frac1{L^d}\log E_{\mu_{L,\beta}}\biggl\{\exp\Bigl(h\sum_{x\in\T_L}g\circ\tau_x\Bigr)\biggr\}
\end{equation}
exists for all~$h\in\R$ and is convex in~$h$.
\item[(2)]
If~$\mu\in\frakG_\beta$ is translation invariant, then
\begin{equation}
\label{plus-minus-bd}
\frac{\partial f}{\partial h^-}\Bigl|_{h=0}\le E_\mu(g)\le \frac{\partial f}{\partial h^+}\Bigl|_{h=0}
\end{equation}\item[(3)]
There exist translation-invariant, ergodic measures~$\mu^\pm\in\frakG_\beta$ such that
\begin{equation}
\label{PMeq}
E_{\mu^\pm}(g)=\frac{\partial f}{\partial h^\pm}\Bigl|_{h=0}
\end{equation}\end{enumerate}
\end{theorem}

\begin{proofsectB}{Proof of (1), main idea}
For compact state-spaces and absolutely-summable interactions, the existence of the limit follows by standard subadditivity arguments. In fact, the limit will exist and will be the same even if we replace~$\mu_{L,\beta}$ in \eqref{MB1.26} by any sequence of Gibbs measures in~$\Lambda_L$ with (even $L$-dependent) boundary conditions. The convexity of~$f$ is a consequence of the H\"older inequality applied to the expectation in \eqref{MB1.26}.
\end{proofsectB}

\begin{proofsectB}{Proof of (2)}
Let~$\mu\in\frakG_\beta$ be translation invariant and abbreviate
\begin{equation}
%\label{}
Z_L(h):=E_\mu\biggl\{\exp\Bigl(h\sum_{x\in\Lambda_L}g\circ\tau_x\Bigr)\biggr\}
\end{equation}
Since~$\log Z_L$ is convex in~$h$ (again, by H\"older) we have for any~$h>0$ that
\begin{equation}
\begin{aligned}
\log Z_L(h)-\log Z_L(0)&\ge\frac\partial{\partial h}\log Z_L(h)\Bigl|_{h=0}\,h
\\
&=hE_\mu\Bigl(\sum_{x\in\Lambda_L}g\circ\tau_x\Bigr)=h|\Lambda_L|E_\mu(g)
\end{aligned}
\end{equation}
Dividing by~$|\Lambda_L|$, passing to~$L\to\infty$ and using that~$f$ is independent of the boundary condition, we get
\begin{equation}
%\label{}
f(h)-f(0)\ge h E_\mu(g)
\end{equation}Divide by~$h$ and let~$h\downarrow0$ to get one half of \eqref{plus-minus-bd}. The other half is proved analogously.
\end{proofsectB}

\begin{proofsectB}{Proof of (3)}
Let~$\frakG_{\beta,h}$ be the set of Gibbs measures for the Hamiltonian $H-(h/\beta)\sum_x g\circ\tau_x$. A variant of the proof of (2) shows that if~$\mu_h\in\frakG_{\beta,h}$ is translation-invariant, then
\begin{equation}
%\label{}
\frac{\partial f}{\partial h^-}\le E_{\mu_h}(g)\le \frac{\partial f}{\partial h^+}
\end{equation}In particular, if~$h>0$ we have
\begin{equation}
%\label{}
E_{\mu_h}(g)\ge\frac{\partial f}{\partial h^-}\,\,\underset{h>0}\ge\,\,\frac{\partial f}{\partial h^+}\Bigl|_{h=0}
\end{equation}
by the monotonicity of derivatives of convex functions. Taking~$h\downarrow0$ and extracting a weak limit from~$\mu_h$, we get a Gibbs measure $\mu^+\in\frakG_\beta$ such that
\begin{equation}
%\label{}
E_{\mu^+}(g)\ge\frac{\partial f}{\partial h^+}\Bigl|_{h=0}
\end{equation}(The expectations converge because~$g$ is a local --- and thus continuous, in the product topology --- function.) Applying~(2) we verify \eqref{PMeq} for~$\mu^+$. 

The measure~$\mu^+$ is translation invariant and so it remains to show that~$\mu^+$ can actually be chosen ergodic. To that end let us first prove that
\begin{equation}
\label{lim-prob}
\frac1{|\Lambda_L|}\sum_{x\in\Lambda_L}g\circ\tau_x\,\underset{L\to\infty}\longrightarrow\, E_{\mu^+}(g),\quad \text{in $\mu^+$-probability}
\end{equation}
The random variables on the left are  bounded by the norm of~$g$ and have expectation $E_{\mu^+}(g)$ so it suffices to prove that the limsup is no larger than the expectation. However, if that were not the case, we would have
\begin{equation}
%\label{}
\mu^+\Bigl(\sum_{x\in\Lambda_L}g\circ\tau_x>\bigl(E_{\mu^+}(g)+\epsilon\bigr)|\Lambda_L|\Bigr)>\epsilon
\end{equation}
for some~$\epsilon>0$ and some sequence of~$L$'s. But then for all~$h>0$,
\begin{equation}
%\label{}
E_{\mu^+}\biggl\{\exp\Bigl(h\sum_{x\in\Lambda_L}g\circ\tau_x\Bigr)\biggr\}
\ge\epsilon \texte^{|\Lambda_L| h[E_{\mu^+}(g)+\epsilon]}
\end{equation}
In light of the independence of the limit in (1) on the measure we use --- as discussed in the sketch of the proof of~(1) --- this would imply
\begin{equation}
%\label{}
f(h)\ge h\bigl(E_{\mu^+}(g)+\epsilon\bigr)
\end{equation}
which cannot hold for all~$h>0$ if the right-derivative of~$f$ at~$h=0$ is to equal $E_{\mu^+}(g)$. Hence \eqref{lim-prob} holds.

By the Pointwise Ergodic Theorem, the convergence in \eqref{lim-prob} actually occurs --- and, by \eqref{lim-prob}, the limit equals $E_{\mu^+}(g)$ --- for $\mu^+$-almost every spin configuration. This implies that the same must be true for any measure in the decomposition of~$\mu^+$ into ergodic components. By classic theorems from Gibbs-measure theory, every measure in this decomposition is also in~$\frakG_\beta$ and so we can choose~$\mu^+$ ergodic.
\end{proofsectB}

The above theorem is very useful for the proofs of phase coexistence. Indeed, one can often prove some estimates that via \eqref{plus-minus-bd} imply that $f$ is not differentiable at~$h=0$. Then one applies \eqref{PMeq} to infer the existence of two distinct, ergodic Gibbs measures saturating the bounds in \eqref{plus-minus-bd}. Examples of this approach will be discussed throughout these notes.

\subsection{Literature remarks}
This section contains only the minimum necessary to understand the rest of the course. For a comprehensive treatment of Gibbs-measure theory, we refer to classic monographs by Ruelle~\cite{Ruelle-redbook}, Israel~\cite{Israel}, Simon~\cite{Simon} and Georgii~\cite{Georgii}. Further general background on statistical mechanics of such systems can be found in Ruelle's ``blue'' book~\cite{Ruelle1Bi}. The acronym DLR derives from the initials of Dobrushin and the team of Lanford \& Ruelle who first introduced the idea of conditional definition of infinite volume Gibbs measures; cf e.g.~\cite{Dobrushin}.

The proof of Theorem~\ref{thm-exist} touches upon the subject of \emph{large deviation theory} \index{large deviations} which provides a mathematical framework for many empirical principles underlying classical thermodynamics. The connection of course appears in various disguises in the textbooks \cite{Israel,Simon,Georgii}; for expositions dealing more exclusively with large-deviation theory we refer to the books by den Hollander~\cite{denHollander}, Dembo and Zeitouni~\cite{Dembo-Zeitouni}, and Deuschel and Stroock~\cite{Deuschel-Stroock}. For the Pointwise Ergodic Theorem and other facts from \emph{ergodic theory} we refer to the textbooks by, e.g., Krengel~\cite{Krengel} and Petersen~\cite{Petersen}.

Stochastic domination and the FKG inequality --- dealing with partial ordering of spin configurations, functions thereof and thus also measures --- are discussed in, e.g., Georgii~\cite{Georgii} or Grimmett~\cite{Grimmett}. The proof of \eqref{MB2.22} can alternatively be based on Griffiths' correlation inequalities (Griffiths~\cite{Griffiths-JMP}). The phase coexistence in the Ising model at large~$\beta$ was first proved by Peierls via a contour argument that now bears his name (see Griffiths~\cite{Griffiths-Peierls}).

Concerning the historical origin of the various model systems; the $O(n)$ model goes back to Heisenberg (who introduced its quantum version), the Ising model was introduced by Lenz and given to Ising as a thesis problem while the Potts model was introduced by Domb and given to Potts as a thesis problem. Ironically, the $O(1)$-model bears Ising's name even though his conclusions about it were quite wrong! Apparently, Potts was more deserving.

An excellent reference for mathematical physics of liquid crystals is the monograph by de Gennes and Prost~\cite{deGennes}; other, more combinatorial models have been considered by Heilmann and Lieb~\cite{Heilmann-Lieb} and Abraham and Heilmann~\cite{Abraham-Heilmann}. The tetrahedral representation of the Potts model can be found in Wu's review article~\cite{Wu}; the matrix representation of the liquid-crystal model is an observation of Angelescu and Zagrebnov~\cite{Angelescu-Zagrebnov}. Gradient fields --- of which the GFF is the simplest example --- have enjoyed considerable attention in recent years; cf the review articles by Funaki~\cite{Funaki}, Velenik~\cite{Velenik} and Sheffield~\cite{Sheffield}. Another name for the GFF is \emph{harmonic crystal}.

%%%%%%%%%%%%%%%%%%%%%%%%%%

\section[Infrared bound \& Spin-wave condensation]{Infrared bound \& Spin-wave condensation}
\label{chap-2}
The goal of this section is to elucidate the significance of the infrared bound --- postponing its proof and connection with reflection positivity until Section~\ref{chap-4} --- and the use thereof in the proofs of symmetry breaking via the mechanism of \emph{spin-wave condensation}. \index{symmetry breaking} The presence, and absence, of symmetry breaking in the $O(n)$-model with certain non-negative two-body interactions will be linked to recurrence vs transience of a naturally induced random walk.

\subsection{Random walk connections}
\label{sec-RW}
Consider the model with the Hamiltonian
\begin{equation}
\label{Ham}
H = -\frac12\sum_{x,y}J_{xy}\,S_x\cdot S_y
\end{equation}
where the spins $S_x$ are \emph{a priori} independent and distributed according to a measure~$\mu_0$ which is supported in a compact set $\Omega\subset\R^\nu$. Assume that the interaction constants satisfy the following requirements:
\settowidth{\leftmargini}{(1111)}
\begin{enumerate}
\item[(I1)]
$J_{xx}=0$ and $J_{x,y}=J_{0,y-x}$
\item[(I2)]
$\sum_x|J_{0,x}|<\infty$ and $\sum_xJ_{0,x}=1$
\end{enumerate}
i.e., the coupling constants are translation invariant, absolutely summable and, for convenience, normalized to have unit strength. We will actually always restrict our attention to the following specific examples:
\settowidth{\leftmargini}{(111)}
\begin{itemize}
\item
\emph{Nearest-neighbor interactions}: 
\begin{equation}
J_{x,y}=\begin{cases}
\frac1{2d},\qquad&\text{if }|x-y|=1,
\\
0,\qquad&\text{otherwise}.
\end{cases}
\end{equation}
\item
\emph{Yukawa potentials}:
\begin{equation}
%\label{}
J_{x,y}=C\texte^{-\mu|x-y|_1}
\end{equation}
with $\mu>0$ and $C>0$.
\item
\emph{Power-law decaying potentials}: 
\begin{equation}
%\label{}
J_{x,y}=\frac C{|x-y|_1^s}
\end{equation}
with $s>d$ and~$C>0$.
\end{itemize}
On top of these, we will also permit:
\begin{itemize}
\item
Any convex combination of the three interactions above (with, of course, positive coefficients).
\end{itemize}
Note that we are using the $\ell_1$-distance (rather than the more natural $\ell_2$-distance). This is dictated by our methods of proof (see Lemma~\ref{lemma-RPint}). Also note that the Yukawa potential is in the class of \emph{Kac models} where the coupling constants take the form $J_{x,y}=\gamma^df(\gamma(x-y))$ for some rapidly decaying function $f\colon\R^d\to[0,\infty)$ with unit $L^1$-norm.

A unifying feature of all three interactions is that $J_{xy}\ge0$ which allows us to interpret the coupling constants as the \emph{transition probabilities} of a random walk on~$\Z^d$. Explicitly, consider a Markov chain~$(X_n)$ on~$\Z^d$ with
\begin{equation}
%\label{}
\cmss P_z(X_{n+1}=y|X_n=x):=J_{xy}
\end{equation}
where~$\cmss P_z$ is the law of the chain started at site~$z$. Of particular interest will be the question whether this random walk is \emph{recurrent} or \emph{transient} --- i.e., whether a walk started at the origin returns there infinitely, or only finitely many times. Here is a criterion to this matter:

\begin{lemma}
%\label{lemma}
Let~$\hat J(k):=\sum_x J_{0,x}\texte^{\texti k\cdot x}$, $k\in[-\pi,\pi]^d$. Then $(X_n)$ is transient if and only if
\begin{equation}
\label{transient}
\int_{[-\pi,\pi]^d}\frac{\textd k}{(2\pi)^d}\frac1{1-\hat J(k)}<\infty
\end{equation}
\end{lemma}

\begin{proofsectB}{Proof}
Recall that a random walk is transient if and only if the first return time to the origin, $\tau_0:=\inf\{n>0\colon X_n=0\}$, is infinite with a positive probability, i.e., $\cmss P_0(\tau_0<\infty)<1$. By the formula $\cmss E_0N=[1-\cmss P_0(\tau_0<\infty)]^{-1}$ --- where $\cmss E_0$ is the expectation with respect to~$\cmss P_0$ --- we thus get that transience is equivalent~$\cmss E_0 N<\infty$. To compute the expectation, we note
\begin{equation}
%\label{}
1_{\{X_n=0\}}=\int_{[-\pi,\pi]^d}\frac{\textd k}{(2\pi)^d}\texte^{\texti k\cdot X_n}
\end{equation}
which via $\cmss E_0\texte^{\texti k\cdot X_n}=[\cmss E_0\texte^{\texti k\cdot X_1}]^n=[\sum_x J_{0,x}\texte^{\texti k\cdot x}]^n=\hat J(k)^n$ implies
\begin{equation}
%\label{}
\cmss P_0(X_n=0)=\int_{[-\pi,\pi]^d}\frac{\textd k}{(2\pi)^d}\hat J(k)^n
\end{equation}
Summing over~$n\ge0$ yields
\begin{equation}
%\label{}
\cmss E_0 N = \sum_{n\ge0}\int_{[-\pi,\pi]^d}\frac{\textd k}{(2\pi)^d}\hat J(k)^n=
\int_{[-\pi,\pi]^d}\frac{\textd k}{(2\pi)^d}\frac1{1-\hat J(k)}
\end{equation}
whereby the claim follows. (A careful proof of the latter identity requires justification of the exchange of the integral with the infinite sum; one has to represent the LHS as a power series, perform the sum and justify limits via appropriate convergence theorems.)
\end{proofsectB}

As to the above examples, we have:
%\settowidth{\leftmargini}{(11)}
\begin{itemize}
\item
\emph{n.n.\,\,\& Yukawa potentials}: As $k\to0$,
\begin{equation}
%\label{}
1-\hat J(k)\sim C|k|^2
\end{equation}
and so $(X_n)$ is transient iff $d\ge3$.
\item
\emph{Power-law potentials}: Here as $k\to0$,
\begin{equation}
1-\hat J(k)\sim C\begin{cases}
|k|^{s-d},\qquad&\text{if }s<d+2,
\\
|k|^2\log\frac1{|k|},\qquad&\text{if }s=d+2,
\\
|k|^2,\qquad&\text{if }s>d+2.
\end{cases}
\end{equation}
Hence $(X_n)$ is transient iff $d\ge3$ OR $s<\min\{d+2,2d\}$. 
\end{itemize}
(Note that the walk with~$s<d+2$ has a stable-law tail with index of stability~$\alpha=s-d$.) A convex combination  of the three coupling constants will lead to a transient walk provided at least one of the interactions involved therein (with non-zero coefficients) is transient.

\subsection{Infrared bound}
The principal claim of this section is that the finiteness of the integral in \eqref{transient} is sufficient for the existence of a symmetry-breaking phase transition in many spin systems of the kind~\eqref{Ham}. The reason is the connection of the above random walk to the Gaussian free field \eqref{H-GFF} (GFF) with $\cmss P(x,y):=J_{xy}$. Indeed, consider the field in a square box $\Lambda$ with, say, zero boundary condition. It turns out that
\begin{equation}
%\label{}
\text{Cov}_\Lambda(\phi_x,\phi_y)=\sum_{n\ge0}\cmss P_x(X_n=y,\,\tau_{\Lambda^\cc}=y) =\colon G_\Lambda(x,y)
\end{equation}
where $\tau_{\Lambda^\cc}$ is the first exit time of the walk from $\Lambda$ and $G_\Lambda$ denotes the so called \emph{Green's function} in~$\Lambda$. \index{Green's function} In particular, we have
\begin{equation}
%\label{}
\Var_\Lambda(\phi_0)=G_\Lambda(0,0)
\end{equation}
which, as we will see, tends to the integral \eqref{transient} as~$\Lambda\uparrow\Z^d$. Since $E_\Lambda(\phi_0)=0$ due to our choice of the boundary condition, we conclude
\begin{equation}
%\label{}
\bigl\{\text{Law$(\phi_0)\colon\Lambda\subset\Z^d\bigr\}$ is tight iff $(X_n)$ is transient}
\end{equation}
Physicists actually prefer to think of this in terms of symmetry breaking: Formally, the Hamiltonian of the GFF is invariant under the transformation $\phi_x\to\phi_x+c$, i.e., the model possesses a global spin-translation symmetry. The symmetry group is not compact and so, to define the model even in finite volume, the symmetry needs to be broken by boundary conditions. The existence of a limit law for~$\phi_0$ can be interpreted as the survival of the symmetry breaking in the thermodynamic limit --- while non-existence means that the invariance is restored in this limit. 

Our goal is to show that qualitatively the same conclusions hold also for the $O(n)$-spin system. Explicitly, we will prove: \index{$O(n)$ model!phase transition in}

\begin{theorem}
\label{thm-On}
Let~$(J_{xy})$ be one of the 3 interactions above. Then:
\begin{equation}
\notag
\begin{aligned}
&\text{Global rotation symmetry}
\\
&\text{of $O(n)$-model is broken}
\\
&\text{at low temperatures}
\end{aligned}
\qquad\Longleftrightarrow\qquad
\begin{aligned}
&\text{Random walk driven}
\\
&\text{by $(J_{xy})$ is transient}
\end{aligned}
\end{equation}
\end{theorem}

We begin with the proof of the implication $\Longleftarrow$. The principal tool will be our next theorem which, for technical reasons, is formulated for torus boundary conditions: 

\begin{theorem}[Infrared bound]
\label{thm-IRB}
\index{infrared bound}
Let $L$ be an even integer and consider the model \eqref{Ham} on torus~$\T_L$ with Gibbs measure~$\mu_{L,\beta}$. Suppose~$(J_{xy})$ is one of the three interactions above and let
\begin{equation}
%\label{}
c_{L,\beta}(x):=E_{\mu_{L,\beta}}(S_0\cdot S_x)
\end{equation}
Define $\hat c_{L,\beta}(k):=\sum_{x\in\T_L}c_{L,\beta}(x)\texte^{\texti k\cdot x}$. Then
\begin{equation}
%\label{}
\hat c_{L,\beta}(k)\le\frac\nu{2\beta}\frac1{1-\hat J(k)},\qquad k\in\T_L^\star\setminus\{0\}
\end{equation}
where~$\nu$ is the dimension of the spin vectors and~$\T_L^\star$ is the reciprocal torus, $\T_L^\star:=\{\frac{2\pi}L(n_1,\dots,n_d)\colon n_i=0,\dots,L-1\}$.
\end{theorem}

The proof will require developing the technique of reflection positivity and is therefore postponed to Section~\ref{chap-4}. 

Note that $c_{L,\beta}(x)$ is the \emph{spin-spin correlation} function which, in light of translation invariance of~$\mu_{L,\beta}$ is a function of only the spatial displacement of the two spins. The result has the following equivalent formulation: For all~$(v_x)\in\C^{\T_L}$ with~$\sum_xv_x=0$,
\begin{equation}
%\label{}
\sum_{x,y\in\T_L}v_x\bar v_yE_{\mu_{L,\beta}}(S_0\cdot S_x)
\le\frac{\nu}{2\beta}\sum_{x,y\in\T_L}v_x\bar v_y G_L(x,y)
\end{equation}
where
\begin{equation}
\label{GLdef}
G_L(x,y):=\frac1{L^d}\sum_{k\in\T_L^\star\setminus\{0\}}\frac{\texte^{\texti k\cdot(x-y)}}{1-\hat J(k)}
\end{equation}
Observe that the latter is the covariance matrix of the GFF on~$\T_L$, projected on the set of configurations with total integral zero (i.e., on the orthogonal complement of constant functions). This is a meaningful object because while the~$\phi_x$ are not really well defined --- due to the absence of the boundary --- the differences~$\phi_y-\phi_x$ are.
(These differences are orthogonal to constant functions, of course.)
A short formulation of the infrared bound is thus: 

\begin{center}
\begin{minipage}{0.95\textwidth}
\emph{The correlation of the spins in models~\eqref{Ham} with one of the three interactions above is dominated --- as a matrix on the orthogonal complement of constant functions in~$L^2(\T_L)$ --- by the covariance of a GFF.}
\end{minipage} 
\end{center}

\noindent
This fact is often referred to as \emph{Gaussian domination}. \index{Gaussian domination}

\subsection{Spin-wave condensation in $O(n)$-model}
Having temporarily dispensed with the IRB, we will continue in our original line of thought. Theorem~\ref{thm-IRB} implies:

\begin{corollary}[Spin-wave condensation]
\label{cor2.4}
\index{spin wave!condensation}
Suppose~$|S_x|=1$. Then
\begin{equation}
\label{SWD}
E_{\mu_{L,\beta}}\biggl(\,\Bigl|\frac1{L^d}\sum_{x\in\T_L}S_x\Bigr|^2\biggr)\ge1-\frac\nu{2\beta}G_L(0,0)
\end{equation}
\end{corollary}

\begin{proofsectB}{Proof}
Let~$\hat S_k:=\sum_{x\in\T_L}S_x\texte^{\texti k\cdot x}$ be the Fourier coefficient of the decomposition of~$(S_x)$ into the so called \emph{spin waves}. The IRB yields
\begin{equation}
\label{2.18}
E_{\mu_{L,\beta}}|\hat S_k|^2\le\frac\nu{2\beta}\,\frac{L^d}{1-\hat J(k)},
\qquad k\in\T_L^\star\setminus\{0\}
\end{equation}
On the other hand, Parseval's identity along with~$|S_x|=1$ implies
\begin{equation}
%\label{}
\sum_{k\in\T_L^\star}|\hat S_k|^2=L^d\sum_{x\in\T_L}|S_x|^2=L^{2d}
\end{equation}
The IRB makes no statement about~$\hat S_0$ so we split it from the rest of the sum:
\begin{equation}
%\label{}
\frac1{L^{2d}}|\hat S_0|^2=1-\frac1{L^{2d}}\sum_{k\in\T_L^\star\setminus\{0\}}|\hat S_k|^2
\end{equation}
Now take expectation and apply \eqref{2.18}:
\begin{equation}
%\label{}
E_{\mu_{L,\beta}}\Bigl(\frac1{L^{2d}}|\hat S_0|^2\Bigr)\ge1-\frac\nu{2\beta}\,
\frac1{L^d}\sum_{k\in\T_L^\star\setminus\{0\}}\frac1{1-\hat J(k)}
\end{equation}
In light of \eqref{GLdef}, this is \eqref{SWD}.
\end{proofsectB}

With \eqref{SWD} in the hand we can apply the same reasoning as for the GFF: In the transient cases,~$G_L(0,0)$ converges to the integral \eqref{transient} and so the right-hand side has a finite limit. By taking $\beta$ sufficiently large, the limit is actually strictly positive. This in turn implies that the zero mode of the spin-wave decomposition is \emph{macroscopically} populated --- very much like the free Bose gas at  \emph{Bose-Einstein condensation}. Here is how we pull the corresponding conclusions from~$\T_L$ onto~$\Z^d$:

\begin{theorem}[Phase coexistence in $O(n)$-model]
%\label{thm}
Consider the $O(n)$-model with $n\ge1$ and one of the three interactions above. Let
\begin{equation}
%\label{}
\beta_0:=\frac n2\int_{[-\pi,\pi]^d}\frac{\textd k}{(2\pi)^d}\frac1{1-\hat J(k)}
\end{equation}
Then for any $\beta>\beta_0$ and any $\theta\in\BbbS^{n-1}$ there exists $\mu_\theta\in\mathfrak G_\beta$ which is translation invariant and ergodic such that
\begin{equation}
\label{2.23}
\frac1{|\Lambda_L|}\sum_{x\in\Lambda_L}S_x\,\underset{L\to\infty}\longrightarrow\,m_\star\,\theta,\qquad\mu_\theta\text{\rm-a.s.}
\end{equation}
for some $m_\star=m_\star(\beta)>0$.
\end{theorem}

Note that \eqref{2.23} implies that the measures~$\mu_\theta$ are mutually singular with respect to one another. Note also that $\beta_0$ is finite --- and the statement is not vacuous --- if and only if the associated random walk is transient.

\begin{proofsectB}{Proof}
Suppose, without loss of generality, that we are in the transient case, i.e., $\beta_0<\infty$. The idea of the proof is quite simple: We use \eqref{SWD} to show that the free energy is not differentiable in an appropriately-chosen external field when this field is set to zero. Then we apply Theorem~\ref{thm-exist} to conclude the existence of the required distinct, ergodic Gibbs measures.

Fix~$\theta\in\BbbS^{n-1}$ and define
\begin{equation}
%\label{}
f(h):=\lim_{L\to\infty}\frac1{L^d}\log E_{\mu_{L,\beta}}\bigl(\texte^{h\theta\cdot\hat S_0}\bigr)
\end{equation}
(The limit exists by Theorem~\ref{thm-exist}.) We want to show that~$\frac{\partial}{\partial h^+}f(0)>0$ (and thus, by symmetry, $\frac{\partial}{\partial h^-}f(0)<0$). Corollary~\ref{cor2.4} yields
\begin{equation}
%\label{}
E_{\mu_{L,\beta}}\bigl(L^{-2d}|\hat S_0|^2\bigr)\ge \frac{\beta-\beta_0}\beta+o(1)
\end{equation}
Since $|\hat S_0|\le L^d$, for any $0<\epsilon<1$ we have
\begin{equation}
%\label{}
E_{\mu_{L,\beta}}\bigl(L^{-2d}|\hat S_0|^2\bigr)\le\epsilon+\mu_{L,\beta}\bigl(|\hat S_0|\ge\epsilon L^d\bigr)
\end{equation}
and so
\begin{equation}
%\label{}
\mu_{L,\beta}\Bigl(|\hat S_0|\ge\frac12\frac{\beta-\beta_0}\beta L^d\Bigr)\ge\frac12\frac{\beta-\beta_0}\beta+o(1)
\end{equation}
By the $O(n)$ symmetry of the torus measures $\mu_{L,\beta}$, the law of~$\hat S_0/L^d$ is rotationally invariant with non-degenerate ``radius'' distribution. This implies
\begin{equation}
%\label{}
\mu_{L,\beta}\Bigl(\theta\cdot\hat S_0\ge\frac14\frac{\beta-\beta_0}\beta L^d\Bigr)\ge C_n\frac{\beta-\beta_0}\beta+o(1)
\end{equation}
where $C_2:=\ffracB16$ and, in general, $C_n>0$ is an explicitly obtainable constant. But this means that the exponent in the definition of~$f$ is at least~$\frac14\frac{\beta-\beta_0}\beta L^d$ with uniformly positive probability and so
\begin{equation}
%\label{}
\frac{\partial f}{\partial h^+}\Bigl|_{h=0}\ge\frac{\beta-\beta_0}{4\beta}
\end{equation}
Applying Theorem~\ref{thm-exist}, for~$\beta>\beta_0$ and any~$\theta\in\CalS^1$ there exists a translation invariant, ergodic Gibbs state $\mu_\theta\in\frakG_\beta$ such that
\begin{equation}
\label{f+}
E_{\mu_\theta}(\theta\cdot S_x)=\frac{\partial f}{\partial h^+}\Bigl|_{h=0}>0
\end{equation}
Next we need to show that the states $\mu_\theta$ are actually distinct. The Ergodic Theorem implies
\begin{equation}
\label{tildetheta}
\frac1{|\Lambda_L|}\sum_{x\in\Lambda_L}S_x\,\underset{L\to\infty}\longrightarrow\,m_\star\tilde\theta,\qquad\mu_\theta\text{\rm-a.s.}
\end{equation}
where $\tilde\theta\in\BbbS^{n-1}$ and where $m_\star>0$ is the magnitude of the derivative. Note that, in light of \eqref{f+} and the translation invariance of $\mu_\theta$,
\begin{equation}
\label{MB-f+1}
m_\star\,\theta\cdot\tilde\theta=\frac{\partial f}{\partial h^+}\Bigl|_{h=0}
\end{equation}
The distinctness of $\mu_\theta$ will follow once we prove \eqref{2.23}, i.e., $\theta=\tilde\theta$. (This is, of course, intuitively obvious because the way we constructed $\mu_\theta$ indicates that the law of~$S_x$ under~$\mu_\theta$ should be biased in the direction of $\theta$.)

Suppose $\tilde\theta\ne\theta$. Find a rotation $A\in O(n)$ such that $A\tilde\theta=\theta$. Let $\tilde\mu$ be the measure such that $E_{\tilde\mu}(f(S)):=E_{\mu_\theta}(f(AS))$
for all local functions~$f$. (The existence of such a measure follows from the Kolmogorov Extension Theorem.) Since both the Hamiltonian and the \emph{a priori} measure are $O(n)$-invariant, we have $\tilde\mu\in\frakG_\beta$. But \eqref{tildetheta} implies $E_{\mu_\theta}(S_x)=m_\star\,\tilde\theta$, and so from \eqref{MB-f+1} we have
\begin{equation}
%\label{}
E_{\tilde\mu}(\theta\cdot S_x)=E_{\mu_\theta}(\theta\cdot AS_x)=m_\star\,|\theta|^2
\,\underset{\tilde\theta\ne\theta}>\,m_\star\,\theta\cdot\tilde\theta=\frac{\partial f}{\partial h^+}\Bigl|_{h=0}
\end{equation}
As $\tilde\mu$ is a Gibbs measure, this contradicts the general bounds in Theorem~\ref{thm-exist}. Hence, we must have $\theta=\tilde\theta$ after all.
\end{proofsectB}

The above statement and proof are formulated for the specific case of the $O(n)$ model. A similar proof will apply the existence of a symmetry-breaking phase transition at low temperatures in the Ising, Potts and the liquid-crystal models in all transient dimensions. As the Ising and Potts model have only a \emph{discrete} set of spin states, a symmetry-breaking transition will occur generally in all dimensions $d\ge2$. However, this has to be proved by different methods than those employed above (e.g., by invoking chessboard estimates).

\index{Mermin-Wagner theorem}
Our next goal is to establish the complementary part of Theorem~\ref{thm-On}, i.e., the implication $\Longrightarrow$, which asserts the \emph{absence} of symmetry breaking in the recurrent cases. This argument predates the other direction by 20 years and bears the name of its discoverers:

\begin{theorem}[Mermin-Wagner theorem]
%\label{thm}
Let~$n\ge2$ and consider the $O(n)$-model with non-negative interactions constants $(J_{x,y})$ satisfying the conditions (I1,I2) from Sect.~\ref{sec-RW}. Suppose the corresponding random walk is recurrent. Then \emph{every} $\mu\in\frakG_\beta$ is invariant under any simultaneous (i.e., homogeneous) rotation of all spins.
\end{theorem}

\begin{proofsectB}{Proof}
We will show that the spins can be arbitrarily rotated at an arbitrary small cost of the total energy. (This is why we need $n\ge2$.) We will have to work with \emph{in}homogeneous rotations to achieve this, so let $\varphi_x$ be a collection of numbers with $\{x\colon\varphi_x\ne0\}$ finite and let $\texti R$ be a unit element of the Lie algebra~$\mathfrak o(n)$, i.e., $\texte^{\texti R\alpha}$ is a rigid rotation of the unit sphere by angle $\alpha$ about a particular axis. Let~$\omega_\varphi$ be the map on configuration space acting on individual spins via
\begin{equation}
%\label{}
\omega_\varphi(S_x):=\texte^{\texti\varphi_x R}S_x,\qquad x\in\Z^d
\end{equation}
To investigate the effect of such an inhomogeneous rotation on the Hamiltonian, note that
\begin{equation}
\begin{aligned}
\omega_\varphi(S_x)\cdot\omega_\varphi(S_y)
&=S_x\cdot\texte^{\texti(\varphi_y-\varphi_x) R}S_y
\\
&=
S_x\cdot S_y-S_x\cdot[1-\texte^{\texti(\varphi_y-\varphi_x) R}]S_y
\end{aligned}
\end{equation}
Hence the energy of a configuration in any block $\Lambda\supset\{x\colon\varphi_x\ne0\}$ transforms~as
\begin{equation}
%\label{}
H_\Lambda(\omega_\varphi(S))=H_\Lambda(S)+\triangle H
\end{equation}
where
\begin{equation}
%\label{}
\triangle H := \frac12\sum_{x,y}J_{xy}\,S_x\cdot[1-\texte^{\texti(\varphi_y-\varphi_x) R}]S_y
\end{equation}
Using that~$\triangle H$ depends only on the portion of the spin configuration in~$\Lambda$, a simple application of the DLR condition shows that, for any local function $f$, 
\begin{equation}
\label{DLR-MWT}
E_\mu(f\circ\omega_\varphi)=E_\mu(f\texte^{-\beta\triangle H})
\end{equation}
We will now let $\varphi_x\to\alpha$ in a specific way that ensures $\triangle H\to0$; this will permit us to extract the desired conclusion by limiting arguments. 

First we will need to control the $\varphi$-dependence of~$\triangle H$, so we expand the exponential:
\begin{equation}
\begin{aligned}
\triangle H=&-\frac\texti2\sum_{x,y}J_{xy}\,(S_x\cdot RS_y)(\varphi_y-\varphi_x)
\\
&+\frac14\sum_{x,y}J_{xy}\,(RS_x\cdot RS_y)(\varphi_y-\varphi_x)^2+\cdots
\end{aligned}
\end{equation}
In the first term we note that the self-adjointness of~$R$ --- valid by the choice of~$\texti R$ as an element of the Lie algebra --- implies that $J_{xy}\,(S_x\cdot RS_y)$ is symmetric under the exchange of~$x$ and~$y$. Since~$(\varphi_y-\varphi_x)$ is antisymmetric and finitely supported, the sum is zero. Estimating the remainder by the quadratic term, we thus get
\begin{equation}
%\label{}
|\triangle H|\le C\sum_{x,y}J_{xy}(\varphi_y-\varphi_x)^2=2C\EE_{1-J}(\varphi,\varphi)
\end{equation}
for some constant~$C<\infty$. Here we used that $(RS_x\cdot RS_y)$ is bounded and recalled the definition of the Dirichlet form $\EE_{1-J}(\cdot,\cdot)$ of the random walk driven by the~$(J_{x,y})$'s. \index{Dirichlet form}

Our next task will be to control the Dirichlet form under the condition that~$\varphi$ tends to $\alpha$ in every finite set. To that end we fix~$0<R<\infty$ and set
\begin{equation}
\label{MB-phi-choice}
\varphi_x:=\alpha\,\cmss P_x(\tau_0<\tau_{\Lambda_R^\cc})
\end{equation}
This function equals $\alpha$ at~$x=0$, zero on~$\Lambda_R^\cc$ and is harmonic (with respect to the generator of the random walk) in~$\Lambda_R\setminus\{0\}$. A calculation shows
\begin{equation}
\begin{aligned}
\EE_{1-J}(\varphi,&\varphi)
=\sum_x\varphi_x\sum_yJ_{x,y}(\varphi_x-\varphi_y)
\\
&\!\!\!\!\underset{\text{zero in }\{0\}^\cc}{\underset{\text{harmonic or}}=}\alpha\,\sum_yJ_{0,y}(\alpha-\varphi_y)
\end{aligned}
\end{equation}
But the recurrence of the associated random walk implies that~$\varphi_y\to\alpha$ as~$R\to\infty$ for every~$y$ and since the $J_{xy}$'s are summable, the right-hand side tends to zero by the Dominated Convergence Theorem. Thus $\triangle H\to0$ as~$R\to\infty$ and so applying $R\to\infty$ to \eqref{DLR-MWT} with the choice \eqref{MB-phi-choice} yields
\begin{equation}
%\label{}
E_\mu(f\circ\omega_\alpha)=E_\mu(f)
\end{equation}
for every continuous local function~$f$. Thereby we conclude that~$\mu$ is invariant under simultaneous rotation of all spins.
\end{proofsectB}

\subsection{Literature remarks}
The content of the entire section is very classical. The Infrared Bound (and its proof based on reflection positivity) was discovered in the seminal work of Fr\"ohlich, Simon and Spencer~\cite{FSS} from 1976 where it was also applied to prove a phase transition in the $O(n)$-model (as well as the isotropic Heisenberg and other models). Dyson, Lieb and Simon~\cite{DLS} showed how to adapt the method to a (somewhat more limited) class of quantum spin models. The technique was further developed and its applications extended in two papers of Fr\"ohlich, Israel, Lieb and Simon~\cite{FILS1,FILS2}.

Thanks to the representation \eqref{MB-Qmatrix}, the proof of a long-range order in the liquid-crystal model, derived by Angelescu and Zagrebnov~\cite{Angelescu-Zagrebnov}, follows the same route as for the $O(n)$ model. However, the type of long-range order that is concluded for the actual spin system is different. Indeed, let~$\mu$ be a (weak) cluster point of the torus states. Then
\begin{equation}
%\label{}
\lim_{L\to\infty}\frac1{|\Lambda_L|}\sum_{x\in\Lambda_L}\bigl[(S_0\cdot S_x)^2-\ffracB1n\bigr]>0
\end{equation}
with a positive probability under~$\mu$. (The limit exists by the Pointwise Ergodic Theorem.) As $\mu$ is $O(n)$-invariant, if $S_x$ were asymptotically independent of~$S_0$ for large~$x$, we would expect $E_\mu(S_0\cdot S_x)^2\to\ffracB1n$ as~$|x|\to\infty$. Apparently, this is not the case, the direction of $S_x$ remains heavily correlated with the direction of~$S_0$ for arbitrary~$x$, i.e., there is an \emph{orientational} long range order. 

Whether or not the $O(n)$ symmetry of the law of~$S_x$ is broken is an open (and important) question. (The law of each individual~$S_x$ is invariant under the flip $S_x\leftrightarrow-S_x$ and so the magnetization is zero in all states.) As noted before, other models of liquid crystals based on dimers on~$\Z^2$ were considered by Heilmann and Lieb~\cite{Heilmann-Lieb} and Abraham and Heilmann~\cite{Abraham-Heilmann} prior to the work~\cite{Angelescu-Zagrebnov}. There an orientational long-range order was proved using chessboard estimates; the question of absence of \emph{complete translational ordering} (i.e., breakdown of translation invariance) remained open. 

The Mermin-Wagner theorem goes back to 1966~\cite{MW}. Various interesting mathematical treatments and extensions followed~\cite{DS1Bi,Pfister,PF1Bi}; the argument presented here is inspired by the exposition in Simon's book~\cite{Simon}. A fully probabilistic approach to this result, discovered by Dobrushin and Shlosman~\cite{DS1Bi}, has the advantage that no regularity conditions need to be posed on the spin-spin interaction provided it takes the form $V(S_x-S_y)$; cf the recent paper by Ioffe, Shlosman and Velenik~\cite{ISV}. Finally, we remark that a beautiful and more in-depth exposition of this material --- including quantum systems --- was presented at the Prague School in 1996 by 
B\'alint T\'oth; his handwritten lecture notes should be available online~\cite{Balint-web}.

The basis of the Mermin-Wagner theorem, as well as its extension, is the \emph{continuum} nature of the spin space. Indeed, in the Ising (and also Potts) model, a low-temperature symmetry breaking occurs even in some recurrent dimensions; e.g., in $d=2$ for the nearest-neighbor interactions. For what determines the presence and absence of symmetry breaking in $d=1$, see the work of Aizenman, Chayes, Chayes and Newman~\cite{ACCN} and references therein.

The connection with random walk is, of course, made possible by our choice to work with non-negative couplings. However, most of the quantitative conclusions of this section hold without reference to random walks. For detailed expositions of the theory of random walks we recommend the monographs by Spitzer~\cite{Spitzer} and Lawler~\cite{Lawler}; the material naturally appears in most graduate probability textbooks (e.g., Durrett~\cite{Durrett}).

It is interesting to note that even in $d=2$, the nearest-neighbor $O(n)$ model exhibits a phase transition when $n=2$. Namely, while the Gibbs state is unique at all $\beta<\infty$, for large~$\beta$ it exhibits power law decay of correlations with~$\beta$-dependent exponents. This regime is (again, after its discoverers) referred to as the \emph{Kosterlitz-Thouless phase}~\cite{Kosterlitz-Thouless}. A rigorous treatment exits, based on renormalization theory and connection with Coulomb gas, thanks to the pioneering work of Fr\"ohlich and Spencer~\cite{Frohlich-Spencer}; see also more recent papers by Dimock and Hurd~\cite{Dimock-Hurd}. This is of much interest in light of recent discovery of new conformally-invariant planar processes --- the Schramm-Loewner evolution (a.k.a.~SLE). No such phenomenon is expected when $n\ge3$ though there is a minor opposition to this (e.g., Patrasciou and Seiler~\cite{Patrasciou-Seiler}).

\section[Infrared bound \& Mean-field theory]{Infrared bound \& Mean-field theory}
\label{chap-3}
In this chapter we will discuss how the infrared bound can be used to control the error in so-called mean-field approximation. Unlike the spin-wave condensation, which is concerned primarily with the infrared --- i.e., small-$k$ or large spatial scale --- content of the IRB, here will make the predominant use of the finite-$k$ --- i.e., short range --- part of the IRB. (Notwithstanding, the finiteness of the integral \eqref{transient} is still a prerequisite.)

\subsection{Mean-field theory}
\label{secMFT}
\index{mean-field theory}
Mean-field theory is a versatile approximation technique frequently used by physicists to analyze realistic physical models. We begin by a simple derivation  that underscores the strengths, and the shortcomings, of this approach. 

Consider a lattice spin model with the usual Hamiltonian \eqref{Ham}. Pick a translation invariant Gibbs measure~$\mu\in\frakG_\beta$ and consider the expectation of the spin at the origin. The conditional definition of Gibbs measures (the DLR condition) allows us to compute this expectation by first conditioning on all spins outside the origin. Indeed, the one-spin Gibbs measure is determined by the (one-spin) Hamiltonian
\begin{equation}
%\label{}
H_{\{0\}}(S)=-\sum_xJ_{0,x}\,S_0\cdot S_x=-S_0\cdot\sum_xJ_{0,x}S_x
=-S_0\cdot M_0
\end{equation}
where we introduced the shorthand
\begin{equation}
%\label{}
M_0:=\sum_xJ_{0,x}S_x
\end{equation}
Thus, by the DLR,
\begin{equation}
%\label{}
E_\mu(S_0)=E_\mu\biggl(\,\frac{E_{\mu_0}(S_0\,\texte^{\,\beta S_0\cdot M_0})}{E_{\mu_0}(\texte^{\,\beta S_0\cdot M_0})}\biggr)
\end{equation}
where, abusing the notation slightly, the ``inner'' expectations are only over $S_0$ --- $M_0$ acts as a constant here --- and the outer expectation is over the spins in~$\Z^d\setminus\{0\}$, and thus over~$M_0$.

So far the derivation has been completely rigorous but now comes an \emph{ad hoc} step: We suppose that the random variable~$M_0$ is strongly concentrated about its average so that we can replace it by this average. Denoting
\begin{equation}
%\label{}
m:=E_\mu(S_0)
\end{equation}
we thus get that~$m$ should be an approximate solution to
\begin{equation}
\label{MFE}
m=\frac{E_{\mu_0}(S\,\texte^{\,\beta S\cdot m})}{E_{\mu_0}(\texte^{\,\beta S\cdot m})}
\end{equation}
This is the so called \emph{mean-field equation} for the magnetization. \index{magnetization}

Besides the unjustified step in the derivation, a serious practical problem with \eqref{MFE} is that it often has multiple solutions. Indeed, for the set of points $(\beta,m)$ that obey this equation, one typically gets a picture like this:

\vglue0.5cm
\centerline{\includegraphics[width=3in]{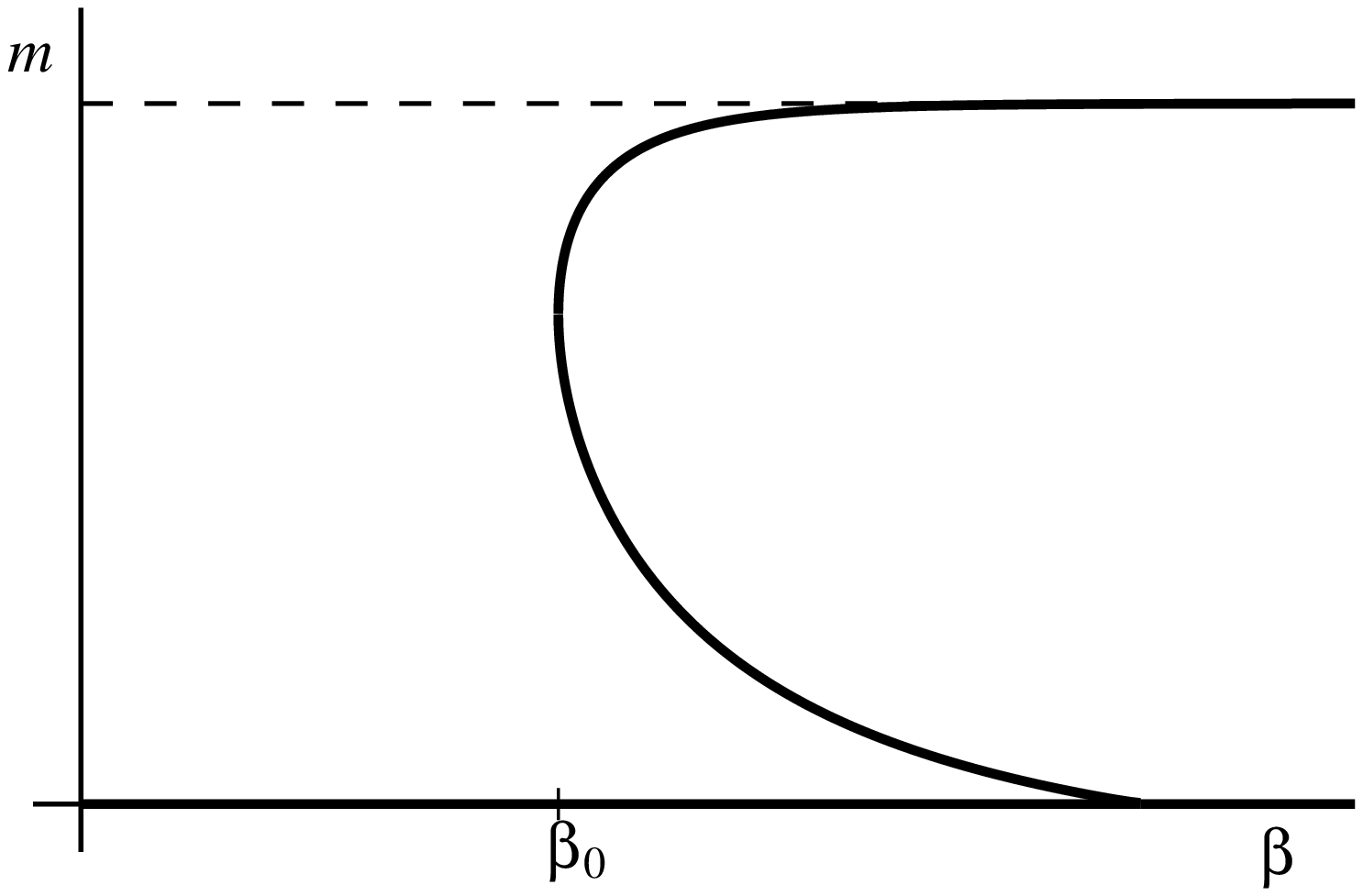}}
\medskip
\noindent
Here, for~$\beta<\beta_0$, the only solution is~$m=0$ --- this is always a solution whenever $E_{\mu_0}(S)=0$ --- but at $\beta=\beta_0$, two new branches appear and coexist over an interval of~$\beta$'s. It is clear that as~$\beta$ varies, the ``physical'' solution must undergo some sort of jump, but it is not possible to tell where this jump occurs on the basis of equation \eqref{MFE} alone. For that one has to go beyond the heuristic derivation presented above. 

As is standard, one comes up with an additional ``selection'' principle that determines which solution is ``physical.'' At the level of classical thermodynamics, this is done by postulating that the solution must minimize an appropriate free energy function. In the choice of this function we will be guided by the fact that there is a proper statistical-mechanical system for which the above derivations can be explicitly validated by way of large-deviation theory. This system is the corresponding model on the \emph{complete graph}.

Consider a graph on~$N$ vertices with each pair of vertices joined by an undirected edge. At each vertex~$x=1,\dots,N$ we have a spin~$S_x$ with i.i.d.~\emph{a priori} law~$\mu_0$. Each spin interacts with every other spin; the interaction Hamiltonian is given by
\begin{equation}
%\label{}
H_N(S):=-\frac1{2N}\sum_{x,y=1}^N S_x\cdot S_y
\end{equation}
The normalization by~$\ffracB1N$ ensures that the energy grows proportionally to~$N$; the~``$2$'' in the denominator compensates for counting each pair of spins twice.

To derive the formula for the free energy function, consider first the cumulant generating function of the measure~$\mu_0$,
\begin{equation}
%\label{}
G(h):=\log E_{\mu_0}\bigl(\texte^{h\cdot S}\bigr),\qquad h\in\R^\nu
\end{equation}
Its Legendre transform,
\begin{equation}
%\label{}
\scrS(m):=\inf_{h\in\R^\nu}\,\bigl[G(h)-h\cdot m\bigr]
\end{equation}
defines the \emph{entropy} which, according to Cram\'er's theorem, is the rate of large-deviation decay in
\begin{equation}
%\label{}
\mu_0\biggl(\,\sum_{x=1}^NS_x\approx mN\biggr)=\texte^{-N\scrS(m)+o(N)}
\end{equation}
(The function is infinite outside $\text{\rm Conv}(\Omega)$, the convex hull of~$\Omega$ and the set of possible values of the magnetization.)
Next we inject the energy into the mix and look at the Gibbs measure. To describe what configurations dominate the partition function, and thus the Gibbs measure, we identify the decay rate of the probability
\begin{equation}
%\label{}
\mu_0\Bigl(\,\texte^{\frac{\beta}{2N}\sum_{x,y=1}^N S_x\cdot S_y}1_{\{\sum_xS_x\,\approx\, mN\}}\Bigr)
=\texte^{-N\Phi_\beta(m)+o(N)}
\end{equation}
Here the rate function
\begin{equation}
\label{MFFEF}
\Phi_\beta(m):=-\frac\beta2|m|^2-\scrS(m)
\end{equation}
is the desired \emph{mean-field free-energy function}. \emph{free energy} The physical solutions are clearly obtained as the \emph{absolute} minima of~$m\mapsto\Phi_\beta(m)$. This is actually completely consistent with \eqref{MFE}:

\begin{lemma}
\label{lemma-min-FEF}
We have
\begin{equation}
%\label{}
\nabla\Phi_\beta(m)=0\qquad\Leftrightarrow\qquad m=\nabla G(\beta m)
\end{equation}
Explicitly, the solutions to \eqref{MFE} are in bijection with the extreme points of~$m\mapsto\Phi_\beta(m)$.
\end{lemma}

\begin{proofsectB}{Proof}
This is a simple exercise on the Legendre transform. First we note that $\nabla\Phi_\beta(m)=0$ is equivalent to~$\beta m=-\nabla \scrS(m)$. The convexity of~$G$ implies that there is a unique~$h_m$ such that~$\scrS(m)=G(h_m)-m\cdot h_m$. Furthermore, $h_m$ depends smoothly on~$m$ and we have $\nabla G(h_m)=m$. It is easy to check that then~$\nabla \scrS(m)=-h_m$. Putting this together with our previous observations, we get that
\begin{equation}
%\label{}
\nabla\Phi_\beta(m)=0\quad\Leftrightarrow\quad\beta m=h_m\quad\Leftrightarrow\quad m=\nabla G(\beta m)
\end{equation}
It remains to observe that $m=\nabla G(\beta m)$ is a concise way to write \eqref{MFE}.
\end{proofsectB}

Lemma~\ref{lemma-min-FEF} shows that the appearance of multiple solutions to \eqref{MFE} coincides with the emergence of secondary local maxima/minima.

\subsection{Example: the Potts model} 
\index{Potts model!mean-field theory of}
It is worthwhile to demonstrate the above general formalism on the explicit example of the Potts model.
We will work with the tetrahedral representation, i.e., on the spin space $\Omega:=\{\hatv_1,\dots,\hatv_q\}$. The mean-field free energy function is best expressed in the parametrization using the \emph{mole fractions}, $x_1,\dots,x_q$, which on the complete graph represent the fractions of all vertices with spins pointing in the directions $\hatv_1,\dots,\hatv_q$, respectively. Clearly,
\begin{equation}
%\label{}
\sum_{i=1}^qx_i=1
\end{equation}
The corresponding magnetization vector is
\begin{equation}
%\label{}
m=x_1\hatv_1+\cdots+x_q\hatv_q.
\end{equation}
In this notation we have
\begin{equation}
%\label{}
\Phi_\beta(m)=\sum_{k=1}^q\Bigl(-\frac\beta2x_k^2+x_k\log x_k\Bigr).
\end{equation}
It is not surprising, but somewhat non-trivial to prove (see~\cite[Lemma~4.4]{BC}) that all interesting behavior of~$\Phi_\beta$ occurs ``on-axes'' that is, the absolute minimizers --- and, in fact, all local extrema --- of $\Phi_\beta$  occur in the directions of one of the spin states. (Which direction we choose is immaterial as they are related by symmetry.) The following picture shows the qualitative look of the function $\mathfrak m\mapsto\Phi_\beta(\mathfrak m\hatv_1)$ at four increasing values of~$\beta$:

\vglue0.2cm
\centerline{\includegraphics[width=4.7in]{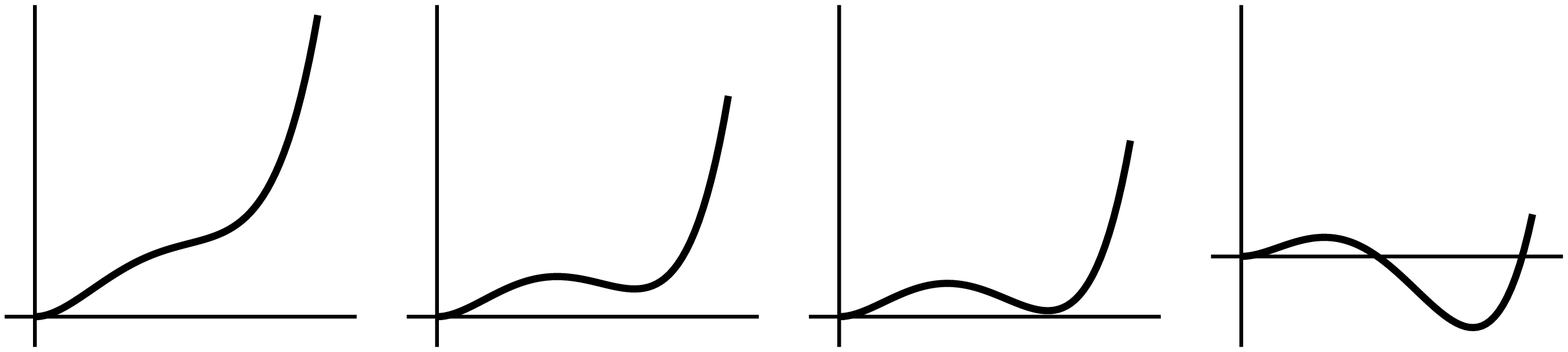}}
\medskip
\noindent
Here the function first starts convex and, as $\beta$ increases, develops a secondary local minimum (plus an inevitable local maximum). For $\beta$ even larger, the secondary minimum becomes degenerate with the one at~$m=0$ and eventually takes over the role of the global minimum. With these new distinctions, the plot of solutions to the mean-field equation for the magnetization becomes:

\vglue0.3cm
\centerline{\includegraphics[width=3in]{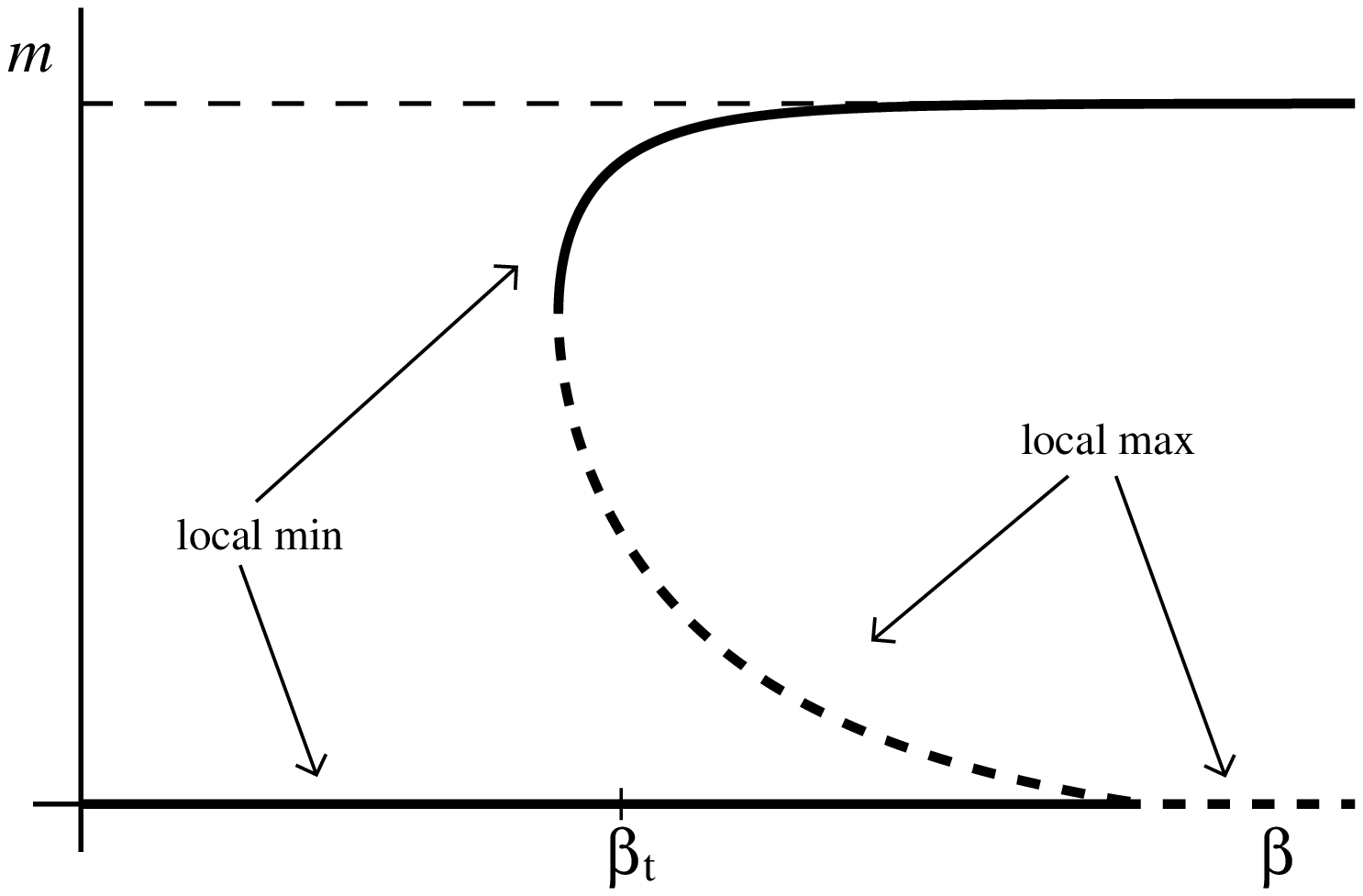}}
\noindent
Note that the local maximum eventually merges with the local minimum at zero --- at which point zero becomes a local maximum. The jump in the position of the global minimum occurs at some $\betat$, which is strictly larger than the point~$\beta_0$ where the secondary minima/maxima first appear.

\subsection{Approximation theorem \& applications}
The goal of this section is to show that, with the help of the IRB, the conclusions of mean-field theory can be given a quantitative form. Throughout we restrict ourselves to interactions of the form \eqref{Ham} and the coupling constants being one of the 3 types above.

\begin{definition}
We say that a measure $\mu\in\frakG_\beta$ is a \emph{torus state} if it is either a (weak) cluster point of measures~$\mu_{L,\beta}$ or can be obtained from such cluster points by perturbing either $\beta$ or $\mu_0$ or the inner product between spins.
\end{definition}

The reason for the second half of this definition is that the ``operations'' thus specified preserve the validity of the IRB. For such states we prove:

\begin{theorem}
\label{thm-MFT}
Suppose $|S_x|\le1$. Let~$\mu\in\frakG_\beta$ be a translation-invariant, ergodic, torus state and define
\begin{equation}
%\label{}
m_\star:=E_\mu(S_0).
\end{equation}
Let~$\Phi_\beta$ be the mean-field free energy function corresponding to this model. Then
\begin{equation}
\label{Phi-error}
\Phi_\beta(m_\star)\le\inf_{m\in\text{\rm Conv}(\Omega)}\Phi_\beta(m)+\frac{\nu\beta}2\II_d
\end{equation}
where
\begin{equation}
%\label{}
\II_d:=\int_{[-\pi,\pi]^d}\frac{\textd k}{(2\pi)^d}\frac{\hat J(k)^2}{1-\hat J(k)}
\end{equation}
\end{theorem}

Note that the integral is finite iff the random walk corresponding to~$(J_{xy})$ is transient. However, unlike for Green's function, $\II_d$ represents the expected number of returns back to the origin after the walk has left the origin. Thus, in strongly transient situations one should expect that~$\II_d$ is fairly small. And, indeed, we have the following asymptotics:
\begin{itemize}
\item
\emph{n.n.\ interactions}:
\begin{equation}
%\label{}
\II_d\sim\frac1{2d},\qquad d\to\infty.
\end{equation}
\item
\emph{Yukawa potentials}: If~$d\ge3$,
\begin{equation}
%\label{}
\II_d\le C\mu^d.
\end{equation}
\item
\emph{Power-law potentials}: If~$d\ge3$ OR $s<\min\{d+2,2d\}$, 
\begin{equation}
%\label{}
\II_d\le C(s-d).
\end{equation}
\end{itemize}
Of course, one is able to make the integral small for interactions with power law tails even when~$s$ is not too close to~$d$: Just take a mixture of Yukawa and power-law with positive coefficients and let~$\mu$ be sufficiently small. Within the class of above models, we can rephrase Theorem~\ref{thm-MFT} as:

%\bigskip
\begin{center}
\it Physical magnetizations nearly minimize
\\the mean-field free energy function
\end{center}
%\bigskip

This is justified because, as it turns out, all relevant magnetizations can be achieved in ergodic torus states. Let us again demonstrate the conclusion on the example of the $q$-state Potts model:

\begin{theorem}
\label{thm-Potts}
\index{Potts model!phase transition in}
Let~$q\ge3$ and suppose that~$\II_d\ll\ffracB 1q$. Then there is $\betat\in(0,\infty)$ and translation-invariant, ergodic measures~$\nu_0,\nu_1,\dots\nu_q\in\frakG_{\betat}$ such that
\begin{equation}
%\label{}
|E_{\nu_0}(S_x)|\ll1
\end{equation}
and
\begin{equation}
\label{MB3.22}
E_{\nu_j}(S_x) = m_\star\,\hatv_j,\qquad j=1,\dots,q,
\end{equation}
where~$m_\star\ge\ffracB12$. In particular, the $3$-state Potts model undergoes a first-order phase transition provided the spatial dimension is sufficiently large.
\end{theorem}

This result is pretty much the consequence of the pictures in Sect.~\ref{secMFT}. Indeed, including the error bound \eqref{Phi-error}, the physical magnetization is confined to the shaded regions:

\vglue0.3cm
\centerline{\includegraphics[width=4.5in]{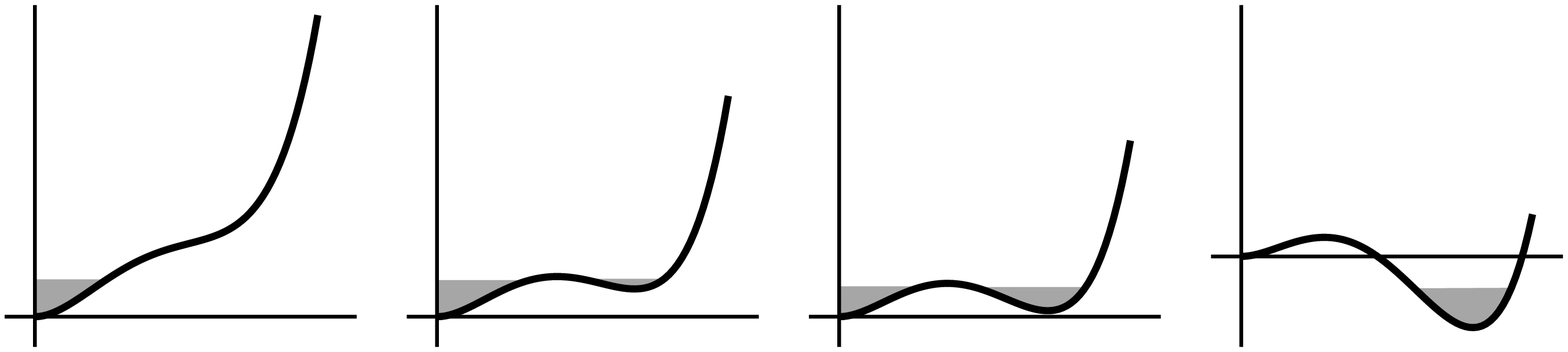}}
\smallskip

\noindent
Thus, once the error is smaller than the ``hump'' separating the two local minima, there is no way that the physical magnetization can change continuously as the temperature varies. This is seen even more dramatically once we mark directly into the mean-field magnetization plot the set of values of the magnetization allowed by the inequality \eqref{Phi-error}:

\vglue0.3cm
\centerline{\includegraphics[width=2.9in]{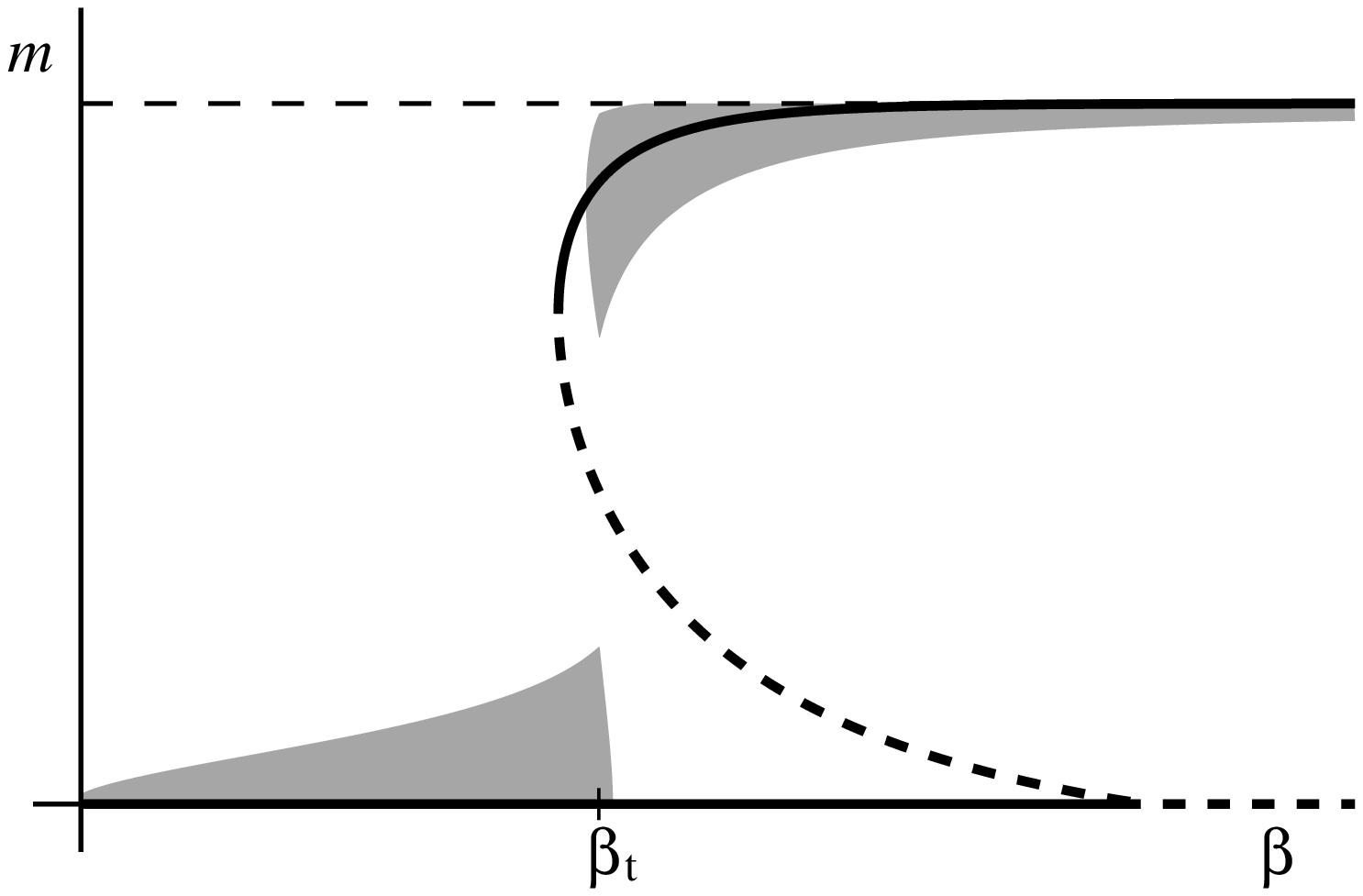}}
\smallskip

\noindent
(To emphasize the effect, the plots are done for the $q=10$ state Potts model rather than the most interesting case of~$q=3$.) Notice that the transition is bound to occur rather sharply and very near the mean-field value of~$\betat$; explicit error bounds can be derived, but there is no need to state them here.

An additional argument is actually needed to provide a full proof of \eqref{MB3.22}. Indeed, we claim that the symmetry breaking happens \emph{exactly} in the direction of one of the spin states while the approximation by mean-field theory only guarantees that the expectation is \emph{near} one of these directions.

\begin{proofsectB}{Proof of \eqref{MB3.22}, sketch}
Consider an ergodic Gibbs state~$\mu$ with $m_\star:=E_\mu(S_x)\ne0$ at inverse temperature~$\beta$. Given a sample~$\sigma=(\sigma_x)$ from~$\mu$, at each \emph{unordered} pair~$\langle x,y\rangle$ of vertices from~$\Z^d$ let
\begin{equation}
%\label{}
\eta_{xy}:=1_{\{\sigma_x=\sigma_y\}}Z_{xy}
\end{equation}
where~$(Z_{xy})$ are \emph{a priori} independent, zero-one valued random variables with
\begin{equation}
%\label{}
\BbbP(Z_{xy}=1)=1-\BbbP(Z_{xy}=0):=1-\texte^{-\beta J_{xy}}
\end{equation}
This defines a coupling of~$\mu$ with a \emph{random cluster measure} --- the distribution of the~$\eta$'s --- which, by the fact that the extension comes from i.i.d.\ random variables, is also ergodic. \index{random-cluster model}

When $m_\star\ne0$, the $\eta$-marginal features a unique infinite connected component of edges $\langle x,y\rangle$ with~$\eta_{xy}=1$ whose (site) density is proportional to~$|m_\star|$. By the construction, the spin variables take a (constant) value on each connected component, which is a.s.~unique (by ergodicity) on the infinite one and uniform on the finite ones. Thus, the bias of the spin distribution comes only from the infinite component and so it points in one of the $q$ spin directions. The claim thus follows.
\end{proofsectB}

\subsection{Ideas from the proofs}
A fundamental technical ingredient of the proof is again provided by the IRB, so throughout we will assume one of the three interactions discussed above. However, we will need the following enhanced version:

\begin{lemma}[IRB enhanced]
%\label{lemma}
Suppose the random walk driven by the $(J_{xy})$ is transient and let~$G(x,y)$ denote the corresponding Green's function on~$\Z^d$. Let~$\mu\in\frakG_\beta$ be a translation-invariant, ergodic, torus state and let us denote~$m_\star:=E_\mu(S_0)$. Then for all~$(v_x)_{x\in\Z^d}\in\C^{\Z^d}$ with finite support,
\begin{equation}
\label{v-IRB}
\sum_{x,y}v_x\bar v_y\,E_\mu\bigl((S_x-m_\star)\cdot(S_y-m_\star)\bigr)\le\frac{\nu}{2\beta}\sum_{x,y}v_x\bar v_y G(x,y).
\end{equation}
\end{lemma}

\begin{proofsectB}{Proof}
The IRB on torus survives weak limits and so we know that, for every~$(w_x)$ with finite support \emph{and} $\sum_xw_x=0$,
\begin{equation}
\label{w-IRB}
\sum_{x,y}w_x\bar w_y\,E_\mu\bigl(S_x\cdot S_y\bigr)\le\frac{\nu}{2\beta}\sum_{x,y}w_x\bar w_y G(x,y)
\end{equation}
where
\begin{equation}
%\label{}
G(x,y):=\lim_{L\to\infty}G_L(x,y)=\int_{[-\pi,\pi]^d}\frac{\textd k}{(2\pi)^d}\frac{\texte^{\texti k\cdot (x-y)}}{1-\hat J(k)}
\end{equation}
What separates \eqref{w-IRB} from \eqref{v-IRB} are the~$m_\star$ terms in the expectation on the left and the absence of the restriction on the sum of~$v_x$. The former is remedied easily; indeed, the restriction $\sum_xw_x=0$ allows us to put the $m_\star$ terms at no additional cost.

To address the latter issue, suppose $(v_x)$ has finite support but let now $\sum_xv_x$ be arbitrary. Let~$\Lambda_L\subset\Z^d$ contain the support of~$(v_x)$. To convert to the previous argument,  let
\begin{equation}
%\label{}
a_L:=\frac1{|\Lambda_L|}\sum_xv_x
\end{equation}
and
\begin{equation}
%\label{}
w_x:=v_x-a_L1_{\Lambda_L}(x)
\end{equation}
Note that $\sum_xw_x=0$. Then
\begin{multline}
\label{3.25}
\sum_{x,y}w_x\bar w_y\,E_\mu\bigl((S_x-m_\star)\cdot(S_y-m_\star)\bigr)
=\sum_{x,y}v_x\bar v_y\,E_\mu\bigl((S_x-m_\star)\cdot(S_y-m_\star)\bigr)\quad
\\\quad-2E_\mu\biggl(\Bigl[a_L\sum_{x\in\Lambda_L}(S_x-m_\star)\Bigr]\,\cdot\,\Bigl[\sum_yv_y(S_y-m_\star)\Bigr]\biggr)
\\\quad+E_\mu\biggl(\Bigl|a_L\sum_{x\in\Lambda_L}(S_x-m_\star)\Bigr|^2\biggr)
\quad
\end{multline}
But ergodicity of~$\mu$ implies that
\begin{equation}
%\label{}
E_\mu\biggl(\Bigl|\frac1{|\Lambda_L|}\sum_{x\in\Lambda_L}(S_x-m_\star)\Bigr|^2\biggr)\,\underset{L\to\infty}\longrightarrow\,0
\end{equation}
and so, by Cauchy-Schwarz, the last two terms in \eqref{3.25} converge to zero as~$L\to\infty$. Now apply \eqref{w-IRB} and pass to the limit $L\to\infty$ there. A direct calculation (and the Riemann-Lebesgue lemma) shows that
\begin{equation}
%\label{}
\frac1{|\Lambda_L|}\sum_{x\in\Lambda_L}G(x,y)\,\underset{L\to\infty}\longrightarrow\,0
\end{equation}
and so the terms involving~$a_L$ on the right-hand side of \eqref{v-IRB} suffer a similar fate. This means that the left-hand sides of \twoeqref{v-IRB}{w-IRB} tend to each other, and same for the right-hand sides. The desired bound \eqref{v-IRB} is thus a limiting version of \eqref{w-IRB}.
\end{proofsectB}

Clearly, the restriction to finitely-supported $(v_x)$ is not necessary; instead, one can consider completions of this set in various reasonable norms. The above formulation has an immediate, but rather fundamental, consequence:

\begin{corollary}[Key estimate]
\label{cor-key-estimate}
Let~$\mu\in\frakG_\beta$ be an ergodic torus state and let $m_\star:=E_\mu(S_x)$. Then we have
\begin{equation}
\label{var-bd}
E_\mu\biggl(\,\Bigl|\sum_x J_{0,x}\,S_x-m_\star\Bigr|^2\biggr)\le\frac{\nu}{2\beta}\,\II_d.
\end{equation}
\end{corollary}

\begin{proofsectB}{Proof}
Choose~$v_x:=J_{0,x}$ and note that with this choice the left-hand side of \eqref{v-IRB} becomes the left-hand side of \eqref{var-bd}. As to the right-hand side of \eqref{v-IRB}, we get
\begin{equation}
%\label{}
\frac\nu{2\beta}\sum_{x,y}\int_{[-\pi,\pi]^d}\frac{\textd k}{(2\pi)^d}\frac{\texte^{\texti k\cdot (x-y)}}{1-\hat J(k)}J_{0,x}J_{0,y}
\end{equation}
Recalling the definition of~$\hat J(k)$, this yields the desired error term.
\end{proofsectB}

This corollary provides a justification of the \emph{ad hoc} step in the derivation of mean-field theory: Indeed, once $\II_d$ is small, the variance of~$M_0$ is small and so~$M_0$ \emph{is} with high probability close to its average. 

The rest of the proof of Theorem~\ref{thm-MFT} is based on inequalities linking the mean-field free energy with the actual magnetization of the system; this part of the proof works for general non-negative coupling constants satisfying conditions (I1-I2) from Sect.~\ref{sec-RW}. The relevant observations are as follows:

\begin{proposition}
\label{prop-convex-ineq}
Let $\mu\in\frakG_\beta$ be translation invariant and let $m_\star:=E_\mu(S_x)$.

\noindent
(1) We have
\begin{equation}
\label{Ebd}
\Phi_\beta(m_\star)\le\inf_{m\in\text{\rm Conv}(\Omega)}\Phi_\beta(m)+\frac\beta2\sum_{x\in\Z^d}J_{0,x}\bigl[E_\mu(S_0\cdot S_x)-|m_\star|^2\bigr]
\end{equation}

\noindent
(2) Suppose also $J_{0,x}\ge0$ and $|S_x|\le1$. Then
\begin{equation}
\label{MVbd}
\sum_{x\in\Z^d}J_{0,x}\bigl[E_\mu(S_0\cdot S_x)-|m_\star|^2\bigr]
\le \beta\,E_\mu\biggl(\,\Bigl|\sum_x J_{0,x}\,S_x-m_\star\Bigr|^2\biggr)
\end{equation}
\end{proposition}

\begin{proofsectB}{Proof of (1)}
The proof is based on convexity inequalities linking the mean-field free energy and the characteristics of the actual system. Fix~$\Lambda\subset\Z^d$ and let~$Z_\Lambda$ be the partition function in~$\Lambda$. A standard example of such convexity inequality is
\begin{equation}
\label{Z-bound}
Z_\Lambda\ge\exp\Bigl\{-|\Lambda|\inf_{m\in\text{\rm Conv}(\Omega)}\Phi_\beta(m)+O(\partial\Lambda)\Bigr\}.
\end{equation}
To prove this we pick $m$ in the (relative) interior of~$\text{\rm Conv}(\Omega)$ and define a \emph{tilted} measure
\begin{equation}
%\label{}
\mu_h(\textd S):=\texte^{h\cdot S-G(h)}\mu_0(\textd S)
\end{equation}
with~$h$ adjusted so that $E_{\mu_h}(S)=m$. (Such~$h$ exists for each~$m$ in the relative interior of~$\text{\rm Conv}(\Omega)$, by standard arguments for the Legendre transform.) We then get
\begin{equation}
%\label{}
Z_\Lambda = E_{\otimes\mu_h}\bigl(\texte^{-\beta H_\Lambda(S)-h\cdot M_\Lambda+|\Lambda|G(h)}\bigr)
\end{equation}
where we introduced the shorthand
\begin{equation}
%\label{}
M_\Lambda:=\sum_{x\in\Lambda}S_x
\end{equation}
Now apply Jensen to get the expectation into the exponent; the product nature of~$\otimes\mu_h$ implies that $E_{\otimes\mu_h}(H_\Lambda(S))=-|\Lambda|\frac12|m|^2+O(\partial\Lambda)$ and so \eqref{Z-bound} follows by noting that $G(h)-h\cdot m=\scrS(m)$ due to our choice of~$h$, and subsequently optimizing over all admissible~$m$.

Now fix a general~$h\in\R^\nu$ and let~$\mu$ be a Gibbs measure as specified in the claim. First we note that the DLR condition implies
\begin{equation}
%\label{}
E_\mu(\texte^{+\beta H_\Lambda+h\cdot M_\Lambda}Z_\Lambda)=E_{\otimes\mu_0}\bigl(\texte^{h\cdot M_\Lambda}\bigr)=\texte^{|\Lambda|G(h)}
\end{equation}
The~$Z_\Lambda$ term can be bounded away via \eqref{Z-bound}; Jensen's inequality then gives
\begin{equation}
%\label{}
\beta\, E_\mu(H_\Lambda)+|\Lambda|\,h\cdot m_\star-|\Lambda|\inf_{m\in\text{\rm Conv}(\Omega)}\Phi_\beta(m)+O(\partial\Lambda)
\le|\Lambda|G(h)
\end{equation}
Next, translation invariance of~$\mu$ yields
\begin{equation}
%\label{}
E_\mu(H_\Lambda)=-|\Lambda|\frac12\sum_x J_{0,x}E_\mu(S_0\cdot S_x)+O(\partial\Lambda)
\end{equation}
and so dividing by~$\Lambda$ and taking~$\Lambda\uparrow\Z^d$ along cubes gets us
\begin{equation}
%\label{}
-\frac\beta2\sum_x J_{0,x}E_\mu(S_0\cdot S_x)\,-\!\inf_{m\in\text{\rm Conv}(\Omega)}\Phi_\beta(m)\le
G(h)-h\cdot m_\star
\end{equation}
Optimizing over~$h$ turns the right-hand side into~$\scrS(m_\star)$. Adding~$\frac12|m_\star|^2$ on both sides and invoking \eqref{MFFEF} now proves the claim.
\end{proofsectB}

\begin{proofsectB}{Proof of (2)}
Let us return to the notation $M_0:=\sum_xJ_{0,x}S_x$. The left-hand side of \eqref{MVbd} can then be written as~$E_\mu(S_0\cdot M_0)-|m_\star|^2$. Since~$J_{0,0}=0$, an application of the DLR condition yields
\begin{equation}
%\label{}
E_\mu(M_0\cdot S_0)=E_\mu\bigl(M_0\cdot\nabla G(\beta M_0)\bigr)
\end{equation}
The DLR condition also implies
\begin{equation}
%\label{}
m_\star=E_\mu(M_0)=E_\mu[\nabla G(\beta M_0)]
\end{equation}
and so we have
\begin{multline}
%\label{}
\qquad
E_\mu(S_0\cdot M_0)-|m_\star|^2
\\=E_\mu\Bigl((M_0-m_\star)\cdot\bigl(\nabla G(\beta M_0)-\nabla G(\beta m_\star)\bigr)\Bigr)
\qquad
\end{multline}
But~$|S_x|\le1$ implies that the Hessian of~$G$ is dominated by the identity, $\nabla\nabla G(m)\le\text{id}$ at any $m\in\text{Conv}(\Omega)$ --- assuming~$J_{x,y}\ge0$ --- and so
\begin{equation}
%\label{}
(M_0-m_\star)\cdot\bigl(\nabla G(\beta M_0)-\nabla G(\beta m_\star)\bigr)\le\beta|M_0-m_\star|^2
\end{equation}
by the Mean-Value Theorem. Taking expectations proves \eqref{MVbd}. 
\end{proofsectB}

Theorem~\ref{thm-MFT} now follows by combining Proposition~\ref{prop-convex-ineq} with Corollary~\ref{cor-key-estimate}.
Interestingly, \eqref{Ebd} gives
\begin{equation}
%\label{}
\sum_{x\in\Z^d}J_{0,x}E_\mu(S_0\cdot S_x)\ge |m_\star|^2
\end{equation}
i.e., the actual energy density always exceeds the mean-field energy density.

\subsection{Literature remarks}
The inception of mean-field theory goes back to Curie~\cite{Curie} and Weiss~\cite{Weiss}. One of the early connections to the models on the complete graph appears in Ellis' textbook on large-deviation theory~\cite{Ellis}. Most of this section is based on the papers of Biskup and Chayes~\cite{BC} and Biskup, Chayes and Crawford~\cite{Biskup-Chayes-Crawford}. The Key Estimate had been used before in some specific cases; e.g., for the Ising model in the paper by Bricmont, Kesten, Lebowitz and Schonmann~\cite{BKLS} and for the $q$-state Potts model in the paper by Kesten and Schonmann~\cite{KS1Bi}. Both these works deal with the limit of the magnetization as~$d\to\infty$; notwithstanding, no conclusions were extracted for the presence of first-order phase transitions in finite-dimensional systems.

The first-order phase transition in the $q$-state Potts model has first been proved by Koteck\'y and Shlosman~\cite{KS1Bi} but the technique works only for extremely large~$q$. The case of small~$q$ has been open. The upshot of the present technique is that it replaces $q$ by $d$ or interaction range in its role of a ``large parameter.'' The price to pay is the lack of explicit control over symmetry: We expect that the measure $\nu_0$ in Theorem~\ref{thm-Potts} is actually ``disordered'' and $E_{\nu_0}(S_x)=0$. This would follow if we knew that the magnetization in the Potts model can be discontinuous only at the percolation threshold --- for the Ising model this was recently proved by Bodineau~\cite{Bodineau} --- but this is so far known only in $d=2$ (or for~$q$ very large). The coupling in the proof of \eqref{MB3.22} is due to Edwards and Sokal \cite{ES1Bi}; for further properties see Grimmett~\cite{Grimmett} or Biskup, Borgs, Chayes and Kotecky~\cite{BBCK}. The uniqueness of the infinite connected component is well known in the nearest-neighbor case from a beautiful argument of Burton and Keane~\cite{Burton-Keane}; for the long-range models it has to be supplied by a percolation bound dominating the number of edges connecting a box of side~$L$ to its complement.

The requirement $\II_d\ll\ffracB1q$ is actually an embarrassment of the theory as the transition should become more pronounced, and thus easier to control, with increasing~$q$. Thus, even for nearest-neighbor case, we still do not have a dimension in which all~$q\ge3$ state Potts models go first order. (It is expected that this happens already in~$d=3$.)
The restriction to transient dimensions is actually not absolutely necessary; cf recent work Chayes~\cite{Chayes}.

It is natural to ask whether one can say anything about the continuum-$q$ extension of the Potts model, the random cluster model; see Grimmett~\cite{Grimmett}. Unfortunately, the main condition for proving the IRB, reflection positivity, holds if \emph{and only if}~$q$ is integer (Biskup~\cite{Biskup}).

Another model for which this method yields a novel result is the liquid-crystal model discussed in Sect.~\ref{sec1.2}. Here Angelescu and Zagrebnov~\cite{Angelescu-Zagrebnov} proved that symmetry breaking (for the order parameter~$\max_\alpha E_\mu[S_x^{(\alpha)}]^2-\ffracB1n$) occurs at low temperatures by exhibiting spin-wave condensation; cf remarks at the end of Chapter~\ref{chap-2}. In~\cite{BC} it has been shown that, for~$n\ge3$, the order parameter undergoes a discontinuous transition at intermediate temperatures; van Enter and Shlosman~\cite{Senya_VE-I,vanEnter-Shlosman} later proved such transitions in highly non-linear cases. Similar ``mean-field driven'' first order phase transitions have also been proved for the cubic model~\cite{BC} and the Blume-Capel model~\cite{Biskup-Chayes-Crawford}. 

Once the general theory is in place, the proof of a phase transition for a specific model boils down to the analysis of the mean-field free energy function. While in principle always doable, in practice this may be quite a challenge even in some relatively simple examples. See, e.g., \cite[Sect.~4.4]{BC} what this requires in the context of the liquid-crystal model.

Finally, we note that the IRB has been connected to mean-field theory before; namely, in the work of Aizenman~\cite{Aizenman} (cf also Fr\"ohlich~\cite{Frohlich} and Sokal~\cite{Sokal}) in the context of lattice field theories and that of Aizenman and Fern\'andez in the context of Ising systems in either high spatial dimensions~\cite{AF1} or for spread-out interactions~\cite{AF2}. A representative result from these papers is that \emph{the critical exponents in the Ising model take mean-field values above 4 dimensions}. The IRB enters as a tool to derive a one-way bound on the critical exponents. Unfortunately, the full conclusions are restricted to interactions that are reflection positive; a non-trivial extension was obtained recently by Sakai~\cite{Sakai} who proved the IRB --- and the corresponding conclusions about the critical exponents --- directly via a version of the lace expansion.

\section{Reflection positivity}
\label{chap-4}

In the last two sections we have made extensive use of the infrared bound. Now is the time to prove it. This will require introducing the technique of reflection positivity which, somewhat undesirably, links long-range correlation properties of the spin models under consideration to the explicit structure of the underlying graph. Apart from the infrared bound, reflection positivity yields also the so called chessboard estimate which we will use extensively in Chapter~\ref{chap-5}.

\subsection{Reflection positive measures}
We begin by introducing the basic setup for the definition of reflection positivity: Consider the torus~$\T_L$ of side~$L$ with~$L$ even. The torus has a natural reflection symmetry along planes orthogonal to one of the lattice directions. (For that purpose we may think of~$\T_L$ as embedded into a continuum torus.) The corresponding ``plane of reflection'' $P$ has two components, one at the ``front'' of the torus and the other at the ``back.'' The plane either passes through the sites of~$\T_L$ or bisects bonds; we speak of reflections \emph{through sites} or \emph{through bonds}, respectively.
The plane splits the torus into two halves, $\T_L^+$ and~$\T_L^-$, which are disjoint for reflections through bonds and obey~$\T_L^+\cap\T_L^-=P$ for reflections through~sites.

Let~$\mathfrak A^\pm$ denote the set of all functions~$f\colon\Omega^{\T_L}\to\R$ that depend only on the spins in~$\T_L^\pm$. Let~$\vartheta$ denote the reflection operator, $\vartheta\colon\mathfrak A^\pm\to\mathfrak A^\mp$, which acts on spins via
\begin{equation}
%\label{}
\vartheta(S_x):=S_{\vartheta(x)}
\end{equation}
Clearly, $\vartheta$ is a morphism of algebra $\mathfrak A^+$ onto $\mathfrak A^-$ and~$\vartheta^2=\text{id}$.

\begin{definition}[Reflection positivity]
\label{def-RP}
\index{reflection positivity}
A measure $\mu$ on~$\Omega^{\T_L}$ is \emph{reflection positive} (RP) with respect to~$\vartheta$ if
\settowidth{\leftmargini}{(11)}
\begin{enumerate}
\item[(1)]
For all~$f,g\in\mathfrak A^+$,
\begin{equation}
\label{fg=gf}
E_\mu(f\,\vartheta g)=E_\mu(g\,\vartheta f)
\end{equation}
\item[(2)]
For all~$f\in\mathfrak A^+$,
\begin{equation}
\label{RP-positive}
E_\mu(f\,\vartheta f)\ge0
\end{equation}
\end{enumerate}
\end{definition}

Note that the above implies that~$f,g\mapsto E_\mu(f\,\vartheta g)$ is a positive-semindefinite symmetric bilinear form. Condition \eqref{fg=gf} is usually automatically true --- it requires only $\vartheta$-invariance of~$\mu$ --- so it is the second condition that makes this concept non-trivial (hence also the name). Here we first note that the concept is not entirely vacuous:

\begin{lemma}
\label{lemma-product-RP}
The product measure, $\mu=\bigotimes\mu_0$, is RP with respect to all reflections.
\end{lemma}
 
\begin{proofsectB}{Proof}
First consider reflections through bonds. Let~$f,g\in\mathfrak A^+$. Since $\T_L^+\cap\T_L^-=\emptyset$, the random variables $f$ and~$\vartheta g$ are independent under~$\mu$. Hence,
\begin{equation}
%\label{}
E_\mu(f\,\vartheta g)=E_\mu(f)E_\mu(\vartheta g)=E_\mu(f)E_\mu(g)
\end{equation}
whereby both conditions in Definition~\ref{def-RP} follow.

For reflections through sites, we note that~$f$ and~$\vartheta g$ are independent conditional on~$S_P$. Invoking the reflection symmetry of $\mu(\cdot|S_P)$, we get
\begin{equation}
%\label{}
E_\mu(f\,\vartheta g|S_P)=E_\mu(f|S_P)E_\mu(\vartheta g|S_P)=E_\mu(f|S_P)E_\mu(g|S_P)
\end{equation}
Again the conditions of RP follow by inspection.
\end{proofsectB}

A fundamental consequence of reflection positivity is the Cauchy-Schwarz inequality
\begin{equation}
\label{CS}
\bigl[E_\mu(f\,\vartheta g)\bigr]^2\le E_\mu(f\,\vartheta f)E_\mu(g\,\vartheta g)
\end{equation}
Here is an enhanced, but extremely useful, version of this inequality:

\begin{lemma}
\label{lemma-CEsimple}
Let~$\mu$ be RP with respect to~$\vartheta$ and let~$A,B,C_\alpha,D_\alpha\in\mathfrak A^+$. Then
\begin{multline}
%\label{}
\quad
\bigl[E_\mu(\texte^{A+\vartheta B+\sum_\alpha C_\alpha\,\vartheta D_\alpha})\bigr]^2
\\
\le\bigl[E_\mu(\texte^{A+\vartheta A+\sum_\alpha C_\alpha\,\vartheta C_\alpha})\bigr]\,
\bigl[E_\mu(\texte^{B+\vartheta B+\sum_\alpha D_\alpha\,\vartheta D_\alpha})\bigr]
\quad
\end{multline}
\end{lemma}

\begin{proofsectB}{Proof}
Clearly, in the absence of the $C_\alpha\,\vartheta D_\alpha$ terms, this simply reduces to \eqref{CS}. To include these terms we use expansion into Taylor series:
\begin{multline}
\label{4.7}
\quad
E_\mu(\texte^{A+\vartheta B+\sum_\alpha C_\alpha\,\vartheta D_\alpha})
\\=\sum_{n\ge0}\frac1{n!}\sum_{\alpha_1,\dots,\alpha_n}E_\mu\bigl(\,(\texte^A C_{\alpha_1}\dots C_{\alpha_n})\,\vartheta(\texte^B D_{\alpha_1}\dots D_{\alpha_n})\bigr)
\quad
\end{multline}
Now we apply \eqref{CS} to the expectation on the right-hand side and then one more time to the resulting sum:
\begin{equation}
\begin{aligned}
E_\mu(&\texte^{A+\vartheta B+\sum_\alpha C_\alpha\,\vartheta D_\alpha})\\
&\le\sum_{n\ge0}\frac1{n!}\sum_{\alpha_1,\dots,\alpha_n}\Bigl[E_\mu\bigl(\,(\texte^A C_{\alpha_1}\dots C_{\alpha_n})\,\vartheta(\texte^A C_{\alpha_1}\dots C_{\alpha_n})\bigr)^{1/2}\\
&\qquad\qquad\times
E_\mu\bigl(\,(\texte^B D_{\alpha_1}\dots D_{\alpha_n})\,\vartheta(\texte^B D_{\alpha_1}\dots D_{\alpha_n})\bigr)^{1/2}
\Bigr]
\\
&\le
\biggl(\sum_{n\ge0}\frac1{n!}\sum_{\alpha_1,\dots,\alpha_n}E_\mu\bigl(\,(\texte^A C_{\alpha_1}\dots C_{\alpha_n})\,\vartheta(\texte^A C_{\alpha_1}\dots C_{\alpha_n})\bigr)\biggr)^{1/2}
\\
&\qquad\times
\biggl(\sum_{n\ge0}\frac1{n!}\sum_{\alpha_1,\dots,\alpha_n}E_\mu\bigl(\,(\texte^B D_{\alpha_1}\dots D_{\alpha_n})\,\vartheta(\texte^B D_{\alpha_1}\dots D_{\alpha_n})\bigr)\biggr)^{1/2}
\end{aligned}
\end{equation}
Resummation via \eqref{4.7} now yields the desired expression.
\end{proofsectB}

The argument we just saw yields a fundamental criterion for proving reflection positivity: \index{reflection positivity!condition for}

\begin{corollary}
\label{lemma-RP-suff}
Fix a plane of reflection~$P$ and let~$\vartheta$ be the corresponding reflection operator. Suppose that the torus Hamiltonian takes the form 
\begin{equation}
\label{torus-H-RP}
-H_L=A+\vartheta A+\sum_\alpha C_\alpha\,\vartheta C_\alpha
\end{equation}
with $A,C_\alpha\in\mathfrak A^+$. Then for all $\beta\ge0$ the torus Gibbs measure, $\mu_{L,\beta}$, is RP with respect to~$\vartheta$.
\end{corollary}

\begin{proofsectB}{Proof}
The proof is a simple modification of the argument in Lemma~\ref{lemma-CEsimple}: Fix $f,g\in\mathfrak A^+$. Expansion of the exponential term in $\sum_\alpha C_\alpha\,\vartheta C_\alpha$ yields
\begin{equation}
\begin{aligned}
&E_{\mu_{L,\beta}}(f\vartheta g)=\frac1{Z_L}E_{\otimes\mu_0}\bigl(f(\vartheta g)\,\texte^{\,\beta(A+\vartheta A+\sum_\alpha C_\alpha\,\vartheta C_\alpha)}\bigr)
\\
&\quad=\frac1{Z_L}\sum_{n\ge0}\frac1{n!}\sum_{\alpha_1,\dots,\alpha_n}
E_{\otimes\mu_0}\bigl(\,(f\texte^{\,\beta A}C_{\alpha_1}\cdots C_{\alpha_n})\,\vartheta(g\texte^{\,\beta A}C_{\alpha_1}\cdots C_{\alpha_n})\bigr)
\end{aligned}
\end{equation}
The conditions of RP for~$\mu_{L,\beta}$ are now direct consequences of the fact that the product measure, $\bigotimes\mu_0$, is itself RP (cf Lemma~\ref{lemma-CEsimple}).
\end{proofsectB}

Now we are ready to check that all 3 interactions that we focused our attention on in previous lectures are of the form in Lemma~\ref{lemma-CEsimple} and thus lead to RP torus Gibbs measures:

\begin{lemma}
\label{lemma-RPint}
For any plane~$P$, the n.n. (ferromagnet) interaction, Yukawa potentials and the power-law decaying potentials, the torus Hamiltonian can be written in the form \eqref{torus-H-RP} for some~$A,C_\alpha\in\mathfrak A^+$.
\end{lemma}

\begin{proofsectB}{Proof}
We focus on reflections through bonds; the case of reflections through sites is analogous. Given~$P$, the terms in the Hamiltonian can naturally be decomposed into three groups: those between the sites in~$\T_L^+$, those between the sites in~$\T_L^-$ and those involving both halves of the torus:
\begin{equation}
%\label{}
-H_L=\,\underbrace{\frac12\sum_{x,y\in\T_L^+}J_{xy}^{(L)}\,S_x\cdot S_y}_{\displaystyle A}\,
+\,\underbrace{\frac12\sum_{x,y\in\T_L^-}J_{xy}^{(L)}\,S_x\cdot S_y}_{\displaystyle \vartheta A}\,
+\sum_{i=1}^d
\underbrace{\sum_{\begin{subarray}{c}
x\in\T_L^+\\y\in\T_L^-
\end{subarray}}
J_{xy}^{(L)}\,S_x^{(i)} S_y^{(i)}}_{\displaystyle R_i}
\end{equation}
where we used the reflection symmetry of the~$J_{xy}^{(L)}$ to absorb the $\ffracB12$ into the sum at the cost of confining~$x$ to~$\T_L^+$ and~$y$ to~$\T_L^-$. The first two terms indentify~$A$ and~$\vartheta A$; it remains to show that the~$R_i$-term can be written as~$\sum_\alpha C_\alpha\,\vartheta C_\alpha$. We proceed on a case-by-case basis:

\smallskip
\emph{Nearest-neighbor interactions}:
Here
\begin{equation}
%\label{}
R_i=\frac1{2d}\sum_{\begin{subarray}{c}
\langle x,y\rangle\\
x\in\T_L^+\\y\in\T_L^-
\end{subarray}}\,S_x^{(i)} S_y^{(i)}
\end{equation}
which is of the desired form since~$S_y=\vartheta(S_x)$ whenever $x$ and~$y$ contribute to the above sum.

\smallskip
\emph{Yukawa potentials}: We will only prove this in~$d=1$; the higher dimensions are harder but similar. Note that if $P$ passes through the origin and~$x\in\T_L^+$ and~$y\in\T_L^-$,
\begin{equation}
%\label{}
J_{xy}^{(L)}=C\sum_{n\ge0}\texte^{-\mu(|x|+|y|+nL)}
\end{equation}
Hence,
\begin{equation}
\label{5.14}
R_i=C\sum_{n\ge0}\texte^{-\mu nL}\Bigl(\sum_{x\in\T_L^+}\texte^{-\mu|x|}S_x^{(i)}\Bigr)
\Bigl(\sum_{y\in\T_L^-}\texte^{-\mu|y|}S_y^{(i)}\Bigr)
\end{equation}
which is of the desired form.

\smallskip
\emph{Power-law potentials}:
Here we note
\begin{equation}
%\label{}
\frac1{|x-y|_1^s}=\int_0^\infty\textd\mu\,\mu^{s-1}\texte^{-\mu|x-y|_1}
\end{equation}
which reduces the problem to the Yukawa case.
\end{proofsectB}

We remark that Corollary~\ref{lemma-CEsimple} allows a minor generalization: \emph{if a torus measure~$\mu$ is RP, and a torus Hamiltonian~$H_L$ takes the form \eqref{torus-H-RP}, then also the measure $\texte^{-\beta H_L}\textd\mu$ is RP}. This may seem to be a useful tool for constructing RP measures; unfortunately, we do not know any natural measures other than product measures for which RP can be shown directly. 

\subsection{Gaussian domination}

Now we are in a position to start proving the infrared bound. First we introduce its integral version known under the name Gaussian domination:

\begin{theorem}[Gaussian domination]
\index{Gaussian domination}
%\label{thm}
Let~$(J_{xy})$ be one of the three interactions above. Fix~$\beta\ge0$ and for $h=(h_x)_{x\in\T_L}\in(\R^\nu)^{\T_L}$ define
\begin{equation}
%\label{}
Z_L(h):=E_{\bigotimes\mu_0}\biggl(\exp\Bigl\{-\beta\sum_{x,y\in\T_L}J_{xy}^{(L)}\,|S_x-S_y+h_x-h_y|^2\Bigr\}\biggr)
\end{equation}
Then
\begin{equation}
\label{GD}
Z_L(h)\le Z_L(0)
\end{equation}
\end{theorem}

\begin{proofsectB}{Proof}
Let~$H_L$ denote the sum in the exponent. It is easy to check that~$H_L$ is of the form
\begin{equation}
%\label{}
-H_L=A+\vartheta B+\sum_\alpha C_\alpha\,\vartheta D_\alpha
\end{equation}
Indeed, for~$h\equiv0$ this is simply Lemma~\ref{lemma-RPint} as the diagonal terms can always by absorbed into the \emph{a priori} measure. To get~$h\not\equiv0$ we replace~$S_x$ by~$S_x+h_x$ at each~$x$. This changes the meaning of the original terms~$A$ and~$C_\alpha$ --- and makes them different on the two halves of the torus --- but preserves the overall structure of the expression.

A fundamental ingredient is provided by Lemma~\ref{lemma-CEsimple} which yields
\begin{equation}
%\label{}
Z_L(h)^2\le Z_L(h_+)Z_L(h_-)
\end{equation}
where~$h_+:=h$ on~$\T_L^+$ and~$h_+:=\vartheta h$ on~$\T_L^-$, and similarly for~$h_-$. Now let us show how this yields~\eqref{GD}: Noting that~$Z_L(h)\to0$ whenever any component of~$h$ tends to $\pm\infty$, the maximum of~$Z_L(h)$ is achieved at some finite~$h$. Let~$h^\star$ be a maximizer for which
\begin{equation}
%\label{}
N(h):=\#\bigl\{\langle x,y\rangle\colon h_x\ne h_y\bigr\}
\end{equation}
is the smallest among all maximizers. We claim that~$N(h^\star)=0$. Indeed, if~$N(h^\star)>0$ then there exists a plane of reflection~$P$ through bonds such that~$P$ intersects at least one bond $\langle x,y\rangle$ with~$h^\star_x\ne h^\star_y$. Observe that then
\begin{equation}
%\label{}
\min\bigl\{N(h_+^\star),\,N(h_-^\star)\bigr\}<N(h^\star)
\end{equation}
Suppose without loss of generality that $N(h_+^\star)<N(h^\star)$. Then the fact that~$h^\star$ was a maximizer implies
\begin{equation}
%\label{}
Z_L(h^\star)^2\le Z_L(h_+^\star)Z_L(h_-^\star)\le Z_L(h_+^\star)Z_L(h^\star)
\end{equation}
which means
\begin{equation}
%\label{}
Z_L(h^\star)\le Z_L(h_+^\star)
\end{equation}
i.e., $h_+^\star$ is also a maximizer. But that contradicts the choice of~$h^\star$ by which~$N(h^\star)$ was already minimal possible. It follows that $N(h^\star)=0$, i.e., $h^\star$ is a constant. Since~$Z(h+c)=Z(h)$ for any constant~$c$, \eqref{GD} follows.
\end{proofsectB}

Now we can finally pay an old debt and prove the infrared bound:\index{infrared bound!proof of}

\begin{proofsectB}{Proof of Theorem~\ref{thm-IRB}}
To ease the notation, we will write throughout
\begin{equation}
%\label{}
\langle \eta,\zeta\rangle:=\sum_{x\in\T_L}\eta_x\zeta_x
\end{equation}
to denote the natural inner product on~$L^2(\T_L)$. First we note that for any~$(\eta_x)\in(\R^\nu)^{\T_L}$,
\begin{equation}
%\label{}
\sum_{x,y\in\T_L}J_{xy}^{(L)}|\eta_x-\eta_y|^2 = \langle\eta,G_L^{-1}\eta\rangle
\end{equation}
where~$G_L$ is as in \eqref{GLdef}. (Indeed, in Fourier components, $\hat G_L^{-1}(k)=1-\hat J(k)$.) As is easy to check,
\begin{equation}
\begin{aligned}
Z_L(h)&=E_{\bigotimes\mu_0}\bigl(\texte^{-\beta\langle S+h,G_L^{-1}(S+h)\rangle}\bigr)
\\
&=Z_L(0)E_{\mu_{L,\beta}}\bigl(\texte^{-2\beta\langle h,G_L^{-1}S\rangle-\beta\langle h,G_L^{-1}h\rangle}\bigr)
\end{aligned}
\end{equation}
where~$\mu_{L,\beta}$ is the torus Gibbs measure. The statement of Gaussian domination \eqref{GD} is thus equivalent to
\begin{equation}
%\label{}
E_{\mu_{L,\beta}}\bigl(\texte^{-2\beta\langle h,G_L^{-1}S\rangle}\bigr)
\le \texte^{\,\beta\langle h,G_L^{-1}h\rangle}
\end{equation}
We will now use invertibility of~$G_L$ to replace~$G_L^{-1}h$ by~$h$. This yields
\begin{equation}
%\label{}
E_{\mu_{L,\beta}}\bigl(\texte^{-2\beta\langle h,S\rangle}\bigr)
\le \texte^{\,\beta\langle h,G_Lh\rangle}
\quad \text{whenever}\quad\sum_{x\in\T_L}h_x=0
\end{equation}
where the latter condition comes from the fact that~$G_L^{-1}$ annihilates constant functions.
Next we expand both sides to quadratic order in~$h$:
\begin{multline}
%\label{}
\qquad
1-2\beta E_{\mu_{L,\beta}}\bigl(\langle h,S\rangle\bigr)+\frac{4\beta^2}2E_{\mu_{L,\beta}}\bigl(\langle h,S\rangle^2\bigr)+\dots
\\
\le1+\beta\langle h,G_Lh\rangle+\dots
\qquad
\end{multline}
Since $E_{\mu_{L,\beta}}(S)$ is constant, $E_{\mu_{L,\beta}}(\langle h,S\rangle)=\langle h, E_{\mu_{L,\beta}}(S)\rangle=0$ and we thus get
\begin{equation}
%\label{}
E_{\mu_{L,\beta}}\bigl(\langle h,S\rangle^2\bigr)\le\frac1{2\beta}\,\langle h,G_Lh\rangle
\end{equation}
Finally, choose~$h_x:=v_x\hate_i$, for some orthonormal basis vectors~$\hate_i$ in~$\R^\nu$. This singles out the $i$-th components of the spins on the left-hand side and has no noticeable effect on the right-hand side (beyond replacing vectors~$h_x$ by scalars~$v_x$). Summing the result over~$i=1,\dots,\nu$ we get the dot product of the spins on the left and an extra factor~$\nu$ on the right-hand side.
\end{proofsectB}

\subsection{Chessboard estimates}
The proof of the infrared bound was based on Lemma~\ref{lemma-CEsimple} which boils down to the Cauchy-Schwarz inequality for the inner product
\begin{equation}
%\label{}
f,g\mapsto E_\mu(f\,\vartheta g)
\end{equation}
 In this section we will systematize the use of the Cauchy-Schwarz inequality to derive bounds on correlation functions. The key inequality --- referred to as the \emph{chessboard estimate} --- will turn out to be useful in the proofs of phase coexistence in specific spin systems (even those to which the IRB technology does not apply).

Throughout we will restrict attention to reflections through planes of sites as this is somewhat more useful in applications (except for quantum systems). Pick two integers, $B<L$, such that~$B$ divides $L$ and~$\ffracB LB$ is even. Fixing the origin of the torus, let~$\Lambda_B$ the block corresponding to~$\{0,1,\dots,B\}^d$ --- i.e., the block of side~$B$ with lower-left corner at the origin. We may cover~$\T_L$ by translates of~$\Lambda_B$,
\begin{equation}
%\label{}
\T_L=\bigcup_{t\in\T_{L/B}}(\Lambda_B+Bt)
\end{equation}
noting that the neighboring translates share the vertices on the adjacent sides. (This is the specific feature of the setup based on reflections through planes of sites.) The translates are indexed by the sites in a ``factor torus'' $\T_{L/B}$.

\begin{definition}
A function~$f\colon\Omega^{\T_L}\to\R$ is called a \emph{$B$-block function} if it depends only on $\{S_x\colon x\in\Lambda_B\}$. An event~$\AA\subset\Omega^{\T_L}$ is called a \emph{$B$-block event} if $1_{\AA}$ is a $B$-block function.
\end{definition}

Given a $B$-block function~$f$, and~$t\in\T_{L/B}$, we define~$\vartheta_t f$ be the reflection of~$f$ ``into'' $\Lambda_B+Bt$. More precisely, for a self-avoiding path on~$\T_{L/B}$ connecting~$\Lambda_B$ to~$\Lambda_B+Bt$, we may sequentially reflect~$f$ along the planes between the successive blocks in the path. The result is a function that depends only on $\{S_x\colon x\in\Lambda_B+Bt\}$. Due to the commutativity of the reflections, this function does not depend on the choice of the path, so we denote it simply by~$\vartheta_t f$. Note that since reflections are involutive, $\vartheta^2=\text{id}$, there are only~$2d$ distinct functions one can obtain from~$f$ modulo translations.

\begin{theorem}[Chessboard estimate]
\label{thm-CE}
\index{chessboard estimate}
Suppose $\mu$ is RP with respect to all reflections between the neighboring blocks of the form~$\Lambda_B+Bt$, $t\in\T_{L/B}$. Then for any $B$-block functions~$f_1,\dots, f_m$, and any \emph{distinct} $t_1,\dots,t_m\in\T_{L/B}$,
\begin{equation}
%\label{}
E_{\mu}\Bigl(\,\prod_{j=1}^m\vartheta_{t_j}f_j\Bigr)\le
\prod_{j=1}^m\biggl[E_{\mu}\Bigl(\,\prod_{t\in\T_{L/B}}\vartheta_{t}f_j\Bigr)\biggr]^{(B/L)^d}
\end{equation}
\end{theorem}

Here is a version of this bound for events: If $\AA_1,\dots,\AA_m$ are $B$-block events and $t_1,\dots,t_m$ are \emph{distinct} elements of~$\T_{L/B}$, then
\begin{equation}
%\label{}
\mu\Bigl(\,\bigcap_{j=1}^m\vartheta_{t_j}(\AA_j)\Bigr)\le
\prod_{j=1}^m\biggl[\mu\Bigl(\,\bigcap_{t\in\T_{L/B}}\vartheta_{t}(\AA_j)\Bigr)\biggr]^{(B/L)^d}
\end{equation}
where
\begin{equation}
%\label{}
\vartheta_t(\AA):=\{\vartheta_t 1_{\AA}=1\}
\end{equation}
Note that the exponent~$(\ffracB BL)^d$ is the reciprocal volume of the torus~$\T_{L/B}$. (This is consistent with the fact that both expressions transform homogeneously under the scaling $f_j\to\lambda_j f_j$ with~$\lambda_j\ge0$.)

\begin{proofsectB}{Proof of Theorem~\ref{thm-CE}}
We will assume throughout that~$E_\mu(f\,\vartheta f)=0$ implies~$f=0$. (Otherwise, one has to factor out the ideal of such functions and work on the factor space.) We will first address the 1D case; the general dimensions will be handled by induction. 

Abbreviate $2n:=\ffracB LB$ and fix a collection of non-zero functions~$f_1,\dots,f_{2n}$. Define a multilinear functional $F$ on the set of $B$-block functions by
\begin{equation}
%\label{}
F(f_1,\dots,f_{2n}) := E_\mu\Bigl(\,\prod_{t=1}^{2n}\vartheta_tf_t\Bigr)
\end{equation}
Noting that~$F(f_j,\dots,f_j)>0$, we also define
\begin{equation}
%\label{}
G(f_1,\dots,f_{2n}):=\frac{F(f_1,\dots,f_{2n})}{\prod_{j=1}^{2n}F(f_j,\dots,f_j)^{\frac1{2n}}}
\end{equation}
These objects enjoy a natural cyclic invariance,
\begin{equation}
%\label{}
F(f_1,\dots,f_{2n})=F(f_{2n},f_1,\dots,f_{2n-1})
\end{equation}
and, similarly,
\begin{equation}
%\label{}
G(f_1,\dots,f_{2n})=G(f_{2n},f_1,\dots,f_{2n-1})
\end{equation}
The definition of~$G$ also implies
\begin{equation}
\label{4.36}
G(f,\dots,f)=1
\end{equation}
Finally, Cauchy-Schwarz along the plane separating~$f_1$ from~$f_{2n}$ and~$f_n$ from~$f_{n+1}$ yields
\begin{multline}
\label{CS-1D}
\qquad
G(f_1,\dots,f_{2n})\le G(f_1,\dots,f_n,f_n,\dots,f_1)^{\ffracB12}
\\\times\,G(f_{2n},\dots,f_{n+1},f_{n+1},\dots,f_{2n})^{\ffracB12}
\qquad
\end{multline}
This will of course be the core estimate of the proof.

The desired claim will be proved if we show that
\begin{equation}
\label{4.37}
G(f_1,\dots,f_{2n})\le1
\end{equation}
i.e., that~$G$ is maximized by $2n$-tuples composed of the same function. We will proceed similarly as in the proof of Gaussian Domination: Given a $2n$-tuple of $B$-block functions, $(f_1,\dots,f_{2n})$, let $(g_1,\dots,g_{2n})$ be such that 
%\settowidth{\leftmargini}{(11)}
\begin{enumerate}
\item[(1)]
$g_i\in\{f_1,\dots,f_{2n}\}$ for each~$i=1,\dots,2n$
\item[(2)]
$G(g_1,\dots,g_{2n})$ maximizes $G$ over all such choices of~$g_1,\dots,g_{2n}$
\item[(3)]
$g_1,\dots,g_{2n}$ is minimal in the sense that it contains the longest block (counted periodically) of the form~$f_i,f_i,\dots,f_i$, for some~$i\in\{1,\dots,2n\}$.
\end{enumerate}
Let~$k$ be the length of this block and, using the cyclic invariance, assume that the block occurs at the beginning of the sequence~$g_1,\dots,g_{2n}$, i.e., we have $g_1,\dots,g_k=f_i$ (with $g_{k+1}\ne f_i$ unless $k=2n$). 

We claim that~$k=2n$. Indeed, in the opposite case, $k<2n$, we must have $g_{2n}\ne f_i$ whereby \eqref{CS-1D} combined with the fact that $(g_1,\dots,g_{2n})$ is a maximizer of~$G$ imply
\begin{equation}
\begin{aligned}
G(g_1,\dots,g_{2n})^2&\le G(g_1,\dots,g_n,g_n,\dots,g_1)
G(g_{2n},\dots,g_{n+1},g_{n+1},\dots,g_{2n})
\\
&\le G(g_1,\dots,g_n,g_n,\dots,g_1)G(g_1,\dots,g_{2n})
\end{aligned}
\end{equation}
i.e.,
\begin{equation}
%\label{}
G(g_1,\dots,g_{2n})\le G(g_1,\dots,g_n,g_n,\dots,g_1)
\end{equation}
This means that~$(g_1,\dots,g_n,g_n,\dots,g_1)$ is also a legitimate maximizer of~$G$ but it has a longer constant block --- namely of length at least~$\min\{2k,2n\}$. This is a contradiction and so we must have~$k=2n$ after all. In light of \twoeqref{4.36}{4.37}, this proves the claim in~$d=1$.

To extend the proof to $d>1$, suppose that~$m=(\ffracB LB)^d$ and assume, without loss of generality, that we have one function~$f_t$ for each block $\Lambda_B+Bt$. Writing
\begin{equation}
%\label{}
\prod_{t\in\T_{L/B}}\vartheta_tf_t = \prod_{j=1}^{2n}\biggl(\,\prod_{\begin{subarray}{c}
t\in\T_{L/B}\\ t_1=j
\end{subarray}}
\vartheta_tf_t\biggr)
\end{equation}
we may apply the 1D chessboard estimate along the product over~$j$. This homogenizes the product over $f_t$ in the first coordinate direction. Proceeding through all directions we eventually obtain the desired claim.
\end{proofsectB}

The chessboard estimate allows us to bound the probability of simultaneous occurrence of distinctly-placed $B$-block events in terms of their \emph{disseminated versions} $\bigcap_{t\in\T_{L/B}}\vartheta_t(\AA)$. The relevant quantities to estimate are thus
\begin{equation}
%\label{}
\fraktura z_L(\AA):=\mu\Bigl(\,\bigcap_{t\in\T_{L/B}}\vartheta_{t}(\AA)\Bigr)^{(B/L)^d}
\end{equation}
The set function $\AA\mapsto\fraktura z_L(\AA)$ is not generally additive. However, what matters for applications is that it is subadditive:

\begin{lemma}[Subadditivity]
\label{lemma-subadditivity}
Let~$\AA$ and~$\AA_1,\AA_2,\dots,$ be a collection of $B$-block events such that
\begin{equation}
\label{AAsubset}
\AA\subset\bigcup_k\AA_k
\end{equation}
Then
\begin{equation}
%\label{}
\fraktura z_L(\AA)\le\sum_k\fraktura z_L(\AA_k)
\end{equation}
\end{lemma}

\begin{proofsectB}{Proof}
First we use the subadditivity of~$\mu$ and \eqref{AAsubset} to get
\begin{equation}
\fraktura z_L(\AA)^{|\T_{L/B}|}
=\mu\Bigl(\,\bigcap_{t\in\T_{L/B}}\vartheta_{t}(\AA)\Bigr)\,\,
\underset{\eqref{AAsubset}}\le\,\,\sum_{(k_t)}\mu\Bigl(\,\bigcap_{t\in\T_{L/B}}\vartheta_{t}(\AA_{k_t})\Bigr)
\end{equation}
Next we apply the chessboard estimate 
\begin{equation}
%\label{}
\mu\Bigl(\,\bigcap_{t\in\T_{L/B}}\vartheta_{t}(\AA_{k_t})\Bigr)\le\prod_{t\in\T_{L/B}}\fraktura z_L(\AA_{k_t})
\end{equation}
to each term on the right hand side. Finally we apply the distributive law for sums and products with the result
\begin{equation}
\begin{aligned}
\fraktura z_L(\AA)^{|\T_{L/B}|}
&\le\sum_{(k_t)}\prod_{t\in\T_{L/B}}\fraktura z_L(\AA_{k_t})
\\
&=\prod_{t\in\T_{L/B}}\sum_k\fraktura z_L(\AA_k)=\biggl(\sum_k\fraktura z_L(\AA_k)\biggr)^{|\T_{L/B}|}
\end{aligned}
\end{equation}
Taking the $|\T_{L/B}|$-th root now yields the desired claim.
\end{proofsectB}

Here is how subadditivity $\fraktura z_L$ is generally used in computations: In order to estimate the $\fraktura z_L$-value of an event, we first cover it by the union of a collection of smaller --- and, as desired, easier to compute-with --- events, then evaluate the $\fraktura z_L$-value for each of them and, finally, add the results. 

In estimates, we often work with the limiting version,
\begin{equation}
%\label{}
\fraktura z(\AA):=\lim_{L\to\infty}\fraktura z_L(\AA)
\end{equation}
of this quantity. We may interpret this as a \emph{partition function per site} restricted to event~$\AA$ on each~$B$-block. The advantage of taking the limit is that it often washes out some annoying finite-size factors and thus provides a more tractable expression to work with. In addition, the limit can be computed using arbitrary  --- not just periodic --- boundary conditions.

\subsection{Diagonal reflections, other lattices}
The above proof of the chessboard estimate is tailored to the underlying setting of the hypercubic lattice, primarily because of its use of the orthogonality between the principal lattice directions. However, some practical problems may lead us to the consideration of other lattices. Some cases generalize directly, e.g., certain instances of the \emph{body-centered cubic} (BCC) or \emph{face-centered cubic} (FCC) lattices, whose unit cells look respectively as follows:

\vglue0.3cm
\centerline{\includegraphics[width=3.0in]{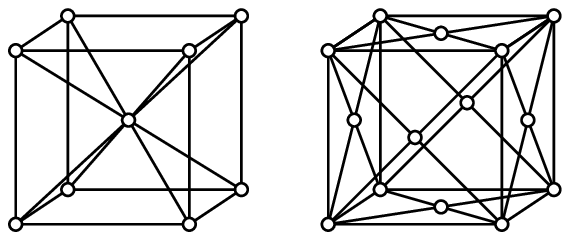}}
\noindent
Both of these are decorations of the cubic lattice in which an extra vertex placed in the center of each unit cube (BCC) or a face (FCC) and is attached by edges to the vertices in its ultimate vicinity. 

Assuming the interaction \eqref{H-spins} with~$J_{xy}$ non-zero and positive only for adjacent (i.e., nearest-neighbor) pairs of vertices, the torus Gibbs measure is reflection positive for reflections both through and between the planes of sites of~$\Z^3$. (A key observation is that the planes between sites of~$\Z^3$ contain some of the added vertices but bisect no additional edges.) The strengths of the interactions across the ``old'' and ``new'' edges may not even be the same.

In $d=2$, a corresponding graph is the lattice with a vertex placed in the middle of each square of~$\Z^2$ and edges from it to each of the four corners thereof. By the same reasoning, the nearest-neighbor ferromagnetic interaction leads to a reflection positive torus Gibbs measure. 

The situation becomes more involved for the \emph{triangular} (two-dimensional) lattice, whose standard embedding into the complex plane~$\C$ has vertices
\begin{equation}
%\label{}
m+n\,\texte^{\texti\pi/3},\qquad m,n\in\Z
\end{equation}
and an edge between any pair of such vertices that differ by a number in the set $\{1,\texte^{\texti\pi/3},\texte^{\texti 2\pi/3}\}$. The principal problem with such graphs is how to place a finite piece of this lattice on a torus in a way that gives rise to reflection positive measures. Here is a convenient choice:
\vglue0.1cm
\centerline{\includegraphics[width=2.5in]{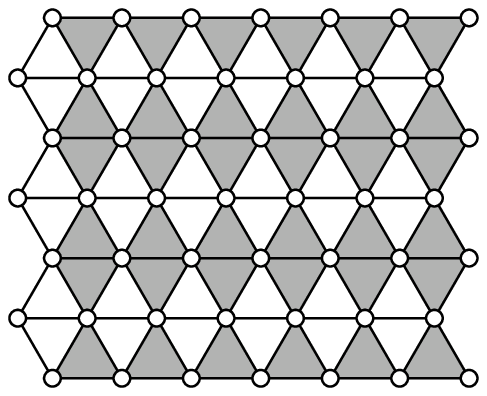}}
\noindent
with the torus obtained by identifying the vertices on the opposite sides. 

The allowed planes of reflection are all horizontal lines (reflections through sites) and the vertical lines (reflections through both sites and bonds). Again, for ferromagnetic nearest-neighbor interactions, the Gibbs measure with interaction \eqref{H-spins} is reflection positive. A minor, though annoying, problem occurs in the application of chessboard estimates because the vertical lines of reflections actually cut through triangles. A solution is to focus only on those events that lie either on white or on gray triangles in the above picture and use reflection only with respect to vertical lines that do not cut through the chosen triangles. 

A completely analogous situation occurs for the \emph{honeycomb} lattice. Here we consider the domain of the form
\vglue0.3cm
\centerline{\includegraphics[width=2.1in]{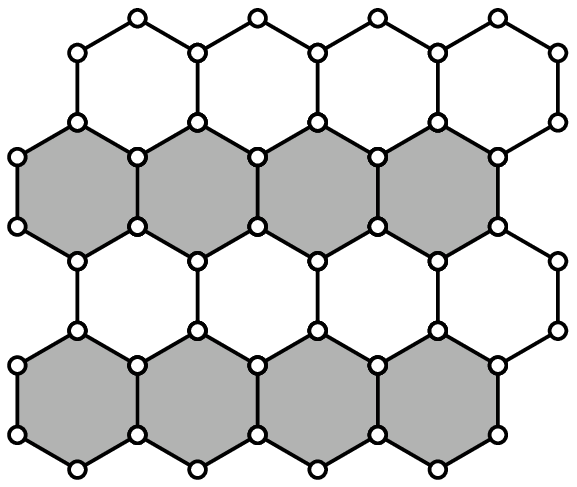}}
\noindent
and wrap it into a torus by identifying the vertices on the opposite side. Again, for nearest-neighbor ferromagnetic interactions, the resulting Gibbs measure is reflection positive with respect to reflections in vertical lines on sites and horizontal lines between sites. In the application of chessboard estimates to a collection of ``hexagon events,'' we only use every other horizontal and vertical reflections to a corresponding subset of these events; e.g., those sitting on the shaded hexagons.

A final case of interest is that of \emph{diagonal} reflections in~$\Z^d$. In~$d=2$, this is achieved by wrapping the domain of the form
\vglue0.3cm
\centerline{\includegraphics[width=1.9in]{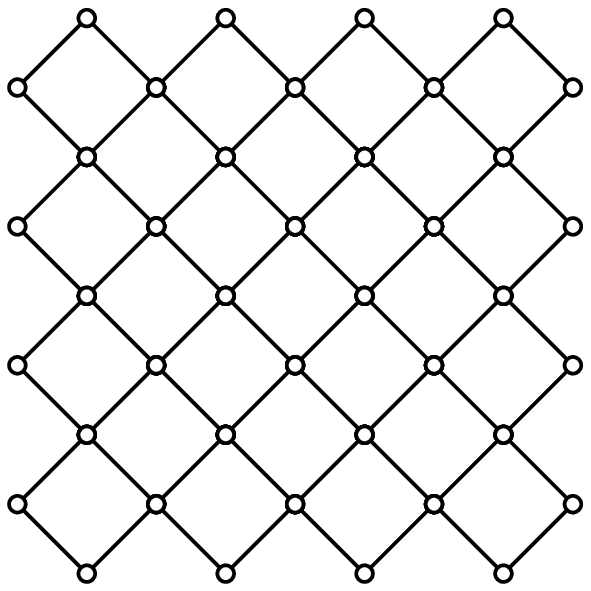}}
\noindent
periodically into a torus. Reflections in the horizontal and vertical lines of sites --- the diagonals --- are now symmetries of this graph; for nearest-neighbor interactions (of any sign) the corresponding torus Gibbs measure is reflection positive.

The advantage of the diagonal torus is that it permits the use of reflection positivity on collections of ``bond events,'' i.e., those associated with pairs of nearest-neighbor spins. Subsequent applications of chessboard estimates disseminate a single-bond event over the entire torus. This, in turn, helps in estimates of the quantity $\fraktura z(\AA)$ whenever~$\AA$ is an event depending on a single square that is itself an intersection of bond events:

\begin{lemma}
\label{lemma-bond-cube}
Given a unit cube in~$\Z^d$, let $\AA_b$, with~$b$ running over all of the $c_d:=d2^{d-1}$ edges in this cube, be a collection of bond events. Then
\begin{equation}
\label{MB-bond-square}
\fraktura z\biggl(\,\bigcap_b\AA_b\biggr)\le
\prod_b\fraktura z(\AA_b)^{1/{c_d}}
\end{equation}
Here $\fraktura z(\AA_b)$ is the partition function per site restricted to configurations such that~$\AA_b$, or its corresponding reflection, occurs at all edges of~$\Z^d$.
\end{lemma}

\begin{proofsectB}{Proof}
Let us first focus on $d=2$. The key fact is that the partition function per site, $\fraktura z(\AA)$, does not depend on what boundary conditions were used to define it. So, in order to compute~$\fraktura z$ of the intersection event, we may first wrap the square lattice into the diagonal torus, and disseminate the bond events before passing to the $L\to\infty$. As there are $c_2=4$ edges in each lattice square, there is an extra power of $\ffracB14$.

In~$d>2$, we perform the same by singling out two lattice directions and wrapping the torus diagonally in these, and regularly in the remaining ones. This homogenizes the event in two lattice directions. Proceeding by induction, the claim follows.
\end{proofsectB}

\subsection{Literature remarks}
\label{sec4.4}
The material of this section is entirely classical; a possible exception is Lemma~\ref{lemma-subadditivity} which seems to have been formulated in the present form only relatively recently~\cite{BCN1}. The use of reflection positivity goes back to the days of constructive quantum field theory (namely, the Osterwalder-Schrader axioms~\cite{OS}) where RP was a tool to obtain a sufficiently invariant --- and natural --- inner product. The use in statistical mechanics was initiated by the work of Fr\"ohlich, Simon and Spencer~\cite{FSS} (infrared bound) and Fr\"ohlich and Lieb~\cite{FL} (chessboard estimates). The theory was further developed in two papers by Fr\"ohlich, Israel, Lieb and Simon~\cite{FILS1,FILS2}. There have been a couple of nice reviews of this material, e.g., by Shlosman~\cite{Senya2} and in Georgii~\cite{Georgii}.

All use of reflection positivity in these notes is restricted to one of the three interactions introduced in Chapter~\ref{chap-2}. Various generalizations beyond these are possible. For instance, the n.n.\ interaction of strength $J$ may be accompanied by a n.n.n.\ interaction of strength~$\lambda$ --- including negative values --- and the result is still RP provided $J\ge2(d-1)|\lambda|$. For reflections through planes of sites, we may even allow any sort of interactions involving the spins in a given lattice cube. (This exhausts all finite range interactions; any longer range RP interactions are automatically infinite range.) Many other examples are discussed, e.g., in~\cite[page~32]{FILS1}. 

Notwithstanding our decision to restrict attention only to three specific interactions, the set of reflection positive interactions is not so small as it may appear. Indeed, in the class of translation and rotation invariant coupling constants, letting
\begin{equation}
%\label{}
F(x_1,\dots,x_d):=J_{0,x}
\end{equation}
we check that a sufficient conditions for RP is that the matrix
\begin{equation}
%\label{}
(x,y)\mapsto F(x_1+y_1,x_2-y_2,\dots,x_d-y_d)1_{\{x_1>0\}}1_{\{y_1>0\}}
\end{equation}
is positive semidefinite. (See \eqref{5.14} for a specific case of this.) By Shur's Theorem --- namely that if $(a_{ij})$ and $(b_{ij})$ are positive semidefinite matrices, then so is $(a_{ij}b_{ij})$ --- we thus know that if $J^{(1)}$ and $J^{(2)}$ are two collections of RP couplings, then also the collection $J^{(1)}_{xy}J^{(2)}_{xy}$ is RP. In particular, the set of RP couplings is closed under taking products.

The situation on other lattices is discussed in~\cite{FILS2}; the use of diagonal reflections goes back to~\cite{Senya2}. We caution the reader that it is rather easy to make a mistake in this context. For instance, the regularly-wrapped ($L\times L$) torus in~$\Z^2$ is also symmetric with respect to all of the diagonal reflections. However, for diagonal reflection on direct torus it is not possible to define \emph{two} components of the ``plane of reflection'' so that the reflection in one leaves the other intact. So we cannot simultaneously use both direct and diagonal reflections, and this prevents a direct proof of \eqref{MB-bond-square} in finite volume. (This error appeared in~\cite[eq.~4.39]{BK2Bi} though, as shown in Lemma~\ref{lemma-bond-cube}, all differences get washed out in the thermodynamic limit.)

Gaussian domination appears in a rather different context as the celebrated \emph{Brascamp-Lieb inequality}. Consider the measure on~$\R^n$ of the form 
\begin{equation}
%\label{}
\mu(\textd x):=Z^{-1}\texte^{-V(x)}\textd x
\end{equation}
with~$V$ smooth and strictly convex. Let $V''(x)$ be the Hessian, i.e., an $n\times n$ matrix of all second derivatives of~$V$. Then for each smooth~$f$ with compact support,
\begin{equation}
%\label{}
E_\mu(f^2)-(E_\mu f)^2\le\int\textd x\,\bigl\langle (V'')^{-1}\nabla f(x),\nabla f(x)\bigr\rangle
\end{equation}
where $\langle\cdot,\cdot\rangle$ denote the $n$-dimensional Euclidean inner product. In particular, if $Q$ is a positive definite $n\times n$ matrix that dominates the Hessian from below at all~$x$, then the correlations of~$\mu$ are dominated by those of the Gaussian measure with covariance~$2Q^{-1}$. This is, unfortunately, not very useful in the analysis of the Gibbs measures for general lattice spin systems as these are generally not of the required form --- e.g., because the restriction to a specific spin-space (a unit sphere for the Heisenberg model) cannot be approximated by convex functions.

\section{Applications of chessboard estimates}
\label{chap-5}

In this section we will apply the technique of chessboard estimates to obtain proofs of phase coexistence in some lattice spin models. The arguments will be carried out in detail only for one rather simple example. For more sophisticated systems we present only the important ideas. Details, anyway, can be found in the corresponding papers.

\subsection{Gaussian double-well model}
\index{Gaussian!double-well model}
Here we will demonstrate the use of chessboard estimates on the model of a Gaussian free-field model in a non-quadratic, double-well on-site potential. The Hamiltonian takes the general form
\begin{equation}
\label{H-orig}
\beta H(\phi):=\beta\sum_{\langle x,y\rangle}(\phi_x-\phi_y)^2+\sum_xV(\phi_x)
\end{equation}
where~$\phi_x\in\R$ with \emph{a priori} measure given by the Lebesgue measure, and~$V$ is a potential. Note that~$\beta$ has been incorporated into the Hamiltonian in such a way that the on-site potential remains independent of it.

The most well known example of such systems is~$V(\phi):=\frac\kappa2\phi^2$ with~$\kappa>0$ which is known as the \emph{massive} Gaussian free field. This case can of course be treated completely explicitly; e.g., on the torus the corresponding Gaussian measure on~$(\phi_x)$ is zero-mean with covariance
\begin{equation}
\label{GFF-cov}
\text{Cov}(\phi_x,\phi_y)=\sum_{k\in\T_L^\star}\frac{\texte^{\texti k\cdot(x-y)}}{\beta\widehat D(k)+\kappa}
\end{equation}
where $\widehat D(k)$ is the Fourier transform of the torus (discrete) Laplacian,
\begin{equation}
%\label{}
\widehat D(k):=\sum_{j=1}^d|1-\texte^{\texti k_j}|^2
\end{equation}
Note that the inclusion of the mass, $\kappa>0$ --- more precisely, $\kappa$ is the mass squared --- makes the covariance regular even for the zero mode $k=0$.

We will look at a modification of this case when~$V$ takes the form
\vglue0.3cm
\centerline{\includegraphics[width=3.1in]{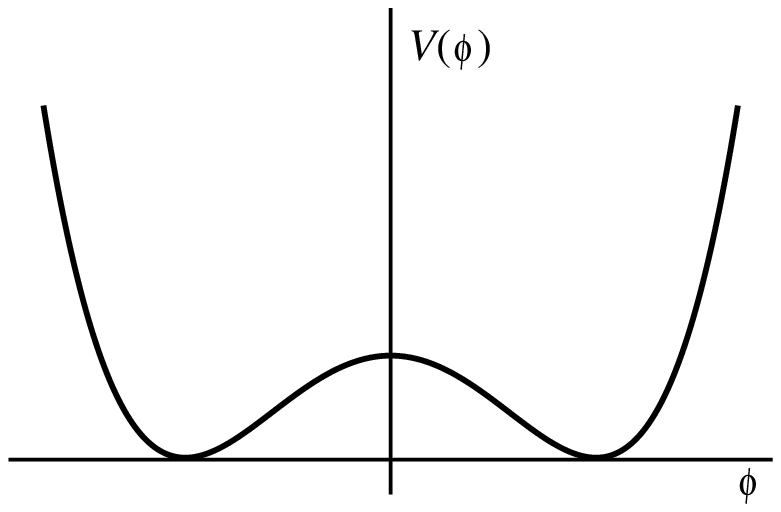}}
\noindent
In fact, we will be even more specific and assume that~$V$ is simply given by
\begin{equation}
\label{V-repr}
\texte^{-V(\phi)}:=\texte^{-\frac\kappa2(\phi-1)^2}+\texte^{-\frac\kappa2(\phi+1)^2}
\end{equation}
It is easy to check that, for~$\kappa$ sufficiently large, $V$ defined using this formula looks as in the figure. The reason for assuming \eqref{V-repr} is the possibility of an \emph{Ising-spin representation}. Indeed, we may rewrite \eqref{V-repr} as
\begin{equation}
%\label{}
\texte^{-V(\phi)}=\sum_{\sigma=\pm1}\texte^{-\frac\kappa2(\phi-\sigma)^2}
=C\sum_{\sigma=\pm1}\texte^{-\frac\kappa2\phi_x-\kappa\phi_x\sigma_x}
\end{equation}
where~$C:=\texte^{-\kappa}$. A product of such terms is thus proportional to
\begin{equation}
%\label{}
\prod_x\texte^{-V(\phi_x)}\,\propto\,\sum_{(\sigma_x)}\prod_x\texte^{-\frac\kappa2\phi_x^2-\kappa\phi_x\sigma_x}
\end{equation}
This means we can write the Gibbs weight of the model as follows
\begin{multline}
%\label{}
\qquad
\texte^{-\beta\sum_{\langle x,y\rangle}(\phi_x-\phi_y)^2-\sum_xV(\phi_x)}
\\\propto\,\sum_{(\sigma_x)}\texte^{-\beta\sum_{\langle x,y\rangle}(\phi_x-\phi_y)^2-\frac\kappa2\sum_x\phi_x^2}\,\texte^{-\kappa\sum_x\phi_x\sigma_x}
\qquad
\end{multline}
If we elevate~$(\sigma_x)$ to genuine degrees of freedom, we get a model on spins~$S_x:=(\phi_x,\sigma_x)$ with \emph{a priori} law Lebesgue on $\R\,\times\,$counting measure on~$\{-1,1\}$ and the Hamiltonian
\begin{equation}
\label{H-extend}
\beta H(\phi,\sigma):=\beta\sum_{\langle x,y\rangle}(\phi_x-\phi_y)^2+\frac\kappa2\sum_x\phi_x^2+\kappa\sum_x\phi_x\sigma_x
\end{equation}
Notice the first two terms on the right-hand side is the Hamiltonian of the massive (centered) Gaussian free field while the interaction between the $\phi$'s and the $\sigma$'s has on-site form. 

Here are some observations whose (simple) proof we leave to the reader:

\begin{lemma}
\label{lemma-marginal}
Let $\mu$ be a Gibbs measure for Hamiltonian \eqref{H-extend} and let~$\nu$ be its $\phi$-marginal. Then~$\nu$ is a Gibbs measure for the Hamiltonian \eqref{H-orig} subject to \eqref{V-repr}. The marginal $\nu$ completely determines $\mu$: For any~$f$ depending only on~$\phi$ and~$\sigma$ in a finite set~$\Lambda$,
\begin{equation}
\label{marginal-equiv}
E_\mu(f)=E_\nu\Bigl(\,\sum_{(\sigma_x)_{x\in\Lambda}}f(\phi,\sigma)\,\prod_{x\in\Lambda}\,\texte^{V(\phi_x)-\frac\kappa2(\phi_x-\sigma_x)^2}\Bigr)
\end{equation}
\end{lemma}

We will use $\mathfrak G_{\beta,\kappa}$ to denote the set of all Gibbs measures for the Hamiltonian \eqref{H-extend} with parameters~$\beta$ and~$\kappa$. The principal result for this model is as follows:

\begin{theorem}
\label{thm-Gauss-DW}
Let~$d\ge2$. For each~$\epsilon>0$ there is~$c>0$ such that if~$\kappa,\ffracB\kappa\beta>c$, then there exist~$\mu^+,\mu^-\in\mathfrak G_{\beta,\kappa}$ which are translation invariant and obey
\begin{equation}
\label{sigma-bd}
\mu^\pm(\sigma_x=\pm1)\ge1-\epsilon
\end{equation}
and
\begin{equation}
\label{phi-bd}
E_{\mu^\pm}\bigl((\phi_x\mp1)^2\bigr)\le\epsilon
\end{equation}
\end{theorem}

In simple terms, at low temperatures and large curvature of the wells of~$V$, the fields prefer to localize in one of the wells. We remark that, while we chose the model as simple as possible, a similar conclusion would follow for with~$V$ given by
\begin{equation}
\label{V-asym}
\texte^{-V(\phi)}=\texte^{-\frac{\kappa_+}2(\phi-1)^2+h}+\texte^{-\frac{\kappa_-}2(\phi+1)^2-h}.
\end{equation}
where~$h$ changes the relative weight of the two minima. Indeed, there exists $h_{\text{t}}$ at which one has two Gibbs measure --- the analogues of $\mu^+$ and~$\mu^-$. Moreover, if~$\kappa_+\gg\kappa_-$, then~$h_{\text{t}}>0$ because, roughly speaking, the well at $-1$ offers ``more room'' for fluctuations.

\def\plaquette#1#2#3#4{
{\mkern4mu\setlength{\unitlength}{7pt}
\begin{picture}(1,1)(0.4,0.4)
\linethickness{0.1\unitlength} 
\put(0,0){$\scriptstyle #1$}
\put(0,1){$\scriptstyle #2$}
\put(1,0){$\scriptstyle #3$}
\put(1,1){$\scriptstyle #4$}
\end{picture}}
\mkern4mu}

\def\bigplaquette#1#2#3#4{
{\setlength{\unitlength}{12pt}
\mkern4mu
\begin{picture}(1,1)(0.4,0.4)
\linethickness{0.1\unitlength} 
\put(0,0){$#1$}
\put(0,1){$#2$}
\put(1,0){$#3$}
\put(1,1){$#4$}
\end{picture}}
\mkern4mu}

\subsection{Proof of phase coexistence}
Here we will prove Theorem~\ref{thm-Gauss-DW}. We will focus on $d=2$; the proof in general dimension is a straightforward, albeit more involved, generalization.

Let us refer to a face of~$\Z^2$ as a \emph{plaquette} (i.e., a plaquette is a square of side one with a vertex of~$\Z^2$ in each corner). Given a spin configuration~$(\sigma_x)$, we say that a plaquette is \emph{good} if all four spins take the same value, and \emph{bad} otherwise. Let~$\BB$ denote the event that the plaquette with lower-left corner at the origin is bad. 

Since the interaction is that of the GFF with a modified single-spin measure, the torus Gibbs measure is RP. The crux of the proof is to show that bad plaquettes are suppressed. Specifically, we want to show that
\begin{equation}
%\label{}
\fraktura z(\BB)\ll1\quad\text{once}\quad\beta,\kappa\gg1
\end{equation}
Appealing to the subadditivity lemma (Lemma~\ref{lemma-subadditivity}) we only need to estimate the $\fraktura z$-value of all possible configurations on the plaquette that constitute~$\BB$. Due to the plus-minus symmetry of the $\sigma$'s, it suffices to examine three patterns:
\begin{equation}
%\label{}
\bigplaquette+--+\quad\qquad\bigplaquette++--\quad\qquad\bigplaquette+-++
\end{equation}
We begin with the most interesting of the three:

\begin{lemma}
\label{lemma-diag-pattern}
For any $\beta,\kappa>0$,
\begin{equation}
%\label{}
\fraktura z\bigl(\plaquette+--+\bigr)\le\texte^{-\frac{4\beta\kappa}{8\beta+\kappa}}
\end{equation}
\end{lemma}

\begin{proofsectB}{Proof}
\index{chessboard estimate!bounds using}
Let $Z_L:=\sum_\sigma\int\,\texte^{-\beta H_L(\phi,\sigma)}\prod_{x\in\T_L}\textd\phi_x$ be the torus partition function. Given a plaquette spin pattern, let~$Z_L(\text{pattern})$ denote the same object with $\sigma$ fixed to the disseminated pattern --- the sole element of~$\bigcap_{t\in\T_L}\vartheta_t(\text{pattern})$. (We are working with~$B=1$.) By the definition of $\fraktura z$ we have
\begin{equation}
%\label{}
\fraktura z_L\bigl(\plaquette+--+\bigr)^{|\T_L|}:=\frac{Z_L\bigl(\plaquette+--+\bigr)}{Z_L}
\le\frac{Z_L\bigl(\plaquette+--+\bigr)}{Z_L\bigl(\plaquette++++\bigr)}
\end{equation}
Now the partition function with all $\sigma$'s restricted to $+$ is given by
\begin{equation}
\begin{aligned}
Z_L\bigl(\plaquette++++\bigr)
&=\int\,\texte^{-\beta\sum_{\langle x,y\rangle}(\phi_x-\phi_y)^2-\frac\kappa2\sum_x\phi_x^2}\,\texte^{-\kappa\sum_x\phi_x}\prod_{x\in\T_L}\textd\phi_x
\\
&=\Bigl(\,\dots\Bigr) E_{\text{GFF}}\bigl(\texte^{-\kappa\sum_x\phi_x}\bigr)
\end{aligned}
\end{equation}
where the expectation is with respect to the massive Gaussian free field and the prefactor denotes the integral of the Gaussian kernel over all~$\phi_x$. Similarly we obtain
\begin{equation}
Z_L\bigl(\plaquette+--+\bigr)
=\Bigl(\,\dots\Bigr) E_{\text{GFF}}\bigl(\texte^{-\kappa\sum_x\phi_x(-1)^{|x|}}\bigr)
\end{equation}
where we noticed that by disseminating the pattern $\plaquette+--+$ we obtain a configuration which is one at even parity~$x$ and minus on at odd parity~$x$. Thus we conclude
\begin{equation}
\label{z-bound}
\fraktura z_L\bigl(\plaquette+--+\bigr)^{|\T_L|}
\le\frac{E_{\text{GFF}}\bigl(\texte^{-\kappa\sum_x\phi_x}\bigr)}{E_{\text{GFF}}\bigl(\texte^{-\kappa\sum_x\phi_x(-1)^{|x|}}\bigr)
}
\end{equation}
i.e., we only need to compute the ratio of the Gaussian expectations, and not the prefactors.

Next we recall a standard formula for Gaussian moment generating functions: If~$X$ is a multivariate Gaussian, then
\begin{equation}
%\label{}
E(\texte^{\lambda\cdot X}) = \texte^{\lambda\cdot EX+\frac12\text{Var}(\lambda\cdot X)}
\end{equation}
Since~$E_{\,\text{GFF}}(\phi_x)=0$, we only need to compute the (diagonal) matrix element of $\text{Cov}(\phi_x,\phi_y)$ against vectors $1=(1,1,\dots)$ and $(-1)^{|x|}$. However, a quick look at \eqref{GFF-cov} will convince us that these functions are eigenvectors of the covariance matrix corresponding to $k=0$ and $k=(\pi,\pi)$, respectively. Since $\widehat D(0)=0$ while $\widehat D(\pi,\pi)=8$, we get
\begin{equation}
%\label{}
\Var_{\,\text{GFF}}\Bigl(\,\sum_x\phi_x\Bigr)=\frac{|\T_L|}\kappa
\end{equation}
\begin{equation}
%\label{}
\Var_{\,\text{GFF}}\Bigl(\,\sum_x\phi_x(-1)^{|x|}\Bigr)=\frac{|\T_L|}{8\beta+\kappa}
\end{equation}
where the factor $|\T_L|$ is the (square of) the $L^2(\T_L)$-norm of the functions under consideration.
Plugging this in \eqref{z-bound} we conclude
\begin{equation}
%\label{}
\fraktura z_L\bigl(\plaquette+--+\bigr)^{|\T_L|}\le\exp\biggl\{\frac12|\T_L|\kappa^2\Bigl(\frac1{8\beta+\kappa}-\frac1\kappa\Bigr)\biggr\}
\end{equation}
from which the claim readily follows.
\end{proofsectB}

Next we attend to the other patterns:

\begin{lemma}
\label{lemma-other-patterns}
For any $\beta,\kappa>0$,
\begin{equation}
\label{++--}
\fraktura z\bigl(\plaquette++--\bigr)\le\texte^{-\frac{2\beta\kappa}{4\beta+\kappa}}
\end{equation}
and
\begin{equation}
\label{+-++}
\fraktura z\bigl(\plaquette+-++\bigr)\le\texte^{-\frac{2\beta\kappa}{8\beta+\kappa}}
\end{equation}
\end{lemma}

\begin{proofsectB}{Proof}
As for \eqref{++--}, dissemination of \,$\plaquette++--$\, leads to alternating stripes of plusses and minuses, i.e., $\sigma_x=(-1)^{|x_1|}$. Again, this is an eigenvector of the covariance matrix \eqref{GFF-cov} with $k=(\pi,0)$. The corresponding $\widehat D$ equals $4$ and so
\begin{equation}
%\label{}
\fraktura z_L\bigl(\plaquette++--\bigr)^{|\T_L|}\le\exp\biggl\{\frac12|\T_L|\kappa^2\Bigl(\frac1{4\beta+\kappa}-\frac1\kappa\Bigr)\biggr\}
\end{equation}
yielding \eqref{++--}.

The pattern \,$\plaquette+-++$\, is more complex because its dissemination will not lead to an eigenvector of the covariance matrix. However, we circumvent this problem by noting that Lemma~\ref{lemma-bond-cube} implies
\begin{equation}
%\label{}
\fraktura z\bigl(\plaquette+-++\bigr)\le \fraktura z\bigl(\plaquette+--+\bigr)^{\ffracB12}\fraktura z\bigl(\plaquette++++\bigr)^{\ffracB12}
\le\fraktura z\bigl(\plaquette+--+\bigr)^{\ffracB12}
\end{equation}
where we used $\fraktura z\bigl(\plaquette++++\bigr)\le1$. Now \eqref{+-++} follows from Lemma~\ref{lemma-diag-pattern}.
\end{proofsectB}

\begin{corollary}
\label{cor-BB-bound}
For each~$\epsilon>0$ there exists~$a>0$ such that if~$\beta,\kappa>a$, then $\fraktura z(\BB)\le\epsilon$.
\end{corollary}

\begin{proofsectB}{Proof}
The event $\BB$ can be written as the union over a finite number of bad patterns. On the basis of Lemmas~\ref{lemma-diag-pattern}--\ref{lemma-other-patterns} the claim holds for~$\BB$ replaced by any fixed bad pattern. The desired bound now follows --- with slightly worse constants --- by invoking Lemma~\ref{lemma-subadditivity}.
\end{proofsectB}

Next we explain our focus on the bad event:

\begin{lemma}
\label{lemma-BB-bound}
There exists a constant $c\in(1,\infty)$ such that if $c\fraktura z(\BB)<\ffracB12$ then for any $x,y\in\T_L$,
\begin{equation}
\label{don't-like}
\mu_L(\sigma_x=1,\,\sigma_y=-1)\le 2c\fraktura z(\BB).
\end{equation}
\end{lemma}

\begin{proofsectB}{Proof}
This is a consequence of a simple Peierls' estimate. Indeed, if $\sigma_x=1$ and $\sigma_y=-1$, then $x$ is separated from~$y$ by a ``circuit'' of bad plaquettes. (Formally, either all plaquettes containing~$x$ are bad or there exists a non-trivial connected component of good --- i.e., \emph{not} bad --- plaquettes containing~$x$. This component cannot cover the whole torus because~$\sigma_y=-1$; the above ``circuit'' is then comprised of the bad plaquettes on the boundary of this component.) This means that
\begin{equation}
%\label{}
\mu_L(\sigma_x=1,\,\sigma_y=-1)\le\sum_{\gamma}\mu_L\Bigl(\,\bigcap_{t\in\gamma}\vartheta_t(\BB)\Bigr)
\le\sum_\gamma\fraktura z(\BB)^{|\gamma|}
\end{equation}
where $|\gamma|$ denotes the maximal number of disjoint bad plaquettes in~$\gamma$ and where we used the chessboard estimates to derive the second bound.
By standard arguments, the number of circuits of ``length'' $n$ surrounding~$x$ or winding around~$\T_L$ at least once is bounded by~$c^n$, for some constant~$c>1$. It follows
\begin{equation}
%\label{}
\mu_L(\sigma_x=1,\,\sigma_y=-1)\le\sum_{n\ge1}c^n\fraktura z(\BB)^n
\end{equation}
Under the condition $c\fraktura z(\BB)<\ffracB12$ this sum is less than twice its first term.
\end{proofsectB}

Finally, we can assemble the ingredients into the desired proof of phase coexistence:

\begin{proofsectB}{Proof of Theorem~\ref{thm-Gauss-DW}}
By symmetry of the torus measure, we have
\begin{equation}
\label{PM-symmetry}
\mu_L(\sigma_x=1)=\ffracB12=\mu_L(\sigma_x=-1).
\end{equation}
Let~$z$ be a site at the back of the torus --- that is distant at least~$\ffracB L2$ from the origin --- and define
\begin{equation}
%\label{}
\mu_L^\pm(-):=\mu_L(-|\sigma_z=\pm1).
\end{equation}
These measures satisfy the DLR condition with respect to any function that depends only on the ``front'' of the torus and so any weak cluster point of these measures will be an infinite-volume Gibbs measure. Extract such measures by subsequential limits and call them $\mu^+$ and~$\mu^-$, respectively. 

We claim that~$\mu^+\ne\mu^-$. Indeed, by Lemma~\ref{lemma-BB-bound} we have
\begin{equation}
%\label{}
\mu_L^+(\sigma_x=-1)\le 2c\fraktura z(\BB)
\end{equation}
once $\fraktura z(\BB)\ll1$ and, by Corollary~\ref{cor-BB-bound}, this actually happens once~$\beta,\kappa\gg1$. Thus if, say,~$2c\fraktura z(\BB)\le\ffracB14$, then $\mu_L^+(\sigma_x=-1)\le\ffracB14$ and, at the same time, $\mu_L^-(\sigma_x=+1)\le\ffracB14$. The same holds for the limiting objects and so $\mu^+\ne\mu^-$. Note that the measures can be averaged over shifts so that they become translation invariant.
\end{proofsectB}

Notice that in the last step of the proof we used, rather conveniently, the plus-minus symmetry of the torus measure. In the asymmetric cases, e.g.,~\eqref{V-asym}, one can either invoke a continuity argument --- choose~$h=h_L$ such that \eqref{PM-symmetry} holds --- or turn \eqref{don't-like} into the proof that~$|\T_L|^{-1}\sum_{x\in\T_L}\sigma_x$ will take values in~$[-1,-1+\epsilon]\,\cup\,[1-\epsilon,1]$ with probability tending to one as~$L\to\infty$. The latter ``forbidden-gap'' argument is rather robust and extends, with appropriate modifications, to all shift-ergodic infinite-volume Gibbs measures. Hence, the empirical magnetization in ergodic measures cannot change continuously with~$h$.

To prove Theorem~\ref{thm-Gauss-DW}, it remains to show the concentration of the~$\phi$'s around the~$\sigma$'s:

\begin{proofsectB}{Proof of \eqref{phi-bd}}
Let~$\mu$ be a Gibbs measure for parameters~$\beta$ and~$\kappa$. Then \eqref{H-extend} shows that, conditional on the $\sigma$'s, the $\phi$'s are Gaussian with mean
\begin{equation}
%\label{}
E_{\mu}(\phi_x|\sigma)=\kappa\bigl((2\beta\Delta+\kappa)^{-1}\sigma\bigr)_x
\end{equation}
and covariance $(2\beta\Delta+\kappa)^{-1}$, where $\Delta$ is the lattice Laplacian. Now, once $\ffracB\beta\kappa\ll1$ we may expand the inverse operator into a power series to get
\begin{equation}
%\label{}
E_{\mu}(\phi_x|\sigma)-\sigma_x=\sum_{n\ge1}\Bigl(\frac{2\beta}\kappa\Bigr)^n (\Delta^n\sigma)_x
\end{equation}
which by the fact that $|\sigma_z|=1$ is $O(\ffracB\beta\kappa)$ independently of~$x$. Since the conditional variance of $\phi_x$ is $O(\ffracB1\kappa)$, we obtain
\begin{equation}
\begin{aligned}
E_\mu\bigl((\phi_x-\sigma_x)^2\big|\sigma\bigr)
&\le
2E_\mu\bigl([\phi_x-E(\phi_x|\sigma)]^2\big|\sigma\bigr) +2\bigl(E(\phi_x|\sigma)-\sigma_x\bigr)^2
\\&=O\bigl((\ffracB\beta\kappa)^2\bigr)+O(\ffracB1\kappa)
\end{aligned}
\end{equation}
with the constants implicit in the $O$'s independent of the $\sigma$'s and~$x$. Thus, if $\kappa\gg1$ and $\ffracB\kappa\beta\gg1$, then \eqref{phi-bd} follows by the fact that $\mu^\pm$ put most of the mass on $\sigma_x=\pm1$.
\end{proofsectB}

\subsection{Gradient fields with non-convex potential}
\label{sec5.3}
\index{gradient field}
Having demonstrated the use of chessboard estimates on a toy model, we will proceed to discuss more complicated systems. We begin with an example which is somewhat similar to the Gaussian double-well model.

\newcommand{\kappaX}[1]{\mathchoice%
	{\kappa_{\text{\rm\fontsize{6.5}{6}\selectfont #1}}}
	{\kappa_{\text{\rm\fontsize{6.5}{6}\selectfont #1}}}
	{\kappa_{\text{\rm\fontsize{5}{5}\selectfont #1}}}
	{\kappa_{\text{\rm\fontsize{5}{5}\selectfont #1}}}
}
\newcommand{\kappaO}{\kappaX{O}}
\newcommand{\kappaD}{\kappaX{D}}
\newcommand{\ord}{\text{\rm ord}}
\newcommand{\dis}{\text{\rm dis}}
\newcommand{\pt}{p_{\text{\rm t}}}

A natural generalization of the massless GFF is obtained by replacing the quadratic gradient interaction by a general, even function of the gradients. The relevant Hamiltonian (again with temperature incorporated in it) is
\begin{equation}
%\label{}
\beta H(\phi):=\sum_{\langle x,y\rangle}V(\phi_x-\phi_y)
\end{equation}
The requirements that we generally put on~$V$ are continuity, evenness and quadratic growth at infinity. Under these conditions one can always define finite-volume Gibbs measures. 

As to the measures in infinite volume, the massless nature of the model may prevent existence of a meaningful thermodynamic limit in low dimensions; however, if one restricts attention to \emph{gradient variables},
\begin{equation}
%\label{}
\eta_b:=\phi_y-\phi_x\quad\text{\ if $b$ is the oriented edge }(x,y),
\end{equation}
then the infinite-volume Gibbs measures exist, and may be characterized by a DLR condition, in all $d\ge1$. We call these \emph{gradient Gibbs measures} (GGM). A non-trivial feature of the GGM is that they obey a host of constraints. Indeed, almost every~$\eta$ is such that
\begin{equation}
%\label{}
\eta_{b_1}+\eta_{b_2}+\eta_{b_3}+\eta_{b_4}=0
\end{equation}
for any plaquette $(b_1,\dots,b_4)$ with bonds listed (and oriented) in the counterclockwise direction.

Surprisingly, the classification of all possible translation-invariant, infinite-volume GGMs can be achieved under the condition that~$V$ is strictly convex:

\begin{theorem}
\label{thm-Funaki-Spohn}
Suppose $V$ is convex, twice continuously differentiable with $V''$ bounded away  from zero and infinity.
Then the shift-ergodic GGMs $\mu$ are in one-to-one correspondence with their tilt, which is a vector~$a\in\R^d$ such that
\begin{equation}
%\label{}
E_\mu(\eta_b)=a\cdot b
\end{equation}
for every (oriented) bond $b$ (we regard $b$ as a unit vector for this purpose).
\end{theorem}

The word \emph{tilt} comes from the interpretation of~$a$ as the slope or the incline of the interface whose height-gradient along bond $b$ is given by~$\eta_b$. The proof of this result --- which is due to Funaki and Spohn --- is based on the use of the Brascamp-Lieb inequality through which the convexity assumption enters in an essential way. It is also known that the large-scale fluctuation structure of the $\eta$'s is that of a Gaussian Free Field.
% (Naddaf and Spencer~\cite{Naddaf-Spencer}, Giacomin, Olla and Spohn~\cite{GOS}).

A natural question to ask is what happens when~$V$ is \emph{not} convex. Specific examples of interest might be~$V$ taking the form of a double-well potential --- kind of like for the Gaussian double-well model --- or $V$'s as in the figure:

\vglue0.3cm
\centerline{\includegraphics[width=4.5in]{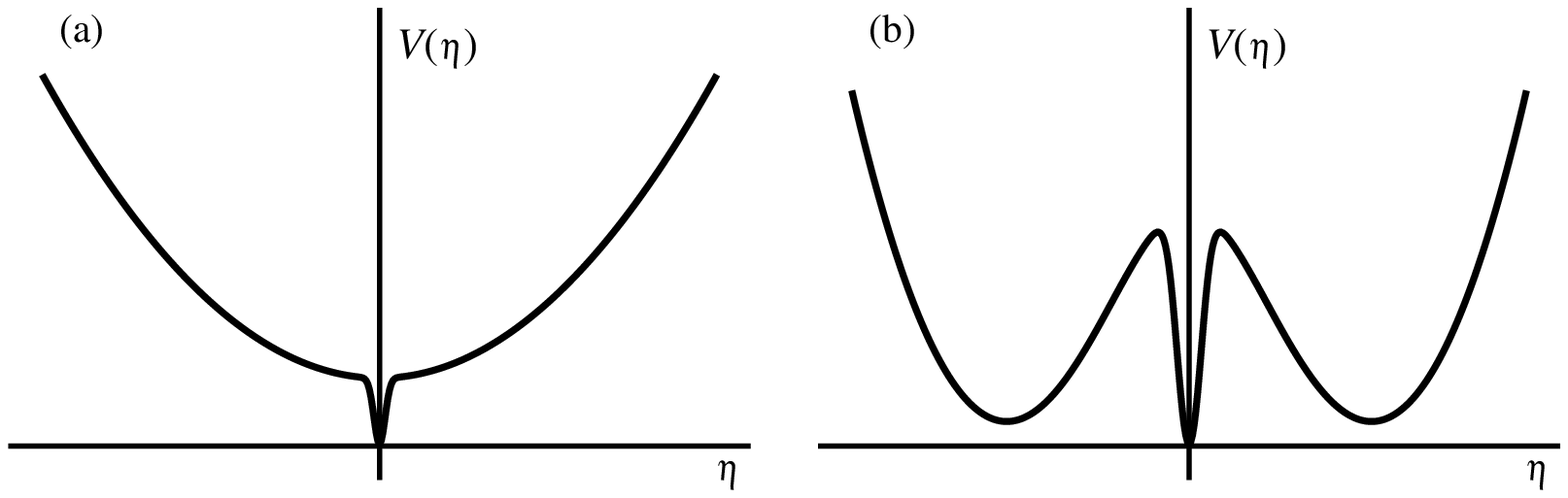}}
\noindent
As it turns out, the double-well case is not quite tractable at the moment --- and most likely behaves like a massless GFF on large scales --- but the other two cases are within reach. We will focus on the case (a) and, as for the Gaussian double-well model, assume a particular form of the potential:
\begin{equation}
\label{V-double}
\texte^{-V(\eta)}:=p\,\texte^{-\kappaO\eta^2/2}+(1-p)\,\texte^{-\kappaD\eta^2/2}
\end{equation}
where~$\kappaO$ and $\kappaD$ are positive numbers and $p\in[0,1]$ is a parameter to be varied. For this system one can prove the following result:

\begin{theorem}
\label{T:main}
Suppose $d=2$ and~$\kappaO\gg\kappaD$. Then there is~$\pt\in(0,1)$ and, for~$V$ with~$p=\pt$, there are
two distinct, infinite-volume, shift-ergodic GGMs~$\mu_\ord$ and~$\mu_\dis$ that are invariant with respect to lattice rotations and have the following properties:
\settowidth{\leftmargini}{(1i)}
\begin{enumerate}
\item[(1)] zero tilt: 
\begin{equation}
%\label{}
\frac1{|\Lambda_L|}\sum_{\begin{subarray}{c}
b=(x,y)\\ x,y\in\Lambda_L
\end{subarray}}
\eta_b\,\underset{L\to\infty}\longrightarrow\,0,
\qquad\mu_\ord,\mu_\dis\text{\rm-a.s.}
\end{equation}
\item[(2)] distinct fluctuation size:
\begin{equation}
%\label{}
E_{\mu_\ord}(\eta_b^2)\ll E_{\mu_\dis}(\eta_b^2)
\end{equation}
\end{enumerate}
\end{theorem}

The upshot of this result is that, once the convexity of~$V$ is strongly violated, the conclusions of Theorem~\ref{thm-Funaki-Spohn} do not apply. While the example is restricted to $d=2$, and to potentials of the form \eqref{V-double}, generalizations to $d\ge2$ and other potentials as in the above figure are possible and reasonably straightforward.

Here are the main steps of the proof. First, as for the Gaussian double-well model, we use \eqref{V-double} to expand the Gibbs weight according to whether the first or the second term in \eqref{V-double} applies. This gives rise to a configuration of coupling strengths $(\kappa_b)$, one for each bond~$b$, which take values in $\{\kappaO,\kappaD\}$. The joint Hamiltonian of the $\eta'$s and the $\kappa$'s is
\begin{equation}
\label{eta-kappa-Ham}
\beta H(\eta,\kappa):=\sum_b\frac{\kappa_b}2\eta_b^2
\end{equation}
The joint measure is RP with respect to reflections through bonds and sites and, conditional on~$(\kappa_b)$, the $\eta$'s are Gaussian. 

For the proof of phase coexistence, we focus on lattice plaquettes and divide these into good and bad according to whether all of the edges have the same coupling~$\kappa$ or not. The dissemination of each bad patterns leads to a Gaussian integral but this time for GFF with inhomogeneous --- yet periodically varying --- couplings. For instance, the pattern with three bonds of type $\kappaO$ and one of type $\kappaD$ disseminates into periodic configuration where the edges on every other vertical line is of type $\kappaD$ and all other edges are of type $\kappaO$. Similarly for all other bad patterns.

The periodic nature of the disseminated events allows the use of Fourier modes --- i.e., pass to the reciprocal torus --- to diagonalize the requisite covariance matrices. For instance, the aforementioned pattern with three $\kappaO$'s and one $\kappaD$ leads to a configuration which is periodic with period two. A calculation shows that the covariance is block diagonal with $2\times 2$ blocks of the form
\begin{equation}
%\label{}
\Pi(k):=\left(\,
\begin{matrix}
\kappaO|a_-|^2+\frac12(\kappaO+\kappaD)|b_-|^2 &
\frac12(\kappaO-\kappaD)
|b_-|^2
\\*[2mm]
\frac12(\kappaO-\kappaD)
|b_-|^2
&
\kappaO|a_+|^2+\frac12(\kappaO+\kappaD)
|b_-|^2
\end{matrix}
\,\right)
\end{equation}
where $a_\pm$ and~$b_\pm$ are defined by
\begin{equation}
%\label{} 
a_\pm=1\pm \texte^{\texti k_1}
\quad\text{and}\quad  b_\pm=1\pm \texte^{\texti k_2}
\end{equation}
with $k:=(k_1,k_2)$ varying through one half of the reciprocal torus $\T_L^\ast$. (The block combines the contribution of both $k$ and $k+\pi\hate_1$, and so we only need half of all $k$'s.) The requisite Gaussian integral then reduces to $\prod_{k\in\T_L^\ast\setminus\{0\}}[\det\Pi(k)]^{-1/4}$ where in the exponent we get $\ffracB14$ instead of the expected $\ffracB12$ to account for double counting of the $k$'s. To estimate the growth rate of this product, we note that
\begin{multline}
%\label{}
\qquad
\prod_{k\in\T_L^\ast\setminus\{0\}}[\det\Pi(k)]^{-1/4}
\\=\exp\biggl\{-|\T_L|\frac14\int\frac{\textd k}{(2\pi)^2}\log\det\Pi(k)+o(|\T_L|)\biggr\}
\qquad
\end{multline}
The integral plays the role of the free energy associated with the Gaussian variables on the background of the specific periodic configuration of the $\kappa$'s. A similar expression --- with different integrand --- applies to each pattern.

Comparing the integrals for all possible arrangements of the two types of bonds around a plaquette, we find that under the condition $\kappaO\gg\kappaD$, the bad patterns are heavily suppressed. Thus bad plaquettes are infrequent and can be regarded as parts of a contour. As it is not possible to pass from all-$\kappaO$ pattern to all-$\kappaD$ pattern without crossing a bad plaquette, the coexistence follows ---  as for the double-well model --- by a standard Peierls' argument and chessboard estimates. Full details of the proof are to be found in a paper by Koteck\'y and the present author.

The two-dimensional model has the special feature that we can actually compute~$\pt$:

\begin{theorem}
\label{T:dual}
Let $d=2$.
If\/ $\ffracB\kappaO\kappaD\gg1$, then~$\pt$ is given by
\begin{equation}
\label{pt-eq}
\frac{\pt}{1-\pt}=\Bigl(\frac{\kappaD}{\kappaO}\Bigr)^{\ffracB14}.
\end{equation}
\end{theorem}

This is a consequence of a duality relation that can be used to exchange the roles of~$\kappaO$ and~$\kappaD$. It is also  interesting to note that, while the one-to-one correspondence between the Gibbs measures and their tilt is violated for non-convex potentials, the large-scale fluctuation structure remains that of a Gaussian Free Field. Indeed, we have:

\begin{theorem}
\label{thm-GFF}
Let~$d=2$. For each translation-invariant, ergodic gradient Gibbs measure~$\mu$ with zero tilt, there exists a positive-definite
$d\times d$ matrix~$q=q(\mu)$ such that for any smooth $f\colon\R^2\to\R$ with compact support and $\int f(x)\textd x=0$, 
\begin{equation}
%\label{}
\int\textd x\,\phi_{\lfloor x/\epsilon\rfloor} f(x)
\,\,\overset{\DD}{\underset{\epsilon\downarrow0}\longrightarrow}\,\,\NN\bigl(0,(f,Q^{-1}f)\bigr)
\end{equation}
where $\NN(0,C)$ denotes a normal random variable with mean zero and covariance~$C$ and $Q$ is the elliptic operator
\begin{equation}
%\label{}
Qf(x):=\sum_{i,j=1}^d q_{ij}\frac{\partial^2}{\partial x_i\partial x_j}f(x)
\end{equation}

\end{theorem}

The basis of this result --- derived in all~$d\ge1$ by Spohn and the present author --- is the fact that, conditional on the $\kappa$'s, the $\phi$'s are Gaussian with mean zero and covariance given by the inverse of the generator of a reversible random walk in random environment. The Gaussian limit is a consequence of an (annealed) invariance principle for such random walks and some basic arguments in homogenization theory. The restriction to zero tilt appears crucially in the proof.

\subsection{Spin-waves vs infinite ground-state degeneracy}
\label{sec5.4}
\index{spin wave}
Next we will discuss a couple of spin models whose distinctive feature is a high degeneracy of their ground state which is removed, at positive temperature, by soft-mode spin-wave fluctuations. The simplest example with such property is as follows:

\bigskip\noindent
\uemph{Orbital compass model}:
\index{orbital compass model}
Here the spins on~$\Z^d$ take values in a unit sphere in~$\R^d$, i.e., $S_x\in\BbbS^{d-1}$ with~$x\in\Z^d$. The Hamiltonian is
\begin{equation}
\label{H-OCM}
H(S):=\sum_x\sum_{\alpha=1}^d(S_x^{(\alpha)}-S_{x+\hate_\alpha}^{(\alpha)})^2
\end{equation}
where $S_x^{(\alpha)}$ denotes the $\alpha$-th Cartesian component of the spin and $\hate_\alpha$ is the unit vector in the $\alpha$-th coordinate direction. 

Despite a formal similarity with the Heisenberg model, note that only one component of the spin is coupled in each lattice direction. Notwithstanding, \emph{every} constant configuration is still a minimum-energy state of~\eqref{H-OCM}. Further ground states may be obtained from the constant ones by picking a coordinate direction $\alpha$ and changing the sign of the $\alpha$-th component of all spins in some of the ``lines'' parallel with~$\hate_\alpha$. In $d=2$ these are all ground states but in $d\ge3$ other operations are possible that preserve the minimum-energy property.

The key question is now what happens with this huge ground-state degeneracy at positive temperatures. Here is a theorem one can prove about the two-dimensional system:

\begin{theorem}
\label{thm-OCM}
For each~$\epsilon>0$ there exist~$\beta_0>0$ and, for each~$\beta\ge\beta_0$, there exist two distinct, shift-ergodic Gibbs measures $\mu_1,\mu_2\in\mathfrak G_\beta$ such that
\begin{equation}
\label{OCM-QQ}
E_{\mu_j}\bigl(|S_x\cdot\hate_j|\bigr)\ge1-\epsilon,\qquad j=1,2
\end{equation}
Moreover, for any $\mu\in\mathfrak G_\beta$ we have
\begin{equation}
%\label{}
E_\mu(S_x)=0
\end{equation}
and there are no shift-ergodic $\mu\in\mathfrak G_\beta$, $\beta\ge\beta_0$, for which we would have $\max_{j=1,2}E_\mu\bigl(|S_x\cdot\hate_j|\bigr)<1-\epsilon$.
\end{theorem}

The main idea underlying the proof is the evaluation of the free energy associated with spin-wave perturbations of the constant ground states; it this expected that only the states with the largest contribution of these fluctuations survive at positive temperatures. Specifically, we need to quantify the growth rate of the torus partition function with all spins constrained to lie within $\Delta$ of a given direction:

\begin{lemma}
\label{lemma-SW-OC}
For each $\epsilon>0$ there is $\delta>0$ such that if $\beta,\Delta$ obey
\begin{equation}
%\label{}
\beta\Delta^2>\frac1\delta\qquad\text{and}\qquad\beta\Delta^3<\delta
\end{equation}
then for every $\hatv_\theta:=(\cos\theta,\sin\theta)\in\BbbS^1$,
\begin{equation}
%\label{}
E_{\otimes\mu_0}\Bigl(\,\texte^{-\beta H_L(S)}\prod_{x\in\T_L}1_{\{|S_x-\hatv_\theta|<\Delta\}}\Bigr)
=\Bigl(\frac{2\pi}{\beta}\Bigr)^{L^2/2}\,\texte^{-L^2[F(\theta)+o(\epsilon)]}
\end{equation}
where
\begin{equation}
\label{SWFE-OC}
F(\theta):=\frac12\int\frac{\textd k}{(2\pi)^2}\log\bigl\{\sin^2(\theta)|1-\texte^{\texti k_1}|^2+\cos^2(\theta)|1-\texte^{\texti k_2}|^2\bigr\}
\end{equation}
\end{lemma}

\newcommand{\BBE}{\BB_{\text{\rm E}}}
\newcommand{\BBSW}{\BB_{\text{\rm SW}}}

The quantity~$F$ has the interpretation of the \emph{spin-wave free energy} where the term ``spin wave'' refers to slowly varying deformations of a constant ground states. A convexity argument --- based on the identity $\sin^2(\theta)+\cos^2(\theta)=1$ --- now shows that~$F$ is minimized by $\theta=0,\ffracB\pi2,\pi,\ffracB{3\pi}2$, i.e., exactly in one of the coordinate directions. This corroborates the intuition that only the configurations with most of the spins aligned in one of these directions will be relevant at low temperatures. However, to extract a proof of phase coexistence, we will have to again invoke a Peierls' argument.

Fix~$\kappa>0$ and let $\Delta:=\beta^{-\frac 5{12}}$ and $B:=\log\beta$ and let $\BBE$ and $\BBSW$ denote the following events:
%\settowidth{\leftmargini}{(11)}
\begin{enumerate}
\item[(1)]
$\BBE$ := $\{$ a pair of neighboring spins in $\Lambda_B$ differ by an angle~$\ge\Delta$ $\}$
\item[(2)]
$\BBSW$ is the set of configurations in the complement of $\BBE$ in which the block~$\Lambda_B$ has all neighboring spins within~$\Delta$ of each other with at least~$\kappa\gg\Delta$ from one of the four coordinate directions
\end{enumerate}
The event $\BBE$ captures the situations when two neighboring spins are not quite close to each other leading to excess energy order $\Delta^2$. As a result of that,
\begin{equation}
%\label{}
\fraktura z(\BBE)\le 3 B^3\texte^{-c_3\beta\Delta^2}
\end{equation}
The event $\BBSW$ collects the configurations where the energy is good but the fluctuations are not sufficiently powerful.
The calculation in Lemma~\ref{lemma-SW-OC} and a simple use of the subadditivity lemma show
\begin{equation}
%\label{}
\fraktura z(\BBSW)\le\frac {c_1}\Delta\texte^{-c_2B^3\kappa^2}
\end{equation}
for some constants~$c_1,c_2>0$. Thus, for our choices of~$\Delta$ and~$B$, once~$\beta\gg1$ the density of blocks where $\BBE\cup\BBSW$ occurs in any typical configuration from the torus measure will be rather small. However, if a block is aligned in one coordinate direction and another block is aligned in a different direction, they must be separated by a ``circuit'' of bad blocks.  Such circuits are improbable which leads to phase separation. Details of these calculations --- which extend even to quantum setting --- can be found in a paper by Chayes, Starr and the present author.

\bigskip\noindent
\uemph{120-degree model}:
\index{120-degree model}
A somewhat more complicated version of the interaction, but with the spins $S_x$ taking values in the unit circle~$\BbbS^1$, can be contrived in $d=3$. The Hamiltonian will actually look just as for the orbital compass model except that~$S_x^{(\alpha)}$ are not Cartesian components but projections on the three third-roots of unity $\hatb_1,\hatb_2,\hatb_3$ in~$\BbbS^1$. Explicitly,
\begin{equation}
%\label{}
H(S):=\sum_x\sum_{\alpha=1,2,3}\bigl(S_x\cdot\hatb_\alpha-S_{x+\hate_\alpha}\cdot\hatb_\alpha\bigr)^2
\end{equation}
Again, all constant configurations are ground states and further ground states may again be obtained by judicious reflections. Fortunately, the number of energy-preserving operations one can perform on ground states is much smaller than for the orbital compass model, and all ground states can thus be classified. Namely, given a ground state configuration, every unit cube in~$\Z^3$ looks as one of the four cubes in the picture

\vglue0.5cm
\centerline{\includegraphics[width=4.4in]{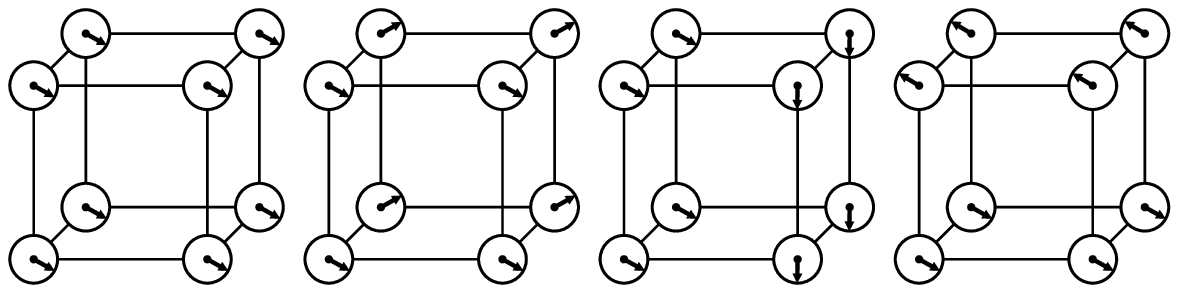}}
\smallskip
\noindent
modulo, of course, a simultaneous rotation of all spins. Here is what we can we say rigorously about this model:

\begin{theorem}
\label{thm-120}
Let $\hatw_1,\dots,\hatw_6\in\BbbS^1$ be the six sixth roots of unity.
For each~$\epsilon>0$ there exist~$\beta_0>0$ and, for each~$\beta\ge\beta_0$, there exist six distinct, shift-ergodic Gibbs measures $\mu_1,\dots,\mu_6\in\mathfrak G_\beta$ such that
\begin{equation}
\label{120M}
E_{\mu_j}\bigl(S_x\cdot\hatw_j\bigr)\ge1-\epsilon,\qquad j=1,\dots,6
\end{equation}
There are no shift-ergodic $\mu\in\mathfrak G_\beta$, $\beta\ge\beta_0$, for which we would have $\max_{j=1,\dots,6}E_\mu\bigl(S_x\cdot\hatw_j\bigr)<1-\epsilon$.
\end{theorem}

The ideas underlying this theorem are quite similar to the orbital compass model. First we find out that the spin-wave free energy for fluctuations about the ground state pointing in direction $\theta$ is given by
\begin{equation}
\label{SWFE-120}
F(\theta):=\frac12\int\frac{\textd k}{(2\pi)^3}\biggl[\,\log\sum_{\alpha=1,2,3}q_\alpha(\theta)|1-\texte^{\texti k_\alpha}|^2\biggr]
\end{equation}
where $q_1:=\sin^2(\theta)$, $q_2:=\sin^2(\theta-120^\circ)$ and~$q_3:=\sin^2(\theta+120^\circ)$. A surprisingly sophisticated argument is then required to show that~$F$ is minimal only for $\theta$ of the form~$\frac\pi3j$, $j=1,\dots,6$. Once we have this information, the rest of the argument follows a route very similar to that for the orbital compass model (including the introduction of the scales $\kappa$ and $\Delta$ and the corresponding events $\BBE$ and~$\BBSW$). Details appeared in a paper by Chayes, Nussinov and the present author.

\bigskip\noindent
\uemph{n.n. and n.n.n.\ antiferromagnet}:
Finally, we will consider a toy model that exemplifies the features of both systems above. Here $d=2$  and the spins take again values in~$\BbbS^1$, but the interaction is \emph{antiferromagnetic} --- that is, with a preference for antialignment --- for both nearest and next-nearest neighbors:
\begin{equation}
%\label{}
H(S):=\gamma\sum_x\bigl[S_x\cdot S_{x+\hate_1}+S_x\cdot S_{x+\hate_2}\bigr]+
\sum_x\bigl[S_x\cdot S_{x+\hate_1+\hate_2}+S_x\cdot S_{x+\hate_1-\hate_2}\bigr]
\end{equation}
Assuming $|\gamma|<2$, the minimum energy state is obtained by first enforcing the n.n.n.\ constraints --- there is an antiferromagnetic, or Ne\'el, order on both even and odd sublattice --- and only then worrying about how to satiate the n.n.\ constraint. But once the sublattices are ordered antiferromagnetically, the net interaction between the sublattices is zero --- and so each of the sublattices can be rotated independently! Here is a configuration of this form:

\vglue0.3cm
\centerline{\includegraphics[width=3.0in]{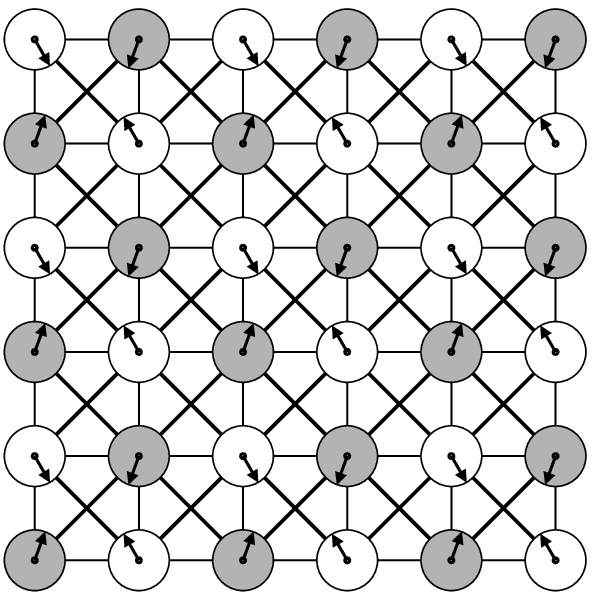}}
\smallskip\medskip
\noindent
For this system we can nevertheless prove the following theorem:

\begin{theorem}
\label{thm-anti}
For each~$\epsilon>0$ there exist~$\beta_0>0$ and, for each~$\beta\ge\beta_0$, there exist two distinct, shift-ergodic Gibbs measures $\mu_1,\mu_2\in\mathfrak G_\beta$ such that
\begin{equation}
\label{O2M1}
-E_{\mu_j}\bigl(S_x\cdot S_{x+\hate_1\pm\hate_2}\bigr)\ge1-\epsilon
\end{equation}
and
\begin{equation}
\label{O2M2}
E_{\mu_j}\bigl(S_x\cdot S_{x+\hate_j}\bigr)\ge1-\epsilon,\qquad j=1,2
\end{equation}
There are no shift-ergodic $\mu\in\mathfrak G_\beta$, $\beta\ge\beta_0$, for which either \eqref{O2M1} or at least one of \eqref{O2M2} does not hold.
\end{theorem}

As for the two models above, everything boils down to a spin-wave calculation. Here the relevant parameter is the relative orientation $\theta$ of the two antiferromagnetically ordered sublattices. The spin-wave free energy is then
\begin{equation}
%\label{3.8}
F(\theta):=\frac12\int_{[-\pi,\pi]^2}\frac{\textd\bk}{(2\pi)^2}\log D_{\bk}(\theta)
\end{equation}
where
\begin{multline}
\label{3.9}
\quad
D_{\bk}(\theta):=|1-\texte^{\texti(k_1+k_2)}|^2+|1-\texte^{\texti(k_1-k_2)}|^2
\\+\gamma\cos(\theta)\bigl(|1-\texte^{\texti k_1}|^2-|1-\texte^{\texti k_1}|^2\bigr)
\quad
\end{multline}
As $D(\theta)=\alpha D(0)+(1-\alpha)D(\pi)$, with $\alpha:=\frac12(1+\cos(\theta))$, Jensen's inequality for the logarithm directly shows that $F$ is minimized by $\theta=0$ or $\theta=\pi$. In spin configurations, the former corresponds to horizontal alignment and vertical antialignment of nearest neighbors, and the latter to horizontal antialignment and vertical alignment, i.e., stripe states. Details of all calculations appeared in a paper by Chayes, Kivelson and the present author.

Notice that, despite the fact that the lattices maintain a specific \emph{relative} orientation at low temperatures, a Mermin-Wagner argument ensures that every Gibbs measure is invariant under a rigid rotation of all spins. 

\subsection{Literature remarks}

The Gaussian double-well model is a standard example which can be treated either by methods of reflection positivity, or by Pirogov-Sinai theory~\cite{Dobrushin-Zahradnik}. Representations of the kind \eqref{V-repr} have been used already before, e.g., by K\"ulske~\cite{Kuelske1,Kuelske2} and Zahradn\'{\i}k~\cite{Zahradnik}. The method of proof presented here draws on the work of Dobrushin, Koteck\'y and Shlosman~\cite{Dobrushin-Shlosman,KS-proceedings,KS1Bi} which was used to control order-disorder transitions in a number of systems; most notably, the $q$-state Potts model with~$q\gg1$~\cite{KS1Bi}. These methods may be combined with graphical representations of Edwards-Sokal~\cite{ES1Bi} (or Fortuin-Kasteleyn~\cite{FK}) to establish rather complicated phase diagrams, e.g.,~\cite{Chayes-Machta,BCK}. Recently, the method has been used to resolve a controversy about a transition can occur in 2D non-linear vector models~\cite{Senya_VE-I,vanEnter-Shlosman}.

Theorem~\ref{thm-Funaki-Spohn} has been proved by Funaki and Spohn~\cite{Funaki-Spohn}. As already mentioned, their proof is based on convexity properties of the potential~$V$ --- by invoking the Brascamp-Lieb inequality as well as certain coupling argument to the natural dynamical version of the model --- and so it does not extend beyond the convex case. (A review of the gradient measures, and further intriguing results, can be found in Funaki~\cite{Funaki}, Velenik~\cite{Velenik} or Sheffield~\cite{Sheffield}.) Theorem~\ref{T:main} was proved by Biskup and Koteck\'y~\cite{BK2Bi}; Theorem~\ref{thm-GFF} was derived by Biskup and Spohn~\cite{Biskup-Spohn}. 

The interest in models in Sect.~\ref{sec5.4} came from a physics controversy about whether orbital ordering in transition-metal oxides exists at low temperatures. On the basis of rigorous work by Biskup, Chayes and Nussinov~\cite{BCN1} (120-degree model) Biskup, Chayes, Nussinov and Starr \cite{Biskup-Chayes-Starr,BCN2} (2D and 3D orbital compass model), it was demonstrated that, at least at the level of classical models, spin-wave fluctuations stabilize certain ground states~\cite{EPL-kompasy}. The conclusions hold also the 2D quantum orbital-compass model with large quantum spins~\cite{Biskup-Chayes-Starr}. The mechanism of entropic stabilization --- or, in physics jargon, \emph{order by disorder} --- is most clearly demonstrated in the n.n.\,\&\,n.n.n.\ antiferromagnet studied by Biskup, Chayes and Kivelson \cite{BCKiv}. This model actually goes back to the papers by Shender~\cite{Shender} and Henley~\cite{Henley} which first spelled out the original order-by-disorder physics arguments.

All three ``phase-coexistence'' theorems in Sect.~\ref{sec5.4} have, apart from an existence clause, also a clause on the \emph{absence} of ergodic states whose local properties deviate from those whose existence was asserted. Actually, these were not the content of the original work \cite{BCN1,BCKiv} because, at that time, the focus on torus measures dictated by reflection positivity was deemed to make it impossible to \emph{rule out} the occurrence of some exotic measures. A passage to such statements was opened by the work of Biskup and Koteck\'y~\cite{BK1Bi}; the non-existence clauses in Theorems~\ref{thm-OCM}, \ref{thm-120} and \ref{thm-anti} are direct consequences of the main result of~\cite{BK1Bi} and the method of proof of the existence part. This technique does not quite apply in the setting of gradient models due to the strong role the boundary conditions play in this case.

\section{Topics not covered}
There are naturally many interesting topics dealing with reflection positivity that have not been covered by these notes. Here we will attempt to at least provide a few relevant comments and give pointers to the literature where an interested reader may explore the subject to the desired level of detail.

The first (and large) area which was neglected is that of \emph{quantum} models. Here one faces the principal difficulty that the spin variables are replaced by operators which, generally, do not commute with one another. Nevertheless, reflection positivity can be proved for reflections through planes between sites under the condition that the Hamiltonian is of the form \eqref{torus-H-RP}. (For reflections through planes of sites the non-commutativity of involved objects makes the above technology largely unavailable.) Thus, chessboard estimates and, by a passage via the Duhamel two-point function, also infrared bound can again be established. This and the resulting applications to proofs of phase transitions in, e.g., the quantum Heisenberg \emph{anti}ferromagnet \index{Heisenberg model!quantum} and XY-model \index{XY-model} constitute the papers of Dyson, Lieb and Simon~\cite{DLS} and Fr\"ohlich and Lieb~\cite{FL}. A pedagogical account of these can be found in the notes by T\'oth~\cite{Balint-web}.

Unlike for the classical models, in the quantum setting reflection positivity appears to be a somewhat peculiar condition. Generally, it requires that the involved operators can be represented by either real or purely imaginary matrices. This is where the technique fails in the case of the quantum Heisenberg ferromagnet (Speer~\cite{Speer}, but see also Kennedy~\cite{Kennedy} and Conlon and Solovej~\cite{Conlon-Solovej}). Notwithstanding, the technique continued to be applied in the quantum world to derive useful conclusions; e.g., to study long range order in two-dimensional antiferromagnets (Kennedy, Lieb and Shastry~\cite{Kennedy-Lieb-Shastry}), to resolve the so called flux phase problem in the Hubbard model (Lieb~\cite{Lieb-flux-phase}, see also Macris and Nachtergaele~\cite{Macris-Nachtergaele}) or to prove uniqueness of the ground state in the half-filled band therein (Lieb~\cite{Lieb-twotheorems}). The latter work invokes \emph{spin-reflection positivity}; a new idea later further exploited by, e.g., Tian~\cite{Tian} and Tasaki~\cite{Tasaki}. Other applications of reflection positivity in itinerant-electron models appear in, e.g., Macris~\cite{Macris} and Macris and Lebowitz \cite{Macris-Lebowitz}.

As already mentioned, one can use RP to develop a rigorous link between the phase transitions in quantum and classical systems (Biskup, Chayes and Starr~\cite{Biskup-Chayes-Starr}). Here the main idea is the conversion of the quantum chessboard estimate to the classical one by means of an extension of Berezin-Lieb inequalitites to matrix elements in the basis of coherent states.

Another topic not sufficiently represented in these notes is that of dimer or other combinatorial models. Here we wish to mention, e.g., the conclusions concerning the six-vertex model and hard-core lattice gasses (Fr\"ohlich, Israel, Lieb and Simon~\cite{FILS2}) or the liquid-crystal models based on interacting dimers (Heilmann and Lieb~\cite{Heilmann-Lieb} and Abraham and Heilmann~\cite{Abraham-Heilmann}). There is also a novel application to characterization of graph homomorphisms (Freedman, Lov\'asz and Schrijver~\cite{Freedman-Lovasz-Schrijver}). 

The origin of reflection positivity lies within the field theory as part of the Osterwalder-Schrader axioms. A reader interested in this direction should employ the relevant search outlets to explore the literature on the subject. For statistical mechanics, interesting applications come in the proofs of phase transitions in Euclidean field theories, e.g., that of quark confinement (Borgs and Seiler~\cite{Borgs-Seiler}) or chiral symmetry breaking (Salmhofer and Seiler~\cite{Salmhofer-Seiler}) in gauge theories. 

Finally, there is the recent clever application of chessboard estimates to control the rigidity of Dobrushin interfaces in the Ising (and some other) three dimensional models (Shlosman and Vignaud~\cite{Shlosman-Vignaud}). This direction will likely be further exploited to study interface states in continuum-spin systems.

\section{Three open problems}
We finish with a brief discussion of three general open problems of the subject covered by these notes which the present author finds worthy of significant research effort.

In Chapters~\ref{chap-2} and~\ref{chap-3} we have shown how useful the infrared bound is in proofs of symmetry breaking and control of the mean-field approximation. Unfortunately, the only way we currently have for proving the IRB is reflection positivity. So our first problem is:

\begin{problem}
Consider models with the Hamiltonian $H=-\sum_{\langle x,y\rangle}S_x\cdot S_y$. Prove the IRB directly without appeal to RP.
\end{problem}

As already mentioned, a successful attempt in this direction has been made by Sakai~\cite{Sakai}, who managed to apply the  lace expansion to a modified random current representation of the Ising model. However, here we have in mind something perhaps more robust which addresses directly the principal reason why we need RP, which is that 

\smallskip
\centerline{the spins $(S_x)$ are not \emph{a priori} independent Gaussian} 

\smallskip
\noindent
Among approaches in this direction is the \emph{spherical approximation} for the $O(n)$ model, in which the constraint $|S_x|=1$ at every spin is replaced by a constraint on~$\sum_x|S_x|^2$.

\smallskip
The IRB is often viewed as a rigorous version of \emph{spin-wave theory}. This theory, initiated in the work of Dyson~\cite{Dyson} and others, describes continuous deformations of the lowest energy states by means of an appropriate Gaussian field theory. In Chapter~\ref{chap-5} we saw that chessboard estimates may be applied \emph{in conjunction} with spin-wave calculations --- which are generally deemed to be the realm of the IRB --- to prove phase transitions. This was possible because spin-waves disqualified all but a finite number of ground states from candidacy for low-temperature states. Notwithstanding, one might be able to do the same even in the presence of infinitely  many low-temperature states:

\begin{problem}
Prove symmetry breaking at low temperatures in systems with continuous internal symmetry --- e.g., the $O(2)$-model --- without the use of the IRB. Chessboard estimates are allowed.
\end{problem}

An interesting resource for thinking about this problem may be the paper of Bricmont and Fontaine~\cite{Bricmont-Fontaine}.

Further motivation to look at this problem comes from quantum theory: The quantum Heisenberg ferromagnet is not RP (see Speer~\cite{Speer}) and so there is no proof of the IRB and, consequently, no proof of low-temperature symmetry breaking. On the other hand, the classical Heisenberg ferromagnet is RP and so the spin-condensation argument applies. However, if we had a more robust proof of symmetry breaking in the classical model, e.g., using chessboard estimates, one might hope to extend the techniques of Biskup, Chayes and Starr~\cite{Biskup-Chayes-Starr} to include also the quantum system.

\smallskip
While the theory described in these notes is not restricted exclusively to ferromagnetic systems, in order to have the IRB one needs a good deal of attractivity in the system. It is actually clear that the IRB cannot hold as stated for antiferromagnets, e.g., hard core lattice gas, which is a model with variables~$n_x\in\{0,1\}$ and the ``Gibbs'' weight proportional to
\begin{equation}
%\label{}
\lambda^{\sum_xn_x}\prod_{\langle x,y\rangle}(1-n_xn_y),
\end{equation}
or the $q$-state Potts antiferromagnet, which is the model in \eqref{H-1.7} with~$J<0$. Indeed, the \emph{staggered} long-range order, which is known to occur in the hard core lattice gas once $\lambda\gg1$, implies that the macroscopically occupied mode is $k=(\pi,\dots,\pi)$ rather than~$k=0$. Nevertheless, we hope that some progress can be made and so we pose:

\begin{problem}
Derive a version of the IRB for the hard-core lattice gas and/or the $q$-state Potts antiferromagnet at zero temperature.
\end{problem}

Solving this problem would, hopefully, also provide an easier passage to the proof that the critical $\lambda$ for the appearance of staggered order tends to zero as~$d\to\infty$ --- in fact, if the mean-field theory is right then one should have~$\lambda_\cc\sim\ffracB cd$ --- and that the 3-coloring of~$\Z^2$ exhibits six distinct extremal measures of maximal entropy. These results have recently been obtained by sophisticated contour-counting arguments~\cite{Galvin-Kahn,Galvin-Randall}.

%\chapter*{References}


\begin{thebibliography}{100}
%\addcontentsline{toc}{chapter}{{\numberline {}References}{}}

\bibitem{Abraham-Heilmann}
D.B~Abraham and O.J.~Heilmann, \textit{Interacting dimers on the simple cubic lattice as a model for liquid crystals}, J.~Phys. A: Math. Gen. \textbf{13} (1980) 1051--1062.

\bibitem{Aizenman}
M.~Aizenman, \textit{Geometric analysis of $\phi^4$ fields and Ising models. I, II.},  Commun. Math. Phys.~\textbf{86} (1982) 1--48.

\bibitem{ACCN}
M.~Aizenman, J.T.~Chayes, L.~Chayes and C.M.~Newman,
\textit{Discontinuity of the magnetization in one-dimensional
\hbox{$1/\vert x-y\vert^2$} Ising and Potts models}, J.~Statist.
Phys. \textbf{50} (1988), no. 1-2, 1--40.

\bibitem{AF1}
M.~Aizenman and R.~Fern\'andez, \textit{On the critical behavior of the magnetization in high-dimensional Ising models},  J.~Statist. Phys.~\textbf{44} (1986) 393--454. 


\bibitem{AF2}
M.~Aizenman and R.~Fern\'andez, \textit{Critical exponents for long-range interactions}
Lett. Math. Phys.~\textbf{16} (1988), no. 1, 39--49.


\bibitem{Angelescu-Zagrebnov}
N.~Angelescu and V.A.~Zagrebnov, \textit{A lattice model of liquid crystals with matrix order parameter}, J.~Phys.~A~\textbf{15} (1982), no.~11, L639--L643.


\bibitem{Biskup}
M.~Biskup, \textit{Reflection positivity of the random-cluster measure invalidated for non-integer~$q$}, J.~Statist. Phys.~\textbf{92} (1998) 369--375.

\bibitem{BBCK}
M. Biskup, C. Borgs, J.T. Chayes and R. Koteck\'y, \textit{Gibbs states of graphical representations of the Potts model with external fields}, J. Math. Phys. \textbf{41} (2000), no. 3, 1170--1210.

\bibitem{BC}
M.~Biskup and L.~Chayes, \textit{Rigorous analysis of discontinuous phase transitions via mean-field bounds}, Commun. Math. Phys.~\textbf{238} (2003), no.~1-2, 53--93.

\bibitem{Biskup-Chayes-Crawford}
M.~Biskup, L.~Chayes and N.~Crawford, \textit{Mean-field driven first-order phase transitions in systems with long-range interactions}, J.~Statist. Phys.~\textbf{122} (2006), no.~6, 1139--1193.

\bibitem{BCKiv}
M.~Biskup, L.~Chayes, and S.A.~Kivelson, 
\textit{Order by disorder, without order, in a two-dimensional spin system with~$O(2)$-symmetry}, 
Ann. Henri Poincar\'e~\textbf{5} (2004), no.~6, 1181--1205.

\bibitem{BCK}
M.~Biskup, L.~Chayes, and R. Koteck\'y, 
\textit{Coexistence of partially disorder\-ed/or\-dered phases in an extended Potts model}, 
J.~Statist. Phys.~\textbf{99} (2000), no.~5/6, 1169--1206.


\bibitem{BCN1}
M.~Biskup, L.~Chayes, and Z.~Nussinov, 
\textit{Orbital ordering in transition-metal compounds: I.~The 120-degree model},
Commun. Math. Phys.~\textbf{255} (2005), no.~2, 253--292.

\bibitem{BCN2}
M.~Biskup, L.~Chayes and Z.~Nussinov, \textit{Orbital ordering in transition-metal compounds: II. The orbital compass model}, in preparation.

\bibitem{Biskup-Chayes-Starr}
M.~Biskup, L.~Chayes and S.~Starr, \textit{Quantum spin systems at positive temperature}, Commun. Math. Phys. \textbf{269} (2007), no. 3, 611-657 

\bibitem{BK1Bi}
M.~Biskup and R.~Koteck\'y, \textit{Forbidden gap argument for phase transitions proved by means of chessboard estimates}, Commun. Math. Phys.~\textbf{264} (2006), no.~3, 631--656.

\bibitem{BK2Bi}
M.~Biskup and R.~Koteck\'y, \textit{Phase coexistence of gradient Gibbs states}, Probab. Theory Rel. Fields \textbf{139} (2007), no. 1-2, 1--39.

\bibitem{Biskup-Spohn}
M.~Biskup and H.~Spohn, \textit{Scaling limit for a class of gradient fields with non-convex potentials}, Ann. Probab. (to appear)

\bibitem{Bodineau}
T.~Bodineau, \textit{Translation invariant Gibbs states for the Ising model}, Probab. Theory Related Fields~\textbf{135} (2006), no.~2, 153--168.

\bibitem{Borgs-Seiler}
C.~Borgs and E.~Seiler, \textit{Quark deconfinement at high temperature: a rigorous proof}, Nucl. Phys.~B \textbf{215} (1983), no. 1, 125--135.

\bibitem{Bricmont-Fontaine}
J.~Bricmont and J.-R.~Fontaine, \textit{Infrared bounds and the Peierls argument in two dimensions}, Commun. Math. Phys. \textbf{87} (1982/83), no. 3, 417--427.

\bibitem{BKLS}
J.~Bricmont, H.~Kesten, J.L.~Lebowitz and R.H.~Schonmann,
\textit{A note on the Ising model in high dimensions}, Commun.
Math.~Phys.~\textbf{122} (1989) 597--607.

\bibitem{Burton-Keane}
R.M.~Burton and M.~Keane, \textit{Density and uniqueness in percolation}, 
Commun. Math. Phys. \textbf{121} (1989), no. 3, 501--505.

\bibitem{Chayes}
L.~Chayes, \textit{Mean field analysis of low dimensional systems}, Commun. Math. Phys. (to appear)

\bibitem{CKS}
L.~Chayes, R.~Koteck{\'y}, and S.~B. Shlosman.
\emph{Staggered phases in diluted systems with continuous spins},
Commun. Math. Phys.~\textbf{189} (1997) 631--640.

\bibitem{Chayes-Machta}
L.~Chayes and J.~Machta, \textit{Graphical representations and cluster algorithms. Part I: Discrete spin systems},
Physica A \textbf{239} (1997) 542--601.

\bibitem{Conlon-Solovej}
J.G.~Conlon and J.P.~Solovej, \textit{On asymptotic limits for the quantum Heisenberg model},
J.~Phys.~A \textbf{23} (1990), no. 14, 3199--3213.

\bibitem{Curie}
P.~Curie, \textit{Propri\'et\'es magn\'etiques des corps  a  diverses temp\'eratures}, Ann. de Chimie et Physique~\textbf{5} (1885) 289; reprinted in \textit{\OE uvres de Pierre Curie} (Gauthier-Villars, Paris, 1908) pp.~232--334.

\bibitem{Dembo-Zeitouni}
A.~Dembo and O.~Zeitouni, \textit{Large Deviations Techniques and Applications} (Springer Verlag, Inc., New York,~1998).

\bibitem{Deuschel-Stroock}
J.-D.~Deuschel and D.W.~Stroock, \textit{Large deviations}, Pure and Applied Mathematics, 137. Academic Press, Inc., Boston, MA, 1989.

\bibitem{Dimock-Hurd}
J.~Dimock and T.R.~Hurd, \textit{Sine-Gordon revisited}, Ann. Henri Poincar\'e \textbf{1} (2000), no. 3, 499--541.

\bibitem{Dobrushin}
R.~Dobrushin, \textit{The description of a random field by means of conditional  probabilities and conditions of its regularity}, Theor. Prob. Appl.~\textbf{13} (1968) 197Ð-224.

\bibitem{Dobrushin-Shlosman}
R.L.~Dobrushin and S.B.~Shlosman, \textit{Phases corresponding to minima of the local energy}, Selecta Math. Soviet.~\textbf{1} (1981), no.~4, 317--338.

\bibitem{DS1Bi}
R.L.~Dobrushin and S.~Shlosman, \textit{Absence of breakdown of continuous symmetry in two-dimensional models of statistical physics}, Commun. Math. Phys.~\textbf{42} (1975), no.~1, 31--40.

\bibitem{Dobrushin-Zahradnik}
R.L.~Dobrushin and M.~Zahradn\'{\i}k,
\textit{Phase diagrams for continuous-spin models: an extension of the  Pirogov-Sina\u{\i} theory},
In: R.L.~Dobrushin (ed.), \textit{Mathematical problems of statistical mechanics and dynamics},  pp.~1--123, Math. Appl. (Soviet Ser.), vol.~6,  Reidel, Dordrecht,  1986. 

\bibitem{Durrett}
R.~Durrett, \textit{Probability: Theory and Examples}, Second edition. Duxbury Press, Belmont, CA, 1996. 

\bibitem{Dyson}
F.J.~Dyson, \textit{General theory of spin-wave interactions}, Phys. Rev.~\textbf{102} (1956), no.~5, 1217--1230.

\bibitem{DLS}
F.J.~Dyson, E.H.~Lieb and B.~Simon, 
\textit{Phase transitions in quantum spin systems with isotropic and nonisotropic interactions}, J.~Statist. Phys.~\textbf{18} (1978) 335--383.

\bibitem{ES1Bi}
R.G.~Edwards and A.D.~Sokal,
\textit{Generalization of the
Fortuin-Kasteleyn-Swen\-dsen-Wang representation and  Monte Carlo algorithm},  Phys.\ Rev.\ {D}
\textbf{38} (1988) 2009-2012.


\bibitem{Ellis}
R.S.~Ellis, \textit{Entropy, Large Deviations, and Statistical Mechanics}, Grundlehren der Mathematischen Wissenschaften, vol.~271 (Springer-Verlag, New York, 1985).

\bibitem{Senya_VE-I}
A.C.D.~van Enter and S.B.~Shlosman,   
\textit{First-order transitions for $n$-vector models in two and more dimensions: Rigorous proof}, Phys. Rev. Lett. ~\textbf{89} (2002) 285702.

\bibitem{vanEnter-Shlosman}
A.C.D.~van~Enter and S.B.~Shlosman,
\emph{Provable first-order transitions for nonlinear vector and gauge models with continuous symmetries},
Commun. Math. Phys.
\textbf{255} (2005) 21--32.

\bibitem{FK}
C.M.~Fortuin and P.W.~Kasteleyn, \textit{On the random cluster model. I.~Introduction and relation to other models}, Physica (Amsterdam)~\textbf{57} (1972) 536--564.

\bibitem{Freedman-Lovasz-Schrijver}
M.~Freedman, L.~Lov\'asz and A.~Schrijver, \textit{Reflection positivity, rank connectivity, and homomorphism of graphs}, J.~Amer. Math. Soc. \textbf{20} (2007), no. 1, 37--51.

\bibitem{Frohlich}
J.~Fr{\"o}hlich, \textit{On the triviality of $\lambda \varphi^{4}_{d}$ theories and the approach to the critical point in $d\raise2pt\hbox{$\underset{(-)}>$}4$ dimensions},  Nucl. Phys.~B~\textbf{200} (1982), no.~2, 281--296. 

\bibitem{FILS1}
J.~Fr{\"o}hlich, R.~Israel, E.H.~Lieb and B.~Simon,
\textit{Phase transitions and reflection positivity. I.~General theory and long-range lattice models}, Commun. Math. Phys.~\textbf{62} (1978), no.~1, 1--34.

\bibitem{FILS2}
J.~Fr{\"o}hlich, R.~Israel, E.H.~Lieb and B.~Simon,
\textit{Phase transitions and reflection positivity. II.~Lattice systems with short-range and Coulomb interactions}, J.~Statist. Phys.~\textbf{22} (1980), no.~3, 297--347.

\bibitem{FL}
J.~Fr{\"o}hlich and E.H.~Lieb,
\textit{Phase transitions in anisotropic lattice spin systems}, Commun. Math. Phys.~\textbf{60}  (1978),  no.~3, 233--267.

\bibitem{PF1Bi}
J.~Fr{\"o}hlich and C.-E.~Pfister, \textit{On the absence of spontaneous symmetry breaking and of crystalline ordering in two-dimensional systems}, Commun. Math. Phys.~\textbf{81} (1981), no.~2, 277--298.


\bibitem{FSS}
J.~Fr{\"o}hlich, B.~Simon and T.~Spencer, 
\textit{Infrared bounds, phase transitions and continuous symmetry breaking},
Commun. Math. Phys.~\textbf{50} (1976) 79--95.

\bibitem{Frohlich-Spencer}
J.~Fr\"ohlich and T.~Spencer, \textit{The Kosterlitz-Thouless transition in two-dimensional abelian spin systems and the Coulomb gas}, Commun. Math. Phys. \textbf{81} (1981), no. 4, 527--602.

\bibitem{Funaki} 
T.~Funaki, \emph{Stochastic Interface Models},
Lecture Notes for the International Probability School at Saint-Flour, 
Lecture Notes in Math., 1869, Springer, Berlin, 2005.

\bibitem{Funaki-Spohn} 
T. Funaki and H. Spohn,  \emph{Motion by mean curvature from the Ginzburg-Landau 
$\nabla \phi$ interface model}, Commun. Math. Phys.~\textbf{185} (1997) 1--36.

\bibitem{Galvin-Kahn}
D.~Galvin and J.~Kahn, \textit{On phase transition in the hard-core model on $\mathbb Z^d$}, 
Combin. Probab. Comput. \textbf{13} (2004), no. 2, 137--164.

\bibitem{Galvin-Randall}
D.~Galvin and D.~Randall, \textit{Torpid mixing of local markov chains on 3-Colorings of the discrete torus}, Proceedings of the Eighteenth Annual ACM--SIAM Symposium on Discrete Algorithms (SODA 2007), to appear.

\bibitem{deGennes}
P.G.~de~Gennes and J.~Prost, \textit{The Physics of Liquid
Crystals}, Oxford University Press, New York, 1993.

\bibitem{Georgii} 
H.-O.~Georgii, \textit{Gibbs Measures and
Phase Transitions}, de~Gruyter Studies in Mathematics, vol.~9
(Walter de Gruyter~\&~Co., Berlin, 1988).

\bibitem{Gobron-Merola}
T.~Gobron and I.~Merola, \textit{First-order phase transition in Potts models with finite-range interactions}, J.~Statist. Phys. \textbf{126} (2007), no. 3, 507--583.

\bibitem{Griffiths-Peierls}
R.B.~Griffiths, \textit{Peierls proof of spontaneous magnetization in a two-dimensional Ising ferromagnet}, Phys. Rev. (2) \textbf{136} (1964) A437--A439.

\bibitem{Griffiths-JMP}
R.B.~Griffiths, \textit{Correlations in Ising ferromagnets}, J.~Math. Phys. \textbf{8} (1967), no.~3, 478--483. 

\bibitem{Grimmett}
G.~Grimmett, \textit{The Random-Cluster Model}, Grundlehren der Mathematischen Wissenschaften, vol.~333, Springer-Verlag, Berlin, 2006

\bibitem{Heilmann-Lieb}
O.J.~Heilmann and E.H.~Lieb, \textit{Lattice models for liquid crystals}, J.~Statist. Phys. \textbf{20} (1979), no.~6, 679--693. 

\bibitem{Henley}
C.L.~Henley, \textit{Ordering due to disorder in a frustrated vector antiferromagnet}, Phys. Rev. Lett.~\textbf{62} (1989) 2056--2059.

\bibitem{denHollander}
F.~den Hollander, \textit{Large deviations}, Fields Institute Monographs, 14. American Mathematical Society, Providence, RI, 2000.

\bibitem{ISV}
D.~Ioffe, S.~Shlosman and Y.~Velenik, \textit{2D models of statistical physics with continuous symmetry: The case of  singular interactions},  Commun. Math. Phys.~\textbf{226}  (2002),  no.~2, 433--454.

\bibitem{Israel}
R.B.~Israel, \textit{Convexity in the theory of lattice gases}, With an introduction by Arthur S. Wightman. Princeton University Press, Princeton, N.J., 1979.

\bibitem{Kennedy}
T.~Kennedy, \textit{Long range order in the anisotropic quantum ferromagnetic Heisenberg model}, {Commun. Math. Phys.} \textbf{100} (1985), no.~3, 447--462.

\bibitem{Kennedy-Lieb-Shastry}
T.~Kennedy, E.H.~Lieb and B.S.~Shastry, \textit{Existence of N\'eel order in some spin-$\frac12$ Heisenberg antiferromagnets}, J. Statist. Phys. \textbf{53} (1988), no. 5-6, 1019--1030,

\bibitem{KS1Bi}
H.~Kesten and R.~Schonmann, \textit{Behavior in large dimensions
of the Potts and Heisenberg models}, Rev.~Math. Phys.~\textbf{1}
(1990) 147-182.


\bibitem{Kosterlitz-Thouless}
M. Kosterlitz and D.J. Thouless, \textit{Ordering, metastability and phase transitions in two-dimensional systems}, J.~ Phys.~C: Solid State Phys., \textbf{6} (1973) 1181--1203.

\bibitem{KS-proceedings}
R.~Koteck\'y and S.B.~Shlosman, \textit{Existence of first-order transitions for Potts models}, In:~S.~Albeverio, Ph.~Combe, M.~Sirigue-Collins (eds.), Proc. of the International Workshop --- Stochastic Processes in Quantum Theory and Statistical Physics, Lecture Notes in Physics~\textbf{173}, 248--253, Sprin\-ger-Ver\-lag, Berlin-Heidelberg-New York, 1982.


\bibitem{Kotecky-Shlosman}
R.~Koteck\'y and S.B.~Shlosman, \textit{First-order phase
transitions in large entropy lattice models}, Commun.~Math.~Phys.
\textbf{83} (1982), no.~4, 493--515.

\bibitem{Krengel}
U.~Krengel, \textit{Ergodic theorems}, de Gruyter Studies in Mathematics, 6. Walter de Gruyter \& Co., Berlin, 1985.


\bibitem{Kuelske1}
C.~K\"ulske, \textit{The continuous spin random field model: Ferromagnetic ordering in $d \ge3$}, 
Rev. Math. Phys. \textbf{11} (1999), no.10, 1269--1314.

\bibitem{Kuelske2}
C.~K\"ulske, \textit{Stability for a continuous SOS-interface model in a ran\-dom\-ly per\-turb\-ed periodic potential},
WIAS Preprint no.~466 (1998); http://www.math.rug.nl/\lower2pt\hbox{\~{}}kuelske/publications.html

\bibitem{Lawler}
G.L.~Lawler, \textit{Intersections of random walks}, Probability and its Applications. Birkh\"auser Boston, Inc., Boston, MA, 1991.

\bibitem{Lieb-twotheorems}
E.H.~Lieb, \textit{Two theorems on the Hubbard model}, 
Phys. Rev. Lett. \textbf{62} (1989), no. 10, 1201--1204. Errata: Phys. Rev. Lett.~\textbf{62} (1989), no. 16, 1927.

\bibitem{Lieb-flux-phase}
E.H.~Lieb, \textit{Flux phase of the half-filled band}, Phys. Rev. Lett. \textbf{73} (1994) 2158-2161.

\bibitem{Macris}
N.~Macris, \textit{Periodic ground states in simple models of itinerant fermions interacting with classical fields},
The nature of crystalline states (Kudowa-Zdr\'oj, 1995). 
Phys.~A \textbf{232} (1996), no. 3-4, 648--656.

\bibitem{Macris-Lebowitz}
N.~Macris and J.~Lebowitz, \textit{Ground states and low-temperature phases of itinerant electrons interacting with classical fields: a review of rigorous results}, Quantum problems in condensed matter physics, J.~Math. Phys. \textbf{38} (1997), no. 4, 2084--210.

\bibitem{Macris-Nachtergaele}
N.~Macris and B.~Nachtergaele, \textit{On the flux phase conjecture at half-filling: an improved proof}, J. Statist. Phys. 85 (1996), no. 5-6, 745--761.

\bibitem{MW}
N.D.~Mermin and H.~Wagner, \textit{Absence of ferromagnetism or antiferromagnetism in one- or two-dimensional isotropic Heisenberg models}, Phys. Rev. Lett.~\textbf{17} (1966), no.~22, 1133--1136.


\bibitem{EPL-kompasy}
Z.~Nussinov, M.~Biskup, L.~Chayes and J.~van den Brink, \textit{Orbital order in classical models of transition-metal compounds}, Europhys. Lett.~\textbf{67} (2004), no.~6, 990--996.

\bibitem{OS}
K.~Osterwalder and R.~Schrader, \textit{Axioms for Euclidean Green's functions}, Commun. Math. Phys. \textbf{31} (1973), 83--112.

\bibitem{Patrasciou-Seiler}
A. Patrasciou and E.~Seiler, \textit{Percolation theory and the existence of a soft phase in 2D spin models}, Nucl.~Phys.~B (Proc. Suppl.) \textbf{30} (1993) 184--191.


\bibitem{Petersen}
K.~Petersen, \textit{Ergodic theory}, Cambridge Studies in Advanced Mathematics, 2. Cambridge University Press, Cambridge, 1989.

\bibitem{Pfister}
C.-E.~Pfister, \textit{On the symmetry of the Gibbs states in two-dimensional lattice systems}, Commun. Math. Phys.~\textbf{79} (1981), no.~2, 181--188.

\bibitem{Ruelle-redbook}
D.~Ruelle, \textit{Thermodynamic formalism. The mathematical structures of equilibrium statistical mechanics}, Second edition. Cambridge Mathematical Library. Cambridge University Press, Cambridge, 200

\bibitem{Ruelle1Bi}
D.~Ruelle, \textit{Statistical mechanics. Rigorous results},
Reprint of the 1989 edition, World Scientific Publishing Co.,
Inc., River Edge, NJ; Imperial College Press, London,~1999.

\bibitem{Sakai}
A.~Sakai, \textit{Lace expansion for the Ising model}, Commun. Math. Phys. \textbf{272} (2007), no. 2, 283--344.

\bibitem{Salmhofer-Seiler}
M. Salmhofer and E. Seiler, \textit{Proof of chiral symmetry breaking in strongly coupled lattice gauge theory}, Commun. Math. Phys.  \textbf{139} (1991), no. 2, 395--432. Errata: \textit{ibid} \textbf{146} (1992), no. 3, 637--638.

\bibitem{Sheffield}
S.~Sheffield, \textit{Random Surfaces}, Asterisque 2005, No. 304, 177pp.

\bibitem{Shender}
E.F.~Shender, \textit{Antiferromagnetic garnets with fluctuationally interacting sublattices},
Sov. Phys. JETP~\textbf{56} (1982) 178--184.

\bibitem{Senya1}
S.~Shlosman, \textit{Phase transitions for two-dimensional models with isotropic short range interactions and continuous symmetry}, Commun. Math. Phys.~\textbf{71} (1980) 207--212.

\bibitem{Senya2}
S.B.~Shlosman, \textit{The method of reflective positivity in the mathematical theory of phase  transitions of the first kind} (Russian), Uspekhi Mat. Nauk~\textbf{41}  (1986), no.~3(249), 69--111, 240.

\bibitem{Shlosman-Vignaud}
S.~Shlosman and Y.~Vignaud, \textit{Dobrushin interfaces via reflection positivity}, Commun. Math. Phys. 276 (2007), no. 3, 827--86.

\bibitem{Simon}
B.~Simon, \textit{The statistical mechanics of lattice gases},
Vol.~I., Princeton Series in Physics (Princeton University Press,
Princeton, NJ, 1993).

\bibitem{Sokal}
A.D.~Sokal, \textit{An alternate constructive approach to the $\varphi^4_3$ quantum field theory, and a possible destructive approach to $\varphi^4_4$},
Ann. Inst. H. Poincar\'e Sect.~A~\textbf{37} (1982), no.~4, 317--398.

\bibitem{Speer}
E.R.~Speer, \textit{Failure of reflection positivity in the quantum Heisenberg ferromagnet},
{Lett. Math. Phys.} \textbf{10} (1985), no.~1, 41--47.

\bibitem{Spitzer}
F.~Spitzer, \textit{Principles of random walks}, Second edition. Graduate Texts in Mathematics, Vol. 34. Springer-Verlag, New York-Heidelberg, 1976.

\bibitem{Tasaki}
H.~Tasaki, \textit{Ferromagnetism in the Hubbard model: a constructive approach}, 
Commun. Math. Phys. \textbf{242} (2003), no. 3, 445--472.

\bibitem{Tian}
G.-S.~Tian, \textit{Lieb's spin-reflection-positivity method and its applications to strongly correlated electron systems}, J.~Statist. Phys. \textbf{116} (2004), no. 1-4, 629--680.

\bibitem{Balint-web}
B.~T\'oth's website:\\\texttt{\small www.math.bme.hu/\lower2pt\hbox{\~{}}balint/oktatas/statisztikus\_fizika/jegyzet/}

\bibitem{Velenik}
Y.~Velenik, \textit{Localization and delocalization of random interfaces}, Probab. Surveys~\textbf{3} (2006) 112--169.

\bibitem{Weiss}
P. Weiss, \textit{L'hypoth\`ese du champ mol\'eculaire et la propri\'et\'e ferromagn\'etique}, 
J.~de Physique~\textbf{6} (1907) 661--689.

\bibitem{Wu}
F.Y.~Wu, \textit{The Potts model}, Rev. Modern Phys.~\textbf{54}
(1982) 235--268.

\bibitem{Zahradnik}
M.~Zahradn\'{\i}k, \textit{Contour methods and Pirogov-Sinai theory for continuous spin lattice  models}, In:~R.A.~Minlos, S.~Shlosman and Yu.M.~Suhov (eds.), \textit{On Dobrushin's way. From probability theory to statistical  physics},   pp.~197--220, Amer. Math. Soc. Transl. Ser.~2, vol.~198, Amer. Math. Soc., Providence, RI, 2000.


\end{thebibliography}
\end{document}